\documentclass[preprint,review,12pt]{elsarticle}

\usepackage{graphicx}
\usepackage{amssymb}
\usepackage{verbatim}
\usepackage{wasysym}
\usepackage{stackrel}

\newcommand{\be}{\begin{equation}}
\newcommand{\ee}{\end{equation}}

\title{A periodic representation of the interface for the
  volume of fluid method}

\author{Joris C.G. Verschaeve}

\address{
Department of Energy and Process Engineering,\\
Norwegian University of Science and Technology,\\
N-7491 Trondheim, Norway\\
joris.verschaeve@ntnu.no
}

\begin{document}

\begin{abstract}

We extend the volume of fluid method for the computation of two-phase flow to
a higher order accurate method in two dimensions. 
The interface reconstruction by the PLIC method 
is thereby replaced by a periodic interface reconstruction.
The advection step is reformulated 
and extended to higher order in order to account
for the present interface representation. 
This periodic interface reconstruction describes 
the interface in terms of higher order periodic B-splines.
Numerical tests verify that the theoretical order of 
convergence is indeed exhibited by the present method. 
\end{abstract}

\begin{keyword}
B-Splines \sep Two-phase Flow \sep VOF \sep High Order Accuracy
\end{keyword}
\maketitle

\section{Introduction}

Two-phase flow can be found in many industrial applications. 
A popular method for the computation of two-phase flow is
the volume of fluid method (VOF) \cite{ScardovelliZaleski1999,Benson2002}.\\
The volume fraction, the central object of the volume of fluid method, 
denotes the ratio of the volume (area in 2D) occupied by one phase in 
a cell of the computational domain to the cell volume (cell area in 2D). 
The volume of fluid method can be subdivided into two steps: the 
interface reconstruction step and the advection step. The interface
reconstruction step computes the interface position at time $ t $ using the 
volume fraction field at time $ t $. The advection step advects the volume 
fraction field from time $ t $ to time $ t + \Delta t $ using the 
reconstructed interface at time $ t $. The volume of fluid method 
has its origins in the works of \cite{DeBar1974} and \cite{NohWoodward1976}. 
Substantial improvement of the interface reconstruction has been achieved 
with the piecewise linear interface computation (PLIC) method by 
Youngs 
\cite{Youngs1982} in 1982. However, the resulting interface is 
approximated by piecewise straight lines which makes it necessary to 
estimate the curvature by additional approximation schemes, 
for example the height function method 
\cite{CumminsFrancoisKothe2005,
FrancoisCumminsDendyKotheSicilianWilliams2006,
FerdowsiBussmann2008,FrancoisSwartz2010}. 
In order to obtain a more smooth interface, Price {\em et al.} in 1998 \cite{PriceReaderRoweBugg1998} derived a method 
replacing the straight lines by parabolas.
Due to the fact 
that a numerical minimization has to be performed in each cell to find all the 
coefficients of the interface parabola, the method enjoys less popularity. More 
recently, in 2004, Lopez {\em et al.} \cite{LopezHernandezGomezFaura2004} 
used a parametric cubic spline interpolation
through the midpoints of the PLIC interface lines and obtained a smoother 
description of the interface. However, although cubic splines are known to 
interpolate a function with fourth order accuracy, their method inherits the 
second order accuracy of the PLIC method for the test cases presented in 
\cite{LopezHernandezGomezFaura2004} since it is based on the same 
approach. A further development of this interface reconstruction 
using splines to improve the interface obtained by the piecewise lines 
of the PLIC method has been presented in \cite{DiwakarDasSundararajan2009}
using quadratic splines. Both approaches are, however, based on the PLIC method
for reconstruction and advection and share therefore also the drawbacks of 
the PLIC method. Another approach replacing this time the PLIC method has been 
presented in \cite{Verschaeve2010Reading,Verschaeve2010Lisboa,VerschaevePart1},
where the interface is divided into segments and each segment is reconstructed
globally. This allowed for a more accurate description of the interface.
A drawback of this method is however the need to choose a division of the
interface into segments. \\
In the present discussion we shall modify the approach presented in 
\cite{Verschaeve2010Reading,Verschaeve2010Lisboa,VerschaevePart1} by deriving a 
periodic description of the interface separating two immiscible 
liquids in two dimensions. The present method is, as the method in
\cite{Verschaeve2010Reading,Verschaeve2010Lisboa,VerschaevePart1}, 
a global method opposed to the PLIC method which uses only local information. 
The interface in the present discussion
is represented indirectly by two functions depending on a periodic parameter.
The actual position or other quantities, such as the normal or
the local curvature at the interface, are then derived from these 
two functions. The advection step is adapted to the present
interface representation. \\
The present discussion is organized as follows: 
The present interface representation is derived in the next section,
section \ref{sec:derivation}. Periodic B-splines are used to approximate
the interface, cf. section \ref{sec:interpolation}. 
In section \ref{sec:advection}, 
the advection step is presented. The numerical verification is done 
in section \ref{sec:verification}. Finally, the present 
discussion is concluded in section \ref{sec:conclusions}.

\section{A periodic representation of the interface \label{sec:derivation}}

\begin{figure}[ht]
\centering
\includegraphics[width=0.4\linewidth]{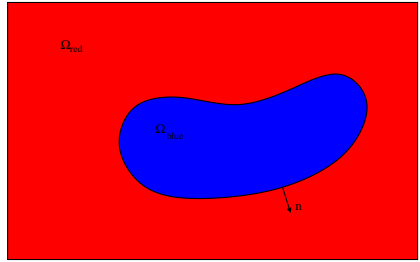}
\vskip-0.2cm
\caption{The two dimensional domain $ \Omega $ with the red fluid 
  occupying the region $ \Omega_{red} $ and the blue fluid occupying the region
  $ \Omega_{blue} $. The common boundary of $ \Omega_{red} $ 
  and $ \Omega_{blue}$ separates both fluids
  and is called the interface. 
  At each point of the interface a normal $ \vec{n} $ can be 
  defined. We define the normal to point into the red domain. 
  In this case where a blue drop is enclosed by red fluid this implies 
  that we transverse the interface of the drop in counter clockwise direction. 
  \label{fig:omega}}
\end{figure}

In the present discussion we treat the case of a two dimensional drop of
blue fluid enclosed in red fluid, cf. figure \ref{fig:omega}. The red fluid
occupies the domain $ \Omega_{red} $, whereas the blue fluid occupies
the domain $ \Omega_{blue} $. These two domains are 
separated by a common boundary, the interface $ I $. 
The central problem of the volume of fluid method is to compute
the temporal evolution of the interface $ I $
when subjecting the fluids to a velocity field $ \vec{u} $:
\be
\vec{u} = \left( { u_x(x,y,t) \atop u_y(x,y,t) } \right). 
\ee
Since we are dealing with incompressible fluids, the velocity is 
solenoidal. In the present discussion
we assume that the interface $ I $ can be described by a periodic line
$ l $. We exclude topological changes in 
the present discussion. In addition any 
third phase should not be present in order to avoid contact points. 
We also assume the line $ l $ to be sufficiently regular. 
As for polygons, cf. \cite{Arvo1991}, the area
of a domain $ \Omega_{blue} $, enclosed by a line $ l $,
can be computed by means of a function $ \vec{F} $:
\be
\vec{F}(\vec{x}) = \frac{1}{2} \vec{x}, \label{eq:F}
\ee
where $ \vec{x} $ is the position vector of a point. 
The divergence of (\ref{eq:F}) is unity, as can be verified straightforwardly. 
Having now a periodic parametrization of the line $ l $:
\be
l : \vec{x}{(s)} = \left( \begin{array}{c} x(s) \\ y(s) \end{array} \right), 
\quad s \in [0,2\pi)
\ee
the area $ V $ of $ \Omega_{blue} $ can be expressed by:
\be
V = 
\int \int \limits_{\Omega_{blue}} \, dx dy =
\int \int \limits_{\Omega_{blue}} \nabla \cdot \vec{F} \, dx dy =
\int \limits_{0}^{2\pi} \vec{F} \cdot \vec{n} \, ds' =
\frac{1}{2} \int \limits_{0}^{2\pi}  x(s')y'(s') - x'(s')y(s') \, ds',
\ee
where the periodicity of the line has, without loss of generality, 
been chosen to be $ 2 \pi $. It is, in addition, implied that the 
tangential on $ l $ points in counter clockwise direction.
We now define two functions $ \alpha(s) $, resp. $ \beta(s) $ by:
\begin{eqnarray}
\alpha(s) & := & \frac{1}{2} \int \limits_{s_0}^s x(s')y'(s') 
- x'(s')y(s') \, ds',
\label{eq:alpha} \\
\beta(s) & := & x(s)y(s). \label{eq:beta}
\end{eqnarray}
The derivatives of $ \alpha(s) $, resp. $ \beta(s) $ are then given by:
\begin{eqnarray}
\alpha'(s) & = & \frac{1}{2} \left( x(s)y'(s) - x'(s)y(s) \right) \\
\beta'(s) & = & x'(s)y(s) + x(s)y'(s).
\end{eqnarray}
Since the position $ \vec{x}(s) $ of a point on 
the interface is periodic in $ s $,
we conclude that the derivatives $ \alpha' $, resp. $ \beta' $ are
also periodic in $ s $. 
The position $ (x(s),y(s)) $ of a point on the interface
on the other hand can then be recovered by the following expressions:
\begin{eqnarray}
\frac{x'}{x} & = & \frac{\frac{1}{2} \beta' - \alpha' }{\beta} = : a(s),
\label{eq:a} \\
\frac{y'}{y} & = & \frac{\frac{1}{2} \beta' + \alpha' }{\beta} = : b(s).
\label{eq:b}
\end{eqnarray}
Integrating equations (\ref{eq:a}) and (\ref{eq:b}) with respect to $ s $
gives us then the final result:
\begin{eqnarray}
x(s) & = & x_0 \exp \int_{s_0}^s a(s') \, ds', \label{eq:a2} \\
y(s) & = & y_0 \exp \int_{s_0}^s b(s') \, ds', \label{eq:b2}
\end{eqnarray}
where $ (x_0,y_0) $ is the position of the interface for $ s = s_0 $. 
In order for equations (\ref{eq:a}) and (\ref{eq:b}) to be well defined we
have to choose a coordinate system having the region $ \Omega_{red} $ in 
the positive quadrant sufficiently far from the origin.
The area $ V $ included by the interface is then given by:
\be
V = \alpha(2\pi). 
\ee
The strategy of the present method is to represent the interface of the
drop by the functions $ \alpha $, resp. $ \beta $ and to obtain 
the position of the interface by formulae \ref{eq:a2}, resp. \ref{eq:b2}.
A normal $ \vec{n} $ on the interface is then given by:
\be
\vec{n} = \left( y'(s) \atop -x'(s) \right), 
\ee
and the curvature $ \kappa $ can be found via
\be
\kappa = \frac{x'y''-y'x''} {\left( x'^2 + y'^2 \right)^{\frac{3}{2}}}.
\ee
%the arc length $ l(s) $ of the interface is given by:
%\be
%l(s) = \int_{0}^s \sqrt{ x'(s^*)^2 + y'(s^*)^2 } \, ds^*.
%\ee

\section{Interpolation by periodic B-Splines \label{sec:interpolation} }

As mentioned above, instead of representing the 
interface position directly by B-splines, 
as for instance done in \cite{YeShyyChung2001} in the 
framework of front-tracking methods or in \cite{LopezHernandezGomezFaura2004}
for the volume of fluid method, we represent the interface
by the two functions $ \alpha $, resp. $ \beta $, equations (\ref{eq:alpha})
resp. (\ref{eq:beta}), defined on the interval $ [0,2\pi] $. 
We use a uniform discretization of the interval
$ [0,2\pi] $, meaning that we choose $ N+1 $ knots $ s_i \in [0,2\pi] $:
\be
s_i = \frac{2\pi i}{N}, \quad i = 0,\ldots,N, \label{eq:discretization}
\ee
dividing the interval $ [0,2\pi] $ into $ N $ 
sections of equal length. Periodic B-splines can actually
handle more flexible discretizations, which could be used to distribute 
points to regions of interest. However, in the
present discussion we restrict us to the uniform case. 
Having now  the knots $ s_i $, $ i = 0,\ldots, N $, 
the periodic basis spline $ B^P_i $ of order $ P $ is
obtained recursively by, see for instance \cite{MiculaMicula1999,Phillips2003}:
\begin{eqnarray}
B^P_i(s) & = & \left( \frac{ s-s_i}{s_{i+P}-s_i} \right) B^{P-1}_i(s) +
\left( \frac{ s_{i+P+1} - s }{s_{i+P+1}-s_{i+1} } \right) B^{P-1}_{i+1}(s) \\
B^0_i(s) & = & \left\{ \begin{array}{cl} 1, & s_i < x \le s_{i+1}, \\
  0, & \mbox{otherwise}. \end{array} \right. 
\end{eqnarray}
The basis spline $ B^P_i $ has finite support $ [s_i,s_{i+P+1}] $. 
Therefore representing the function $ f (s)$ to interpolate
as a linear combination $ R(s) $ of the periodic basis splines $ B^P_i $:
\be
f(s) \approx R(s) = 
\sum \limits_{i=-P}^{N-P-1}c_i B_i^P(s), \label{eq:interpolation}
\ee
leads to a cyclic banddiagonal system with bandwidth $ P $ for
the unknown coefficients $ c_i $. In the present discussion we will 
only use odd order B-splines. The function values $ f_i $ for interpolation
are then taken at the knots $ s_i $ \cite{Phillips2003}:
\be
f_i = f(s_i). 
\ee
For the resulting interpolation we have the following bound, see for 
instance \cite{MiculaMicula1999}. 
If $ f \in \mathcal{C}^{P+1}([0,2\pi]) $ and if 
$ f $ and the interpolating spline $ R $ of order $ P $ 
are periodic on $ [0,2\pi] $,
the following bound holds:
\be
|| f-R ||_{L^2} \le C^2(P) h^{P+1} ||f||_{H^{P+1}}, \label{eq:bound}
\ee
where $ h = \max_{1\le i \le N-1} \left( s_i - s_{i-1} \right) $ and the
constant $ C $ is given by:
\be
C(P) = 2^{-\frac{3}{4} \left( P+1 \right) - 1 } \left( \frac{P+1}{2} + 2 \right)!.
\ee
This implies that if choosing B-splines of order $P $ the interpolation
will have an order of accuracy $ P + 1 $ with respect to the grid spacing. 
The cyclic banddiagonal system resulting from (\ref{eq:interpolation})
can be solved efficiently by means of a
banddiagonal solver in combination with the Woodbury formula 
\cite{PressTeukolskyVetterlingFlannery2002}. \\

The function $ \beta $, equation (\ref{eq:beta}), is periodic and can thus
directly be interpolated by periodic B-splines. However, the 
function $ \alpha $ is not periodic but takes different
values at the right and left boundary of the interval $ [0,2\pi] $:
\be
\alpha(0) = 0 \quad \alpha(2\pi) = V,
\ee
where $ V $ is the area of the drop. 
The derivative $ \alpha' $ is, however, periodic. 
In order to use periodic B-splines to interpolate the function $ \alpha $,
we define a periodic function $ \alpha^* $ by:
\be
\alpha^*(s) = \alpha(s) - \frac{V}{2\pi}s,
\ee
which is then interpolated using periodic B-splines. Once we have 
an approximation to the functions $ \alpha $, resp. $ \beta $, equations
(\ref{eq:alpha}) resp. (\ref{eq:beta}), we evaluate the integrals
\begin{eqnarray}
A(s_0,s) & := & \int \limits_{s_0}^s a(s')\,ds'  \label{eq:integralA}\\
B(s_0,s) & := & \int \limits_{s_0}^s b(s')\,ds', \label{eq:integralB}
\end{eqnarray}
from equations (\ref{eq:a}), resp. (\ref{eq:b}) by Gaussian quadrature 
\cite{QuarteroniSaccoSaleri2000}, 
in order to compute the position by means of 
equations (\ref{eq:a2}), resp. (\ref{eq:b2}). 
This will, however, introduce additional numerical error. 
A consequence of this is that the integrals in 
equations (\ref{eq:integralA}), resp. (\ref{eq:integralB}),
might numerically not evaluate to zero for $ s_0 = 0 $ and $ s = 2\pi $. 
Therefore we determine first the total quadrature error $ \epsilon_{a} $, 
resp. $ \epsilon_{b} $ by
\begin{eqnarray}
\epsilon_{a} & = & \sum \limits_{i=0}^{N-1} q(A,s_i,s_{i+1}), 
\label{eq:epsilonA} \\
\epsilon_{b} & = & \sum \limits_{i=0}^{N-1} q(B,s_i,s_{i+1}), 
\label{eq:epsilonB} \\
\end{eqnarray}
where the symbol $ q(A,s_i,s_{i+1}) $ means taking the
numerical quadrature of the integral $ A $, equation (\ref{eq:integralA}),
from $ s_i $ to $ s_{i+1} $. 
Since B-splines are discontinuous in the $ P^{th} $ derivative
across the knots $ s_i $, we perform a Gaussian quadrature on each 
subinterval $ [s_i,s_{i+1}] $. 
In order to assure that the position, given by 
equations (\ref{eq:a2}), resp. (\ref{eq:b2}) is itself 
a periodic function of $ s $ we replace the argument $ a $, resp. $ b $ 
of the integrals in equations (\ref{eq:integralA}), resp. (\ref{eq:integralB})
by $ a^* $ and $ b^* $ defined the following way:
\begin{eqnarray}
a^*(s) &=& a(s) - \frac{\epsilon_a}{2\pi}, \\
b^*(s) &=& b(s) - \frac{\epsilon_b}{2\pi}. \\
\end{eqnarray}

\section{Advection step \label{sec:advection}}

\begin{figure}[ht]
\centering
\includegraphics[height=5.5cm]{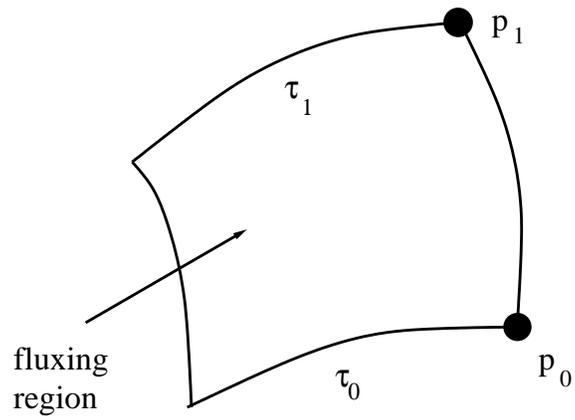}
\caption{The fluxing region is the set of points fluxed through the line 
with end points $ p_0 $, resp. $ p_1 $,
from time $ t $ to $ t + \Delta t $. It is bounded by the trajectories
$ \tau_0 $, resp. $ \tau_1 $ of the points $ p_0 $, resp. $ p_1 $ when tracing
them back in time, the line from $ p_0 $ to $ p_1 $ and the image of this line
when tracing the line back in time. }
\label{fig:flux}
\end{figure}

\begin{figure}[ht]
\centering
\includegraphics[height=5.5cm]{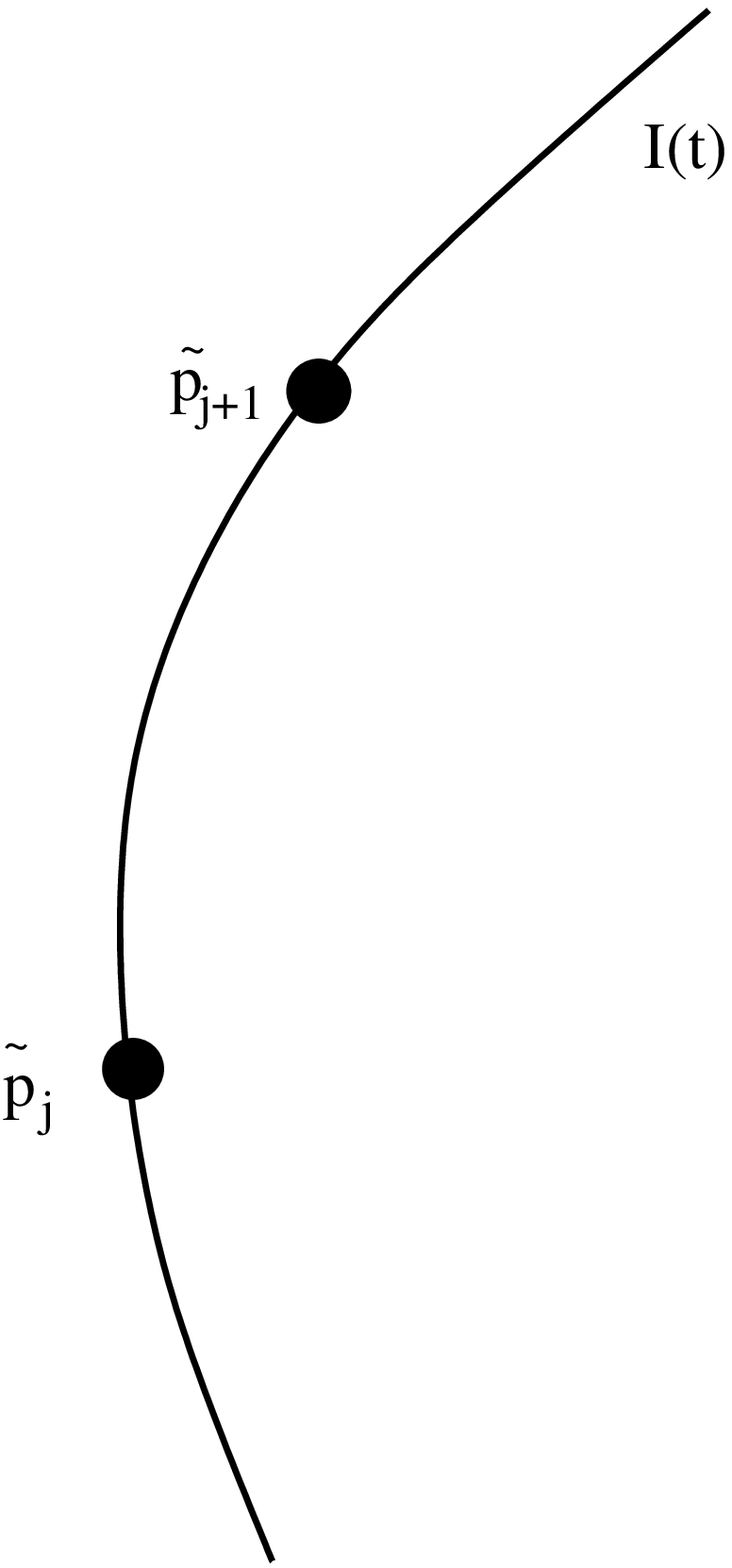}\\
\includegraphics[height=5.5cm]{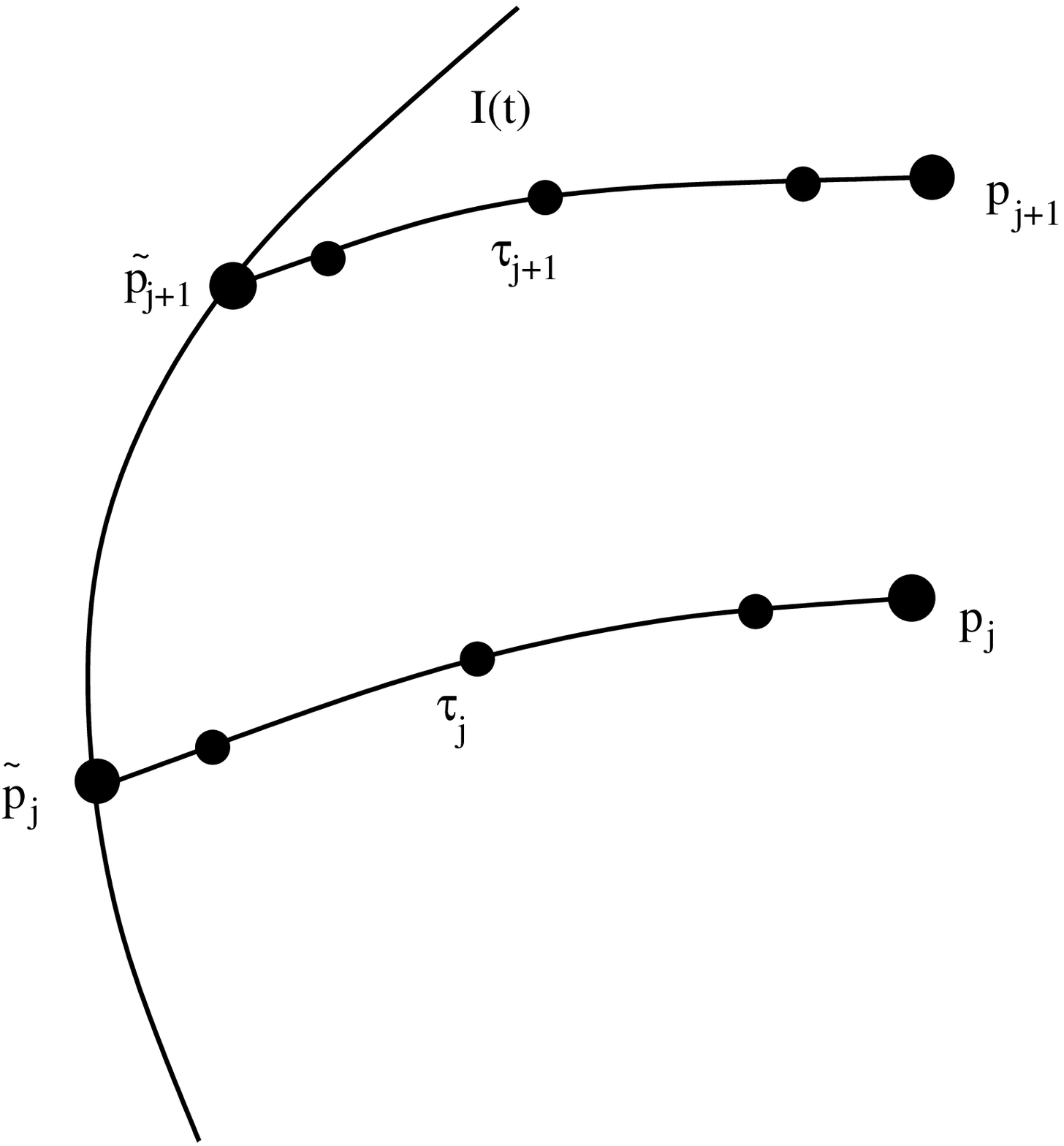}\\
\includegraphics[height=5.5cm]{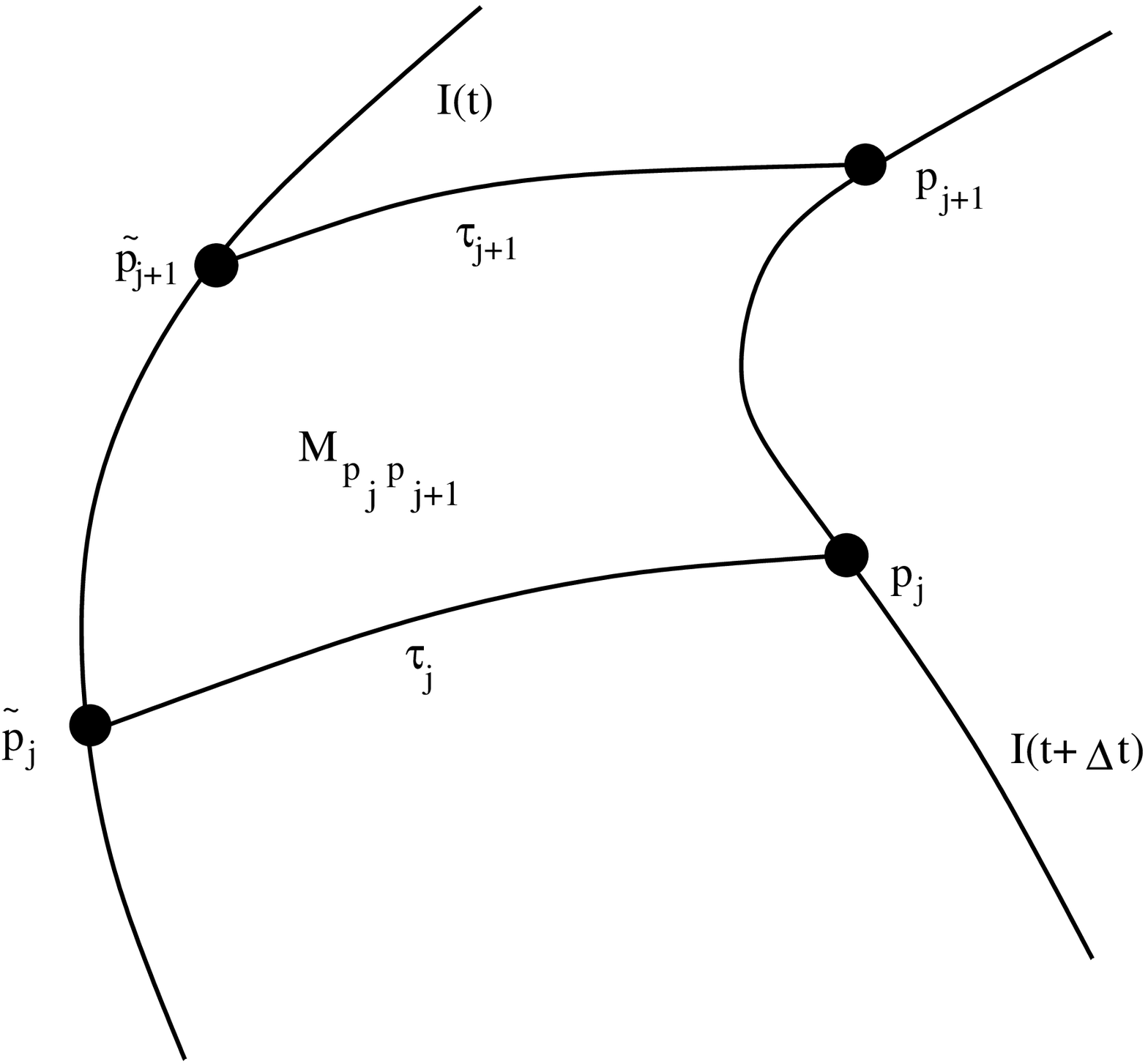}
\vskip-0.2cm
\caption{Sketch of the advection step. TOP: 
Going along the interface $ I(t) $ at time $ t $ 
we sample a sequence of points $ \tilde{p}_j $. 
MIDDLE: These points are traced forward in time along their trajectories
$ \tau_j $ to their positions $ p_j $ at time $ t +\Delta t$. At the
Gauss Labatto Legendre nodes of the interval $ [t, t +\Delta t ] $,
we record the positions of the point $ p_j $ which are then 
used to find an interpolation
approximating the trajectory $ \tau_j $. BOTTOM: 
The flux $ M_{p_j,p_{j+1}} $ can be interpreted as the signed area of the
region bounded by the interface $ I(t) $ at time $ t $ the interface
$ I(t+\Delta t ) $ at time $ t + \Delta t $ and the trajectories 
$ \tau_j $ and $ \tau_{j+1} $.}
\label{fig:advection}
\end{figure}

\begin{figure}[ht]
\centering
\includegraphics[height=0.4\linewidth]{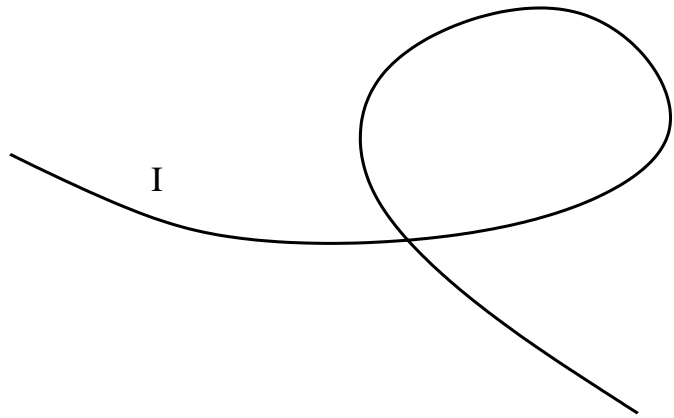}
\vskip-0.2cm
\caption{Self intersection of the interface. Errors can lead to 
self intersection of the interface. }
\label{fig:selfintersection}
\end{figure}

Once we are given an interface $ I(t) $ at time $ t $ represented
by the two functions $ \alpha_t $ and $ \beta_t $, equations (\ref{eq:alpha}),
resp. (\ref{eq:beta}), we need to formulate an advection scheme which 
allows to compute the interface $ I(t+ \Delta t ) $ at time $ t + \Delta t$.
Before going over to the actual derivation
we introduce the notions of flux and fluxing regions. \\
In the present discussion we assume that the flux $ M_{p_0,p_1} $
through a section with end points $ p_0 = ( x_0,y_0) $, 
resp. $ p_1 = (x_1,y_1) $ can be computed for arbitrary points
$ p_0 $ and $ p_1 $. The
flux $ M_{p_0,p_1} $ is given by 
\be
M_{p_0,p_1} = \int \limits_{t}^{t+\Delta t} Q_{p_0,p_1}(t') \, dt' 
= \int \limits_{t}^{t+\Delta t} \psi(x_1,y_1,t') - \psi(x_0,y_0,t') \, dt',
\ee
where $ Q $ is the volume flow and $ \psi $ the stream function.
In the present discussion the flux $ M_{p_0,p_1} $ will be computed
analytically, since we are given the analytical stream function $ \psi $
for the benchmark tests in section \ref{sec:verification}. The flux
$ M_{p_0,p_1} $ has a geometrical interpretation, cf. figure 
\ref{fig:flux}. It can be seen as the signed area of the region of
points passing through a line with end points $ p_0 $, resp. $ p_1 $
from $ t $ to $ t + \Delta t $, the fluxing region. 
This region is bounded by the line
from $ p_0 $ to $ p_1 $, its image at $ t $, when tracing the line
back from $ t + \Delta t $ to $ t $ and the trajectories $ \tau_0 $, 
resp. $ \tau_1 $
of the points $ p_0 $, resp. $ p_1 $. In order to trace
a point $ p = \vec{x}_0^T $ from $ t $ to $ t+\Delta t $ we have to 
solve the following differential equation:
\be
\frac{ d \vec{x}}{d t} = \vec{u}(\vec{x},t), \label{eq:RK}
\ee
with initial condition $ \vec{x}_0 $,
where $ \vec{u} = \left(-\partial_y \psi, \partial_x \psi \right)^T $ is
the velocity field, which is for the present benchmark tests,
cf. section \ref{sec:verification}, given
analytically. We solve equation (\ref{eq:RK}) by the classical 
four stage Runge-Kutta method. For the present advection scheme it
is necessary to approximate the trajectory $ \tau $ 
from time $ t $ to $ t + \Delta t $ of a point $ p = \vec{x}^T_0 $. 
This is done by 
means of Lagrange polynomials on the Gauss Labatto Legendre (GLL) nodes,
cf. for instance \cite{QuarteroniSaccoSaleri2000}, 
on the interval $ [t,t+\Delta t] $.  
If $ n \geq 2 $ is the number of nodes chosen, the 
interval $ [ t , t +\Delta t ] $ is divided
by the $ n $ GLL nodes into $ n-1 $ 
sections $ \delta t_i $, $ i = 1,\ldots,n-1$. 
Solving equation (\ref{eq:RK}) successively
for each point in time $ t_j = t + \sum_{i = 1}^j \delta t_i $, 
$ j = 0,\ldots,n-1$ will give us a set of interpolation points:
\be
(t_j,\vec{x}(t_j)=(x(t_j),y(t_j))^T ),
\ee
where $ \vec{x}(t_0) = \vec{x}_0 $. The trajectory $ \tau $ can 
then be approximated by:
\be
\tau : \vec{x}(t) = \sum \limits_{j=0}^{n-1} \vec{x}({t_j}) L_j(t),
\ee
where $ L_j(t) $ is the $ j^{th} $ Lagrange polynomial. \\
The advection step is sketched schematically in figure \ref{fig:advection}. 
Having the interface $ I(t) $ at time $ t $ represented by the
functions $ \alpha_t $, resp. $ \beta_t $, equations (\ref{eq:alpha}), resp. 
(\ref{eq:beta}), we choose a sequence of points on the interface $ I(t) $,
as depicted in figure \ref{fig:advection}. Several criteria might be
possible according to which the points might be chosen 
\cite{BjontegaardRonquistTrasdahl2010}. 
However, in the present discussion we take the points $ p_j $ at the 
parameter values $ s_j $, the nodes chosen for the discretization, equation 
(\ref{eq:discretization}). The points are computed by equations (\ref{eq:a2}), 
resp. (\ref{eq:b2}):
\be
p_j = \vec{x}^T_j = 
\vec{x}^T(s_j) = \left( x(s_j) , y(s_j) \right) ,\quad j = 0, \ldots, N-1.
\ee
It is known that during simulation the points on the interface 
can cluster in regions of the interface if always the same points on the 
interface are traced forward \cite{BjontegaardRonquist2009}. However, in 
the present discussion we will not treat this problem but instead focus
on the general method itself. \\
Now that we have chosen a sequence of points $ p_j(t) $ on the interface
at time $ t $,
we trace $ p_j(t) $ forward in time from $ t $ to $ t + \Delta t $ by  
solving (\ref{eq:RK}), as depicted in figure \ref{fig:advection}, giving us
the point $ p_j(t+\Delta t) $. In the following the point
$ p_j(t) $ at time $ t $ will be written with a tilde $ \tilde{p}_j $ to 
indicate that this quantity is at time $ t $. Its image 
$ p_j(t+\Delta t) $ at time $ t + \Delta t $ will 
be written without tilde $ p_j$. 
Focusing now on two consecutive points
$ p_j $ and $ p_{j+1} $ at time $ t + \Delta $, 
we know that the area bounded by the interface $I(t)$ at time $ t $, 
by the trajectories $ \tau_j $ and $ \tau_{j+1} $ of the points $ p_j $, resp.
$ p_{j+1} $, and the interface $ I(t+\Delta t ) $ at time $ t + \Delta t $
must equal $ M_{p_jp_{j+1}} $, as depicted in figure \ref{fig:advection}. 
The flux $ M_{p_j,p_{j+1}} $ can thus be written as a sum of integrals 
along the bounding lines of the fluxing region:
\be
M_{p_j,p_{j+1}} = -S_{I_{t,j}} + S_{\tau_j} - S_{\tau_{j+1}} + S_{I_{t+\Delta t,j}},
\label{eq:balance}
\ee
where $ S_{I_{t,j}} $ is the 
integral along the interface $ I(t) $ from $ s_j $ to $ s_{j+1} $,
\begin{eqnarray}
S_{I_{t,j}} & = & \int \limits_{s_j}^{s_{j+1}} \vec{F}(s') \cdot \vec{n}(s') \, ds' \\
& = & \alpha_t(s_{j+1}) - \alpha_t(s_j), 
\end{eqnarray}
since we are representing the interface $ I(t) $ by means of $ \alpha_t $ and
$ \beta_t $, cf. equation (\ref{eq:alpha}), resp. (\ref{eq:beta}). 
The integral $ S_{\tau} $ on a trajectory $ \tau $ is given by
\be
S_{\tau} = \int \limits_{t}^{t + \Delta t} \vec{F}(t') \cdot \vec{n}(t') \, dt'
= \frac{1}{2} \sum_{k=0}^n\sum_{m=0}^n x_ky_m \int 
\limits_{t}^{t + \Delta t}L_k(t')L_m'(t') - L_k'(t')L_m(t') \, dt'.
\label{eq:Stau}
\ee
Since $ L_kL_m' $ is a polynomial of order $ 2(n-1)-1 $ in $ t $, 
Gaussian quadrature can be used to compute the integral 
(\ref{eq:Stau}) exactly.
Finally the integral $ S_{I_{t+\Delta t,j}}$ on the 
interface at time $ t + \Delta t$ can be written as
\begin{eqnarray}
S_{I_{t+\Delta t,j}}& = & \int \limits_{s_j}^{s_{j+1}} \vec{F}(s') \cdot \vec{n}(s') \, ds' \\
& = & \alpha_{t+\Delta t} (s_{j+1}) - \alpha_{t+\Delta t}(s_j), 
\end{eqnarray}
where $ \alpha_{t+\Delta t} $ is the unknown function $ \alpha $ representing
the interface at time $ t + \Delta t $. We can solve equation 
(\ref{eq:balance}) for $ \alpha_{t+\Delta t}(s_{j+1}) $:
\begin{eqnarray}
\alpha_{t+\Delta t}(s_{j+1}) & = & M_{p_j,p_{j+1}} + \alpha_t(s_{j+1}) - 
\alpha_t(s_j) - S_{\tau_j} + S_{\tau_{j+1}} + \alpha_{t+\Delta t} (s_{j})\\
& = & \alpha_{t+\Delta t}(s_0 ) + \sum \limits_{k=0}^{j} M_{p_k,p_{k+1}} 
+ \alpha_t(s_{k+1}) - \alpha_t(s_{k}) - S_{\tau_k} + S_{\tau_{k+1}} \\
& = & \alpha_t(s_{j+1}) + S_{\tau_{j+1}} - S_{\tau_0} 
+ \sum \limits_{k=0}^{j} M_{p_k,p_{k+1}}, 
\end{eqnarray}
since we have chosen $ \alpha_t(s_0) = \alpha_{t+1} (s_0 ) = 0 $. 
A kind of similar idea is used for surface marker particles 
in order to correct for area loss during advection, cf. 
\cite{YeShyyChung2001}. In the present method it is, however, not used
as a correction but as the principle behind the advection step. \\
The interpolation points
\be
(s_{j} , \alpha_{t+\Delta t}(s_{j}) ), \quad \mbox{resp. }
(s_{j} , x_j y_j ), \quad j = 0,\ldots,N, 
\ee
are then interpolated as mentioned in section \ref{sec:interpolation}
to give an interpolant of $ \alpha $ resp. $ \beta $ for the interface 
$ I(t + \Delta t) $ at time $ t + \Delta t $. 
The area of the drop $ V $ is conserved,
since $ \alpha(s_N = 2\pi) = V $ for all time steps. However, this
conservation property should be understood in a less strict sense,
since the quadrature errors $ \epsilon_a $, resp. $ \epsilon_b $
in equations (\ref{eq:epsilonA}), resp. (\ref{eq:epsilonB}) will lead to 
the fact that the area bounded by the actual interface given by the
points computed by equations (\ref{eq:a2}) and (\ref{eq:b2}) can be different
from $ V $. In addition, errors can lead to the development of self 
intersections of the interface, cf. figure (\ref{fig:selfintersection}). 
Area conservation should rather be understood 
as the underlying principle of the method which 
is implemented through the function $ \alpha $ which, 
together with $ \beta $, can be seen as a kind of 
generating function for the position. 

\clearpage
\section{Numerical verification \label{sec:verification}}

The numerical verification of the present third order volume of fluid 
method is done by three classical benchmark tests, the reversed single vortex
test by Rider and Kothe \cite{RiderKothe1998}, Zalesak's slotted disk
test \cite{Zalesak1979} and the deformation field test 
\cite{Smolarkiewicz1982}. However, before going over to the 
numerical verification, we have a glance at the definition 
of the numerical error.

\subsection{Numerical Error}

\begin{figure}
\centering
\scalebox{1}{
\rotatebox{0}{\includegraphics[width=0.5\linewidth]
{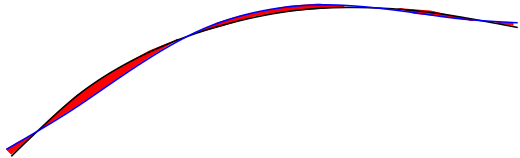}}
}
\caption{The error norm $ E_1 $ measures the total area (marked
by red color) included between the numerical interface (marked by
a blue line) and the exact interface (marked by a black line). 
}
\label{fig:errorNorm1}
\end{figure}

The error norm $ E $ in the present discussion measures
the area difference between the numerical interface and the exact interface
as depicted in figure \ref{fig:errorNorm1}. 
The error is computed using the position of the interface computed by 
means of equations (\ref{eq:a2}), resp. (\ref{eq:b2}) and not by means of
the function $ \alpha $, equation (\ref{eq:alpha}). The integration between
the numerical solution and the exact interface in order to compute the
area difference is done by means of Gaussian quadrature, where we ensured
that the quadrature error is negligible compared to the numerical error 
of the method. The order of convergence $ O $ between
two resolutions $ n $ and $ 2n$, is computed using $ E $:
\be
O = \frac{ \mathrm{ln} \left( E(n)/E(2n) \right)}{\mathrm{ln}\, 2 }. 
\ee

\subsection{Numerical verification part 1}

\begin{figure}
\centering
\rotatebox{270}{\includegraphics[height=7cm]{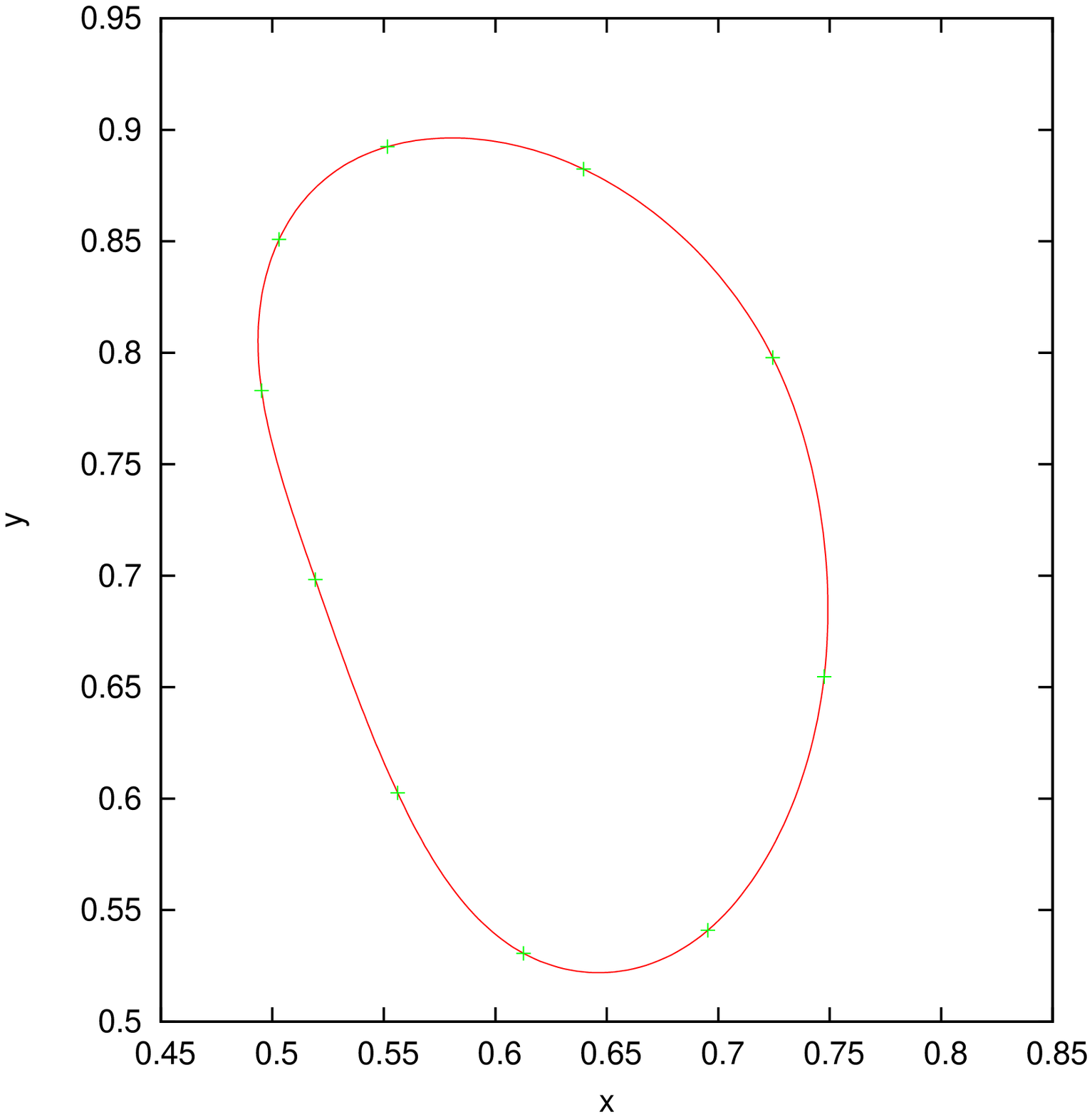}} \nolinebreak
\rotatebox{270}{\includegraphics[height=7cm]{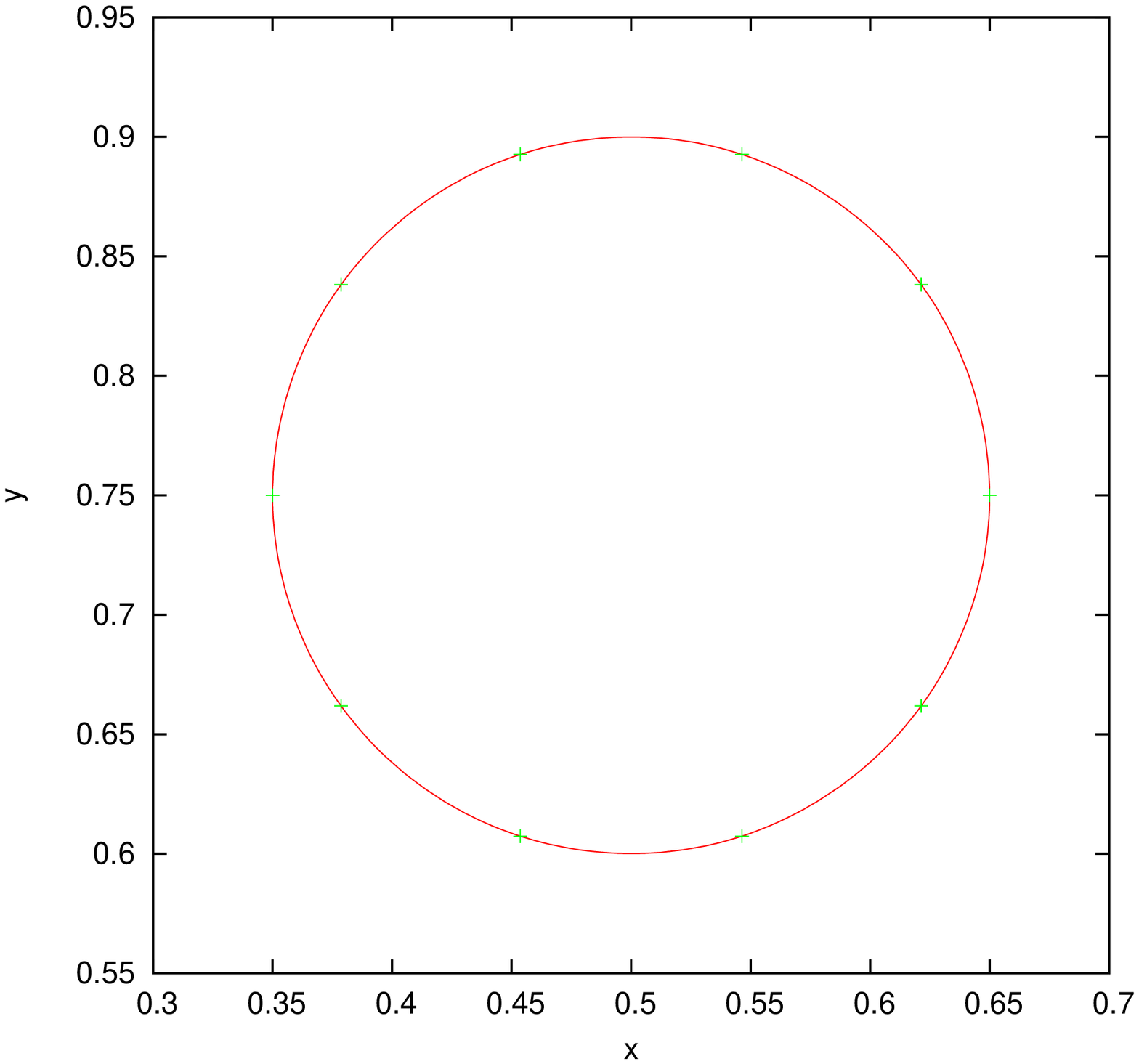}}
\caption{Result of the Rider-Kothe single vortex benchmark test
for $ T = 1/2 $. The resolution was $ N = 10 $ and $ P = 3$. 
LEFT: Position of the interface at $ t = T/2 $. 
RIGHT: Position of the interface at $ t = T $. }
\label{fig:rkT05}
\end{figure}

\begin{figure}
\centering
\rotatebox{270}{\includegraphics[height=10cm]{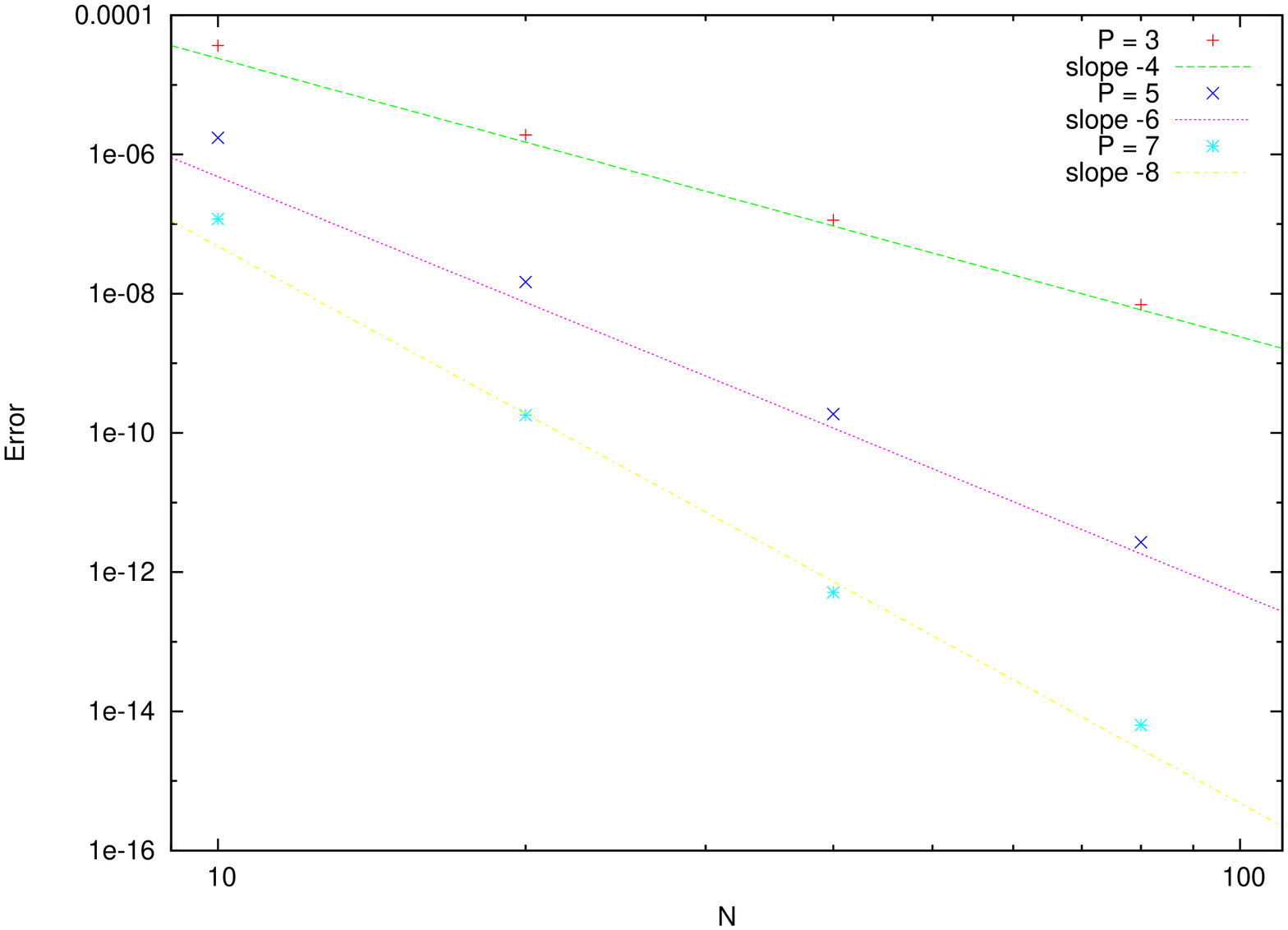}}
\caption{Error decrease for the Rider-Kothe single vortex benchmark test
for $ T = 1/2 $, for increasing resolution $ N $ using B-splines
of different order $ P $.}
\label{fig:rkT05error}
\end{figure}

\begin{figure}
\centering
\rotatebox{270}{\includegraphics[height=7cm]{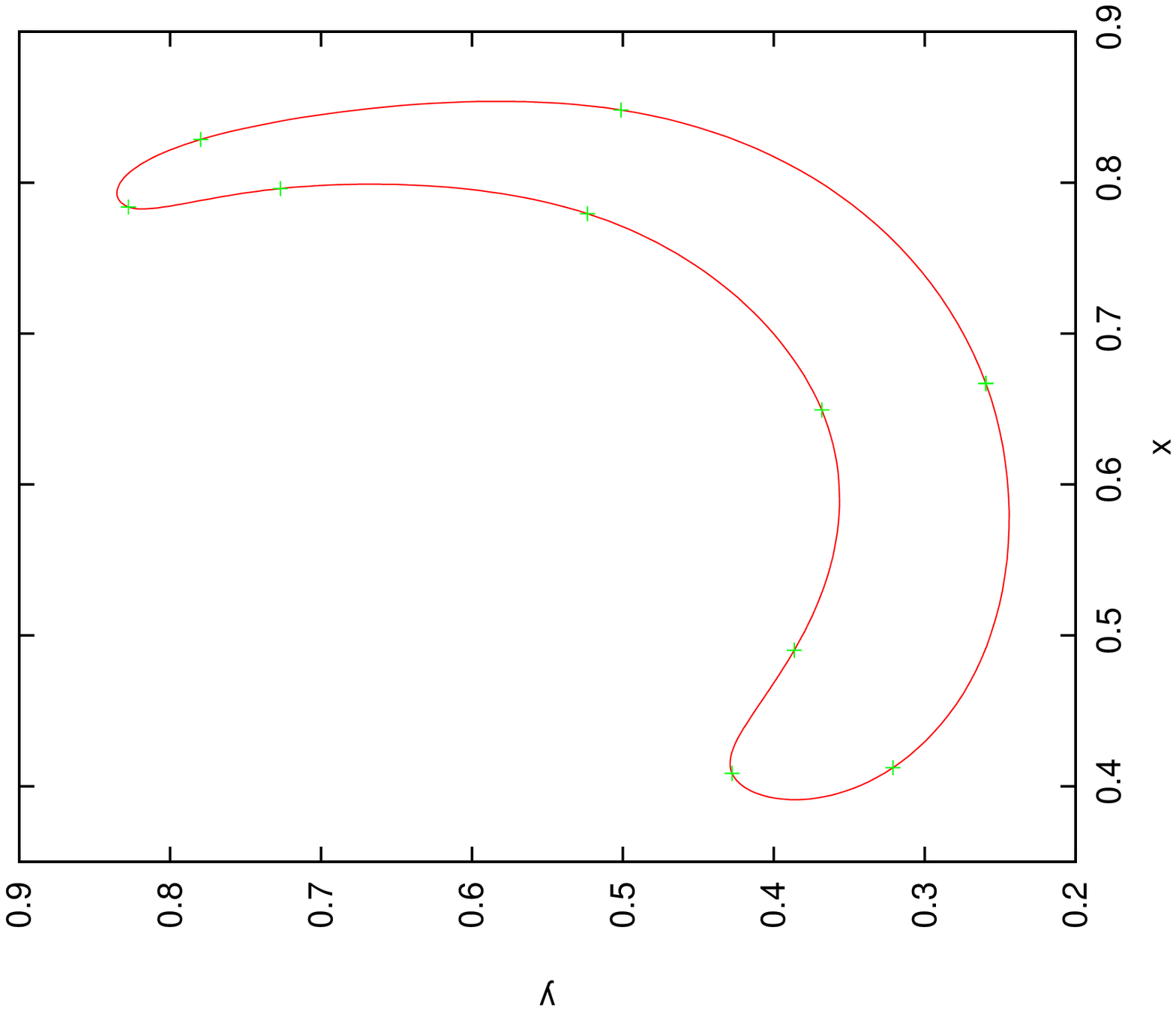}}\nolinebreak
\rotatebox{270}{\includegraphics[height=7cm]{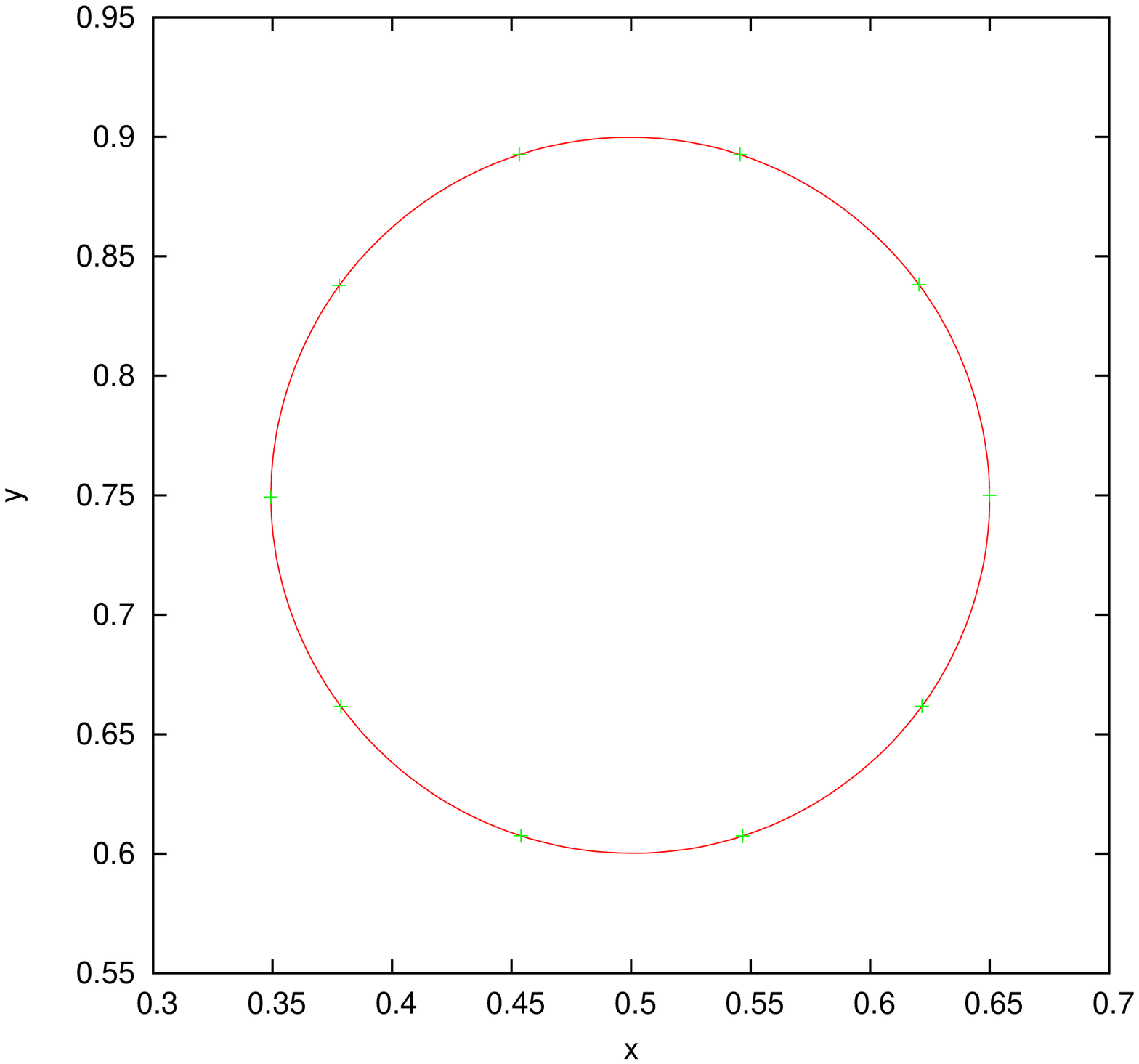}}
\caption{Result of the Rider-Kothe single vortex benchmark test
for $ T = 2 $. The resolution was $ N = 10 $ and $ P = 3$.
LEFT: Position of the interface at $ t = T/2 $. 
RIGHT: Position of the interface at $ t = T $. }
\label{fig:rkT2}
\end{figure}

\begin{figure}
\centering
\rotatebox{270}{\includegraphics[height=10cm]{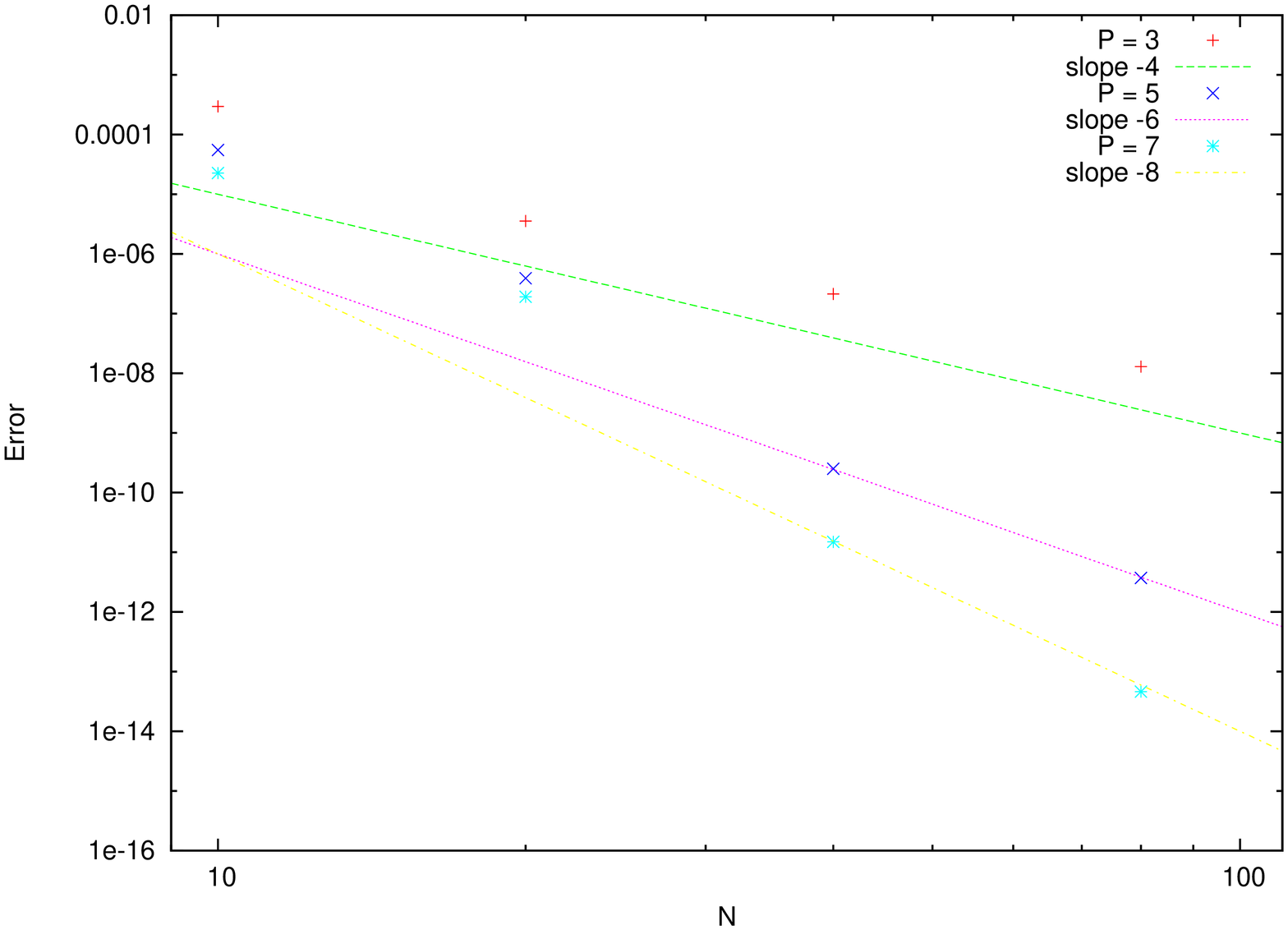}}
\caption{Error decrease for the Rider-Kothe single vortex benchmark test
for $ T = 2 $, for increasing resolution $ N $ using B-splines
of different order $ P $.}
\label{fig:rkT2error}
\end{figure}

\begin{figure}
\centering
\rotatebox{270}{\includegraphics[height=7cm]{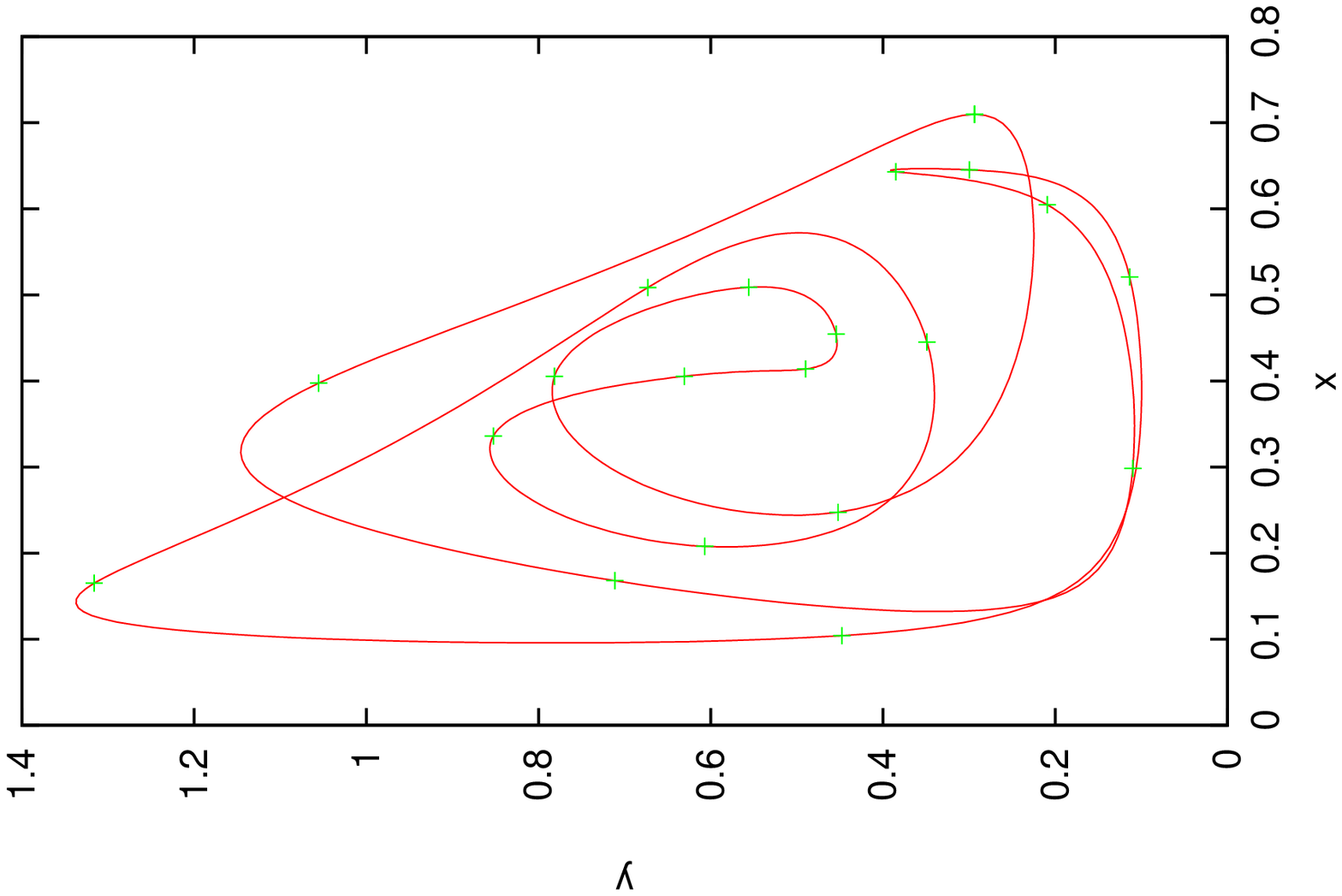}}\nolinebreak
\rotatebox{270}{\includegraphics[height=7cm]{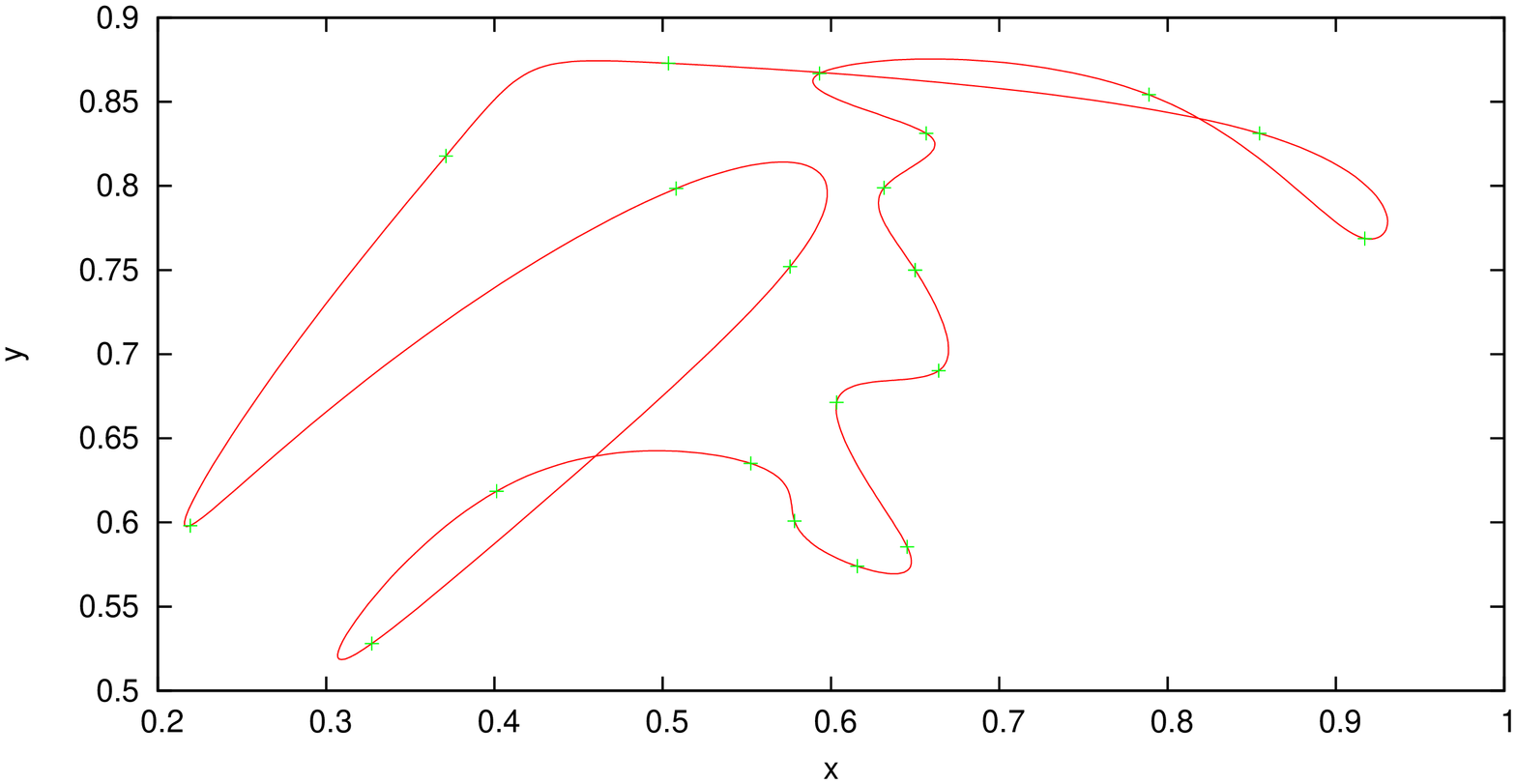}}
\caption{Result of the Rider-Kothe single vortex benchmark test
for $ T = 8 $. The resolution was $ N = 20 $ and $ P = 3 $. 
LEFT: Position of the interface at $ t = T/2 $. 
RIGHT: Position of the interface at $ t = T $. 
In this case the resolution chosen was too coarse leading to a break down of
the present method.}
\label{fig:rkT8miss}
\end{figure}

\begin{figure}
\centering
\rotatebox{270}{\includegraphics[height=7cm]{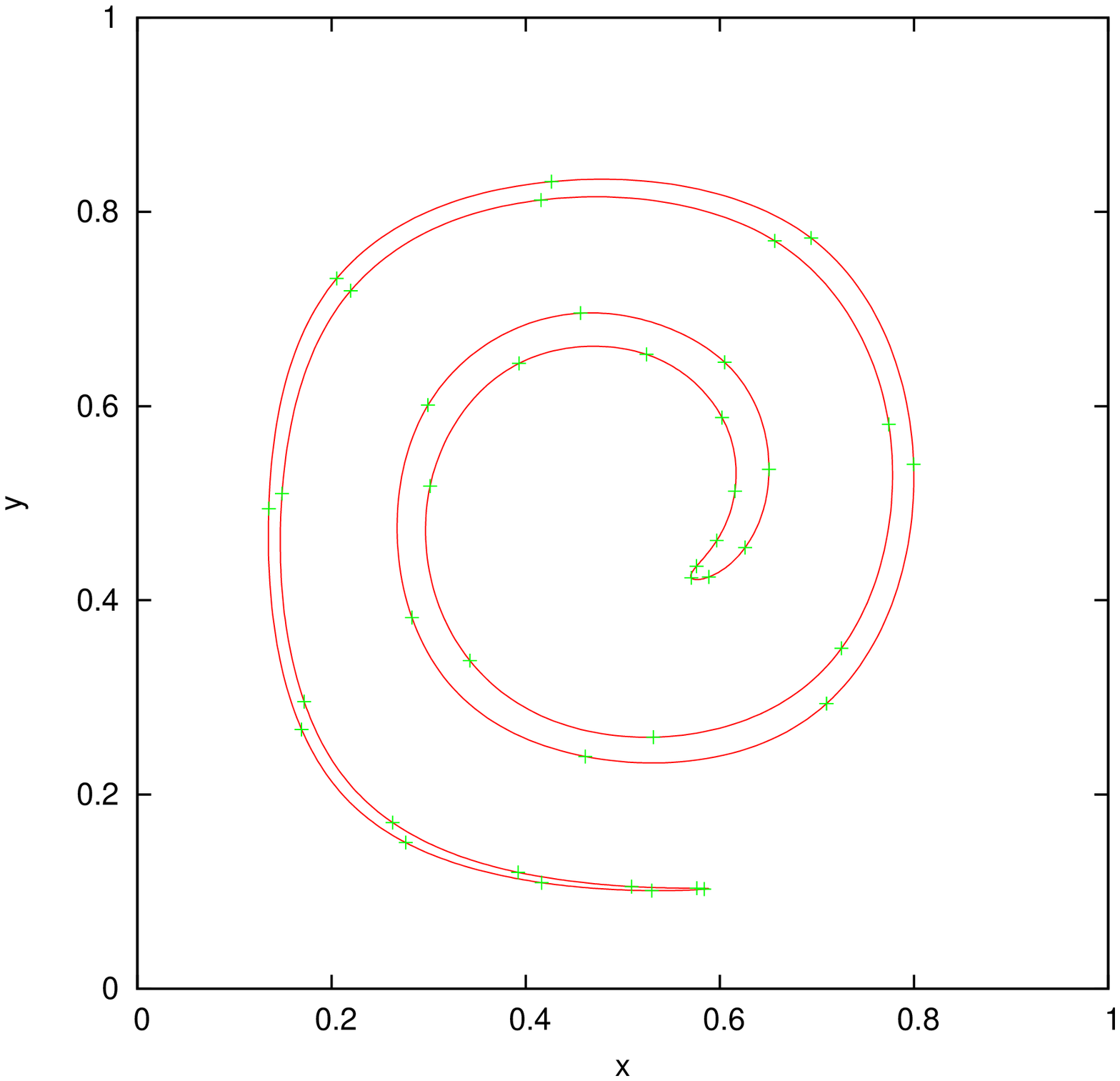}}\nolinebreak
\rotatebox{270}{\includegraphics[height=7cm]{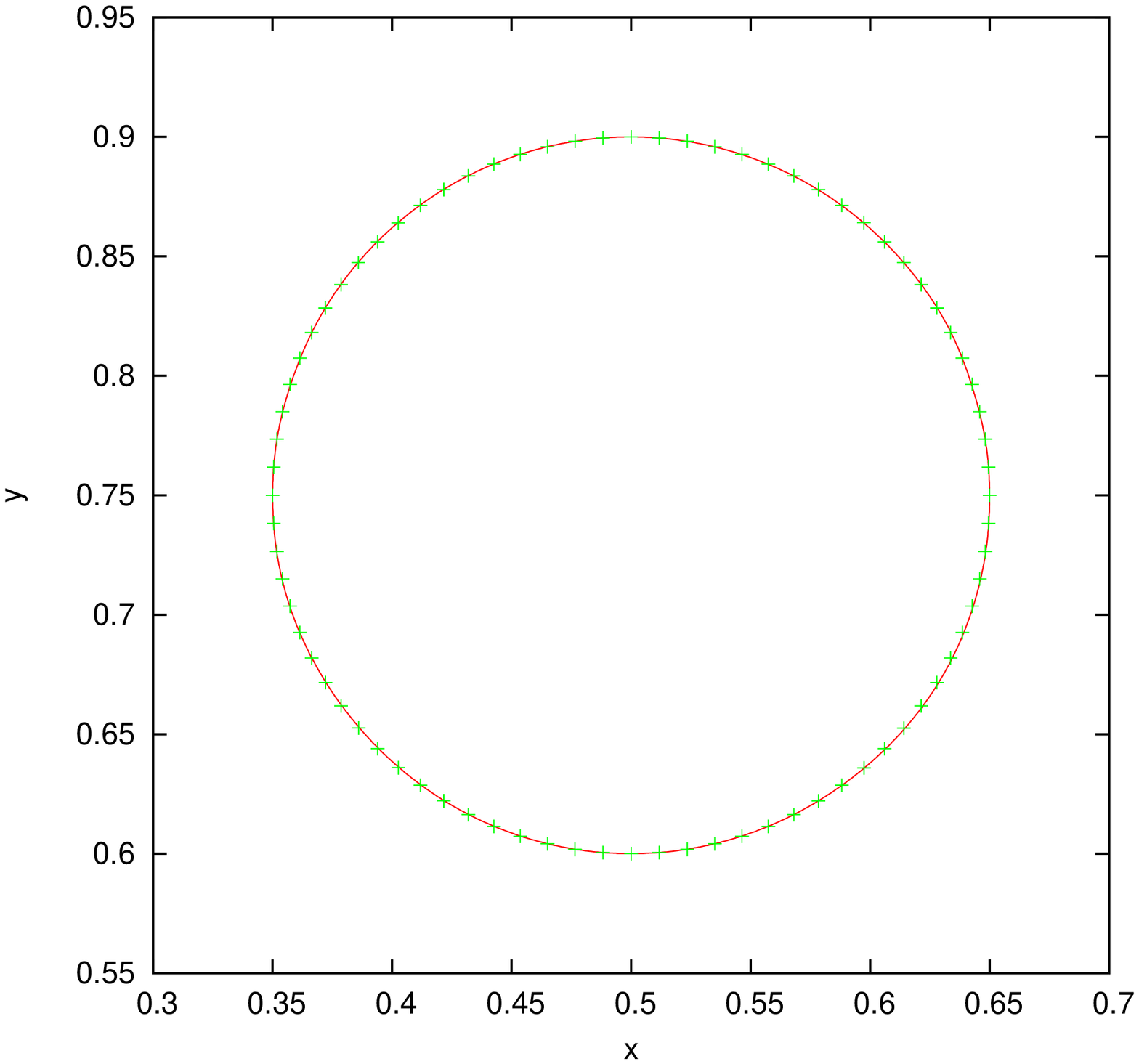}}
\caption{Result of the Rider-Kothe single vortex benchmark test
for $ T = 8 $. The resolution was $ N = 40 $ and $ P = 3$.
LEFT: Position of the interface at $ t = T/2 $. 
RIGHT: Position of the interface at $ t = T $. }
\label{fig:rkT8}
\end{figure}

\begin{figure}
\centering
\rotatebox{270}{\includegraphics[height=10cm]{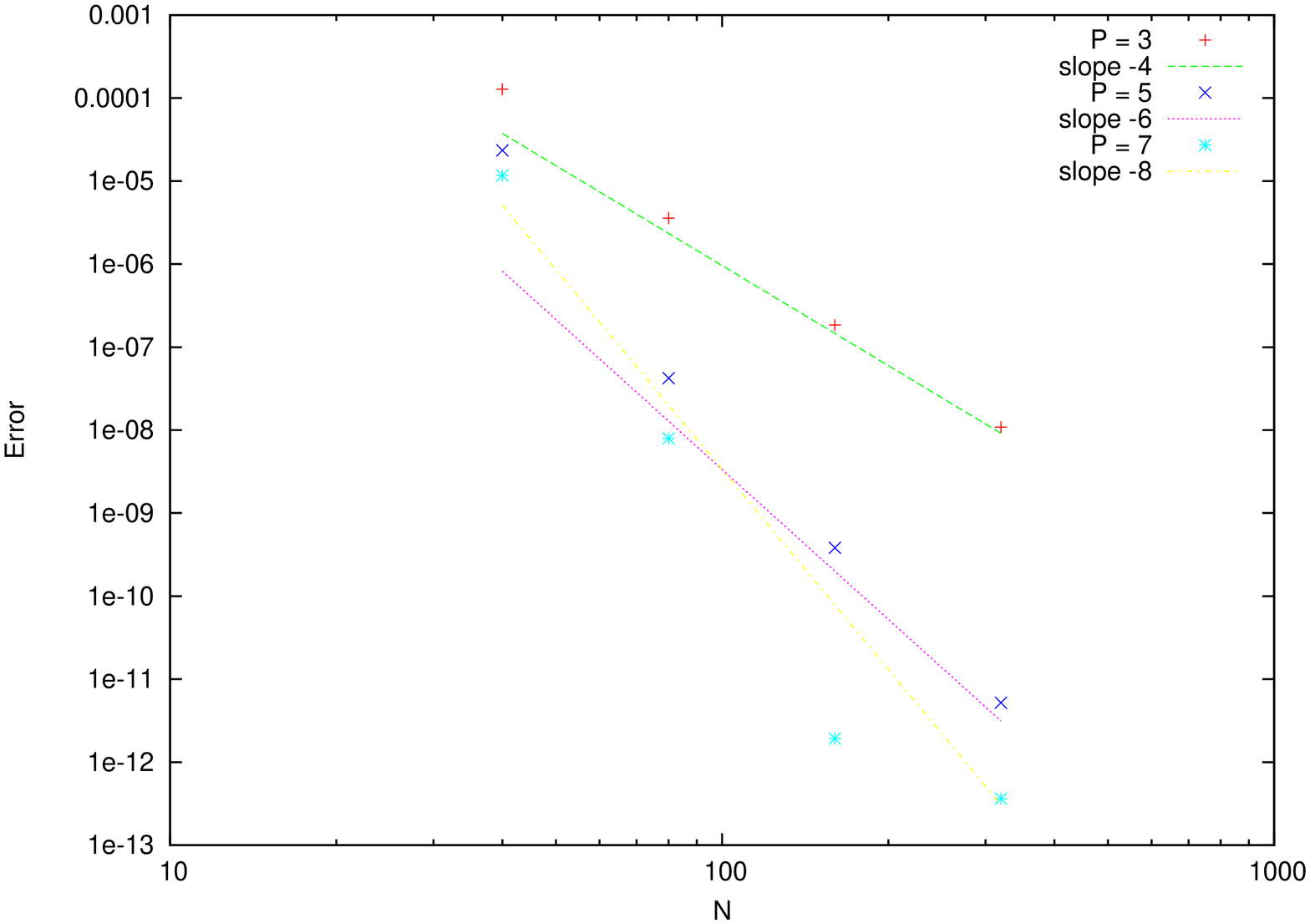}}
\caption{Error decrease for the Rider-Kothe single vortex benchmark test
for $ T = 8 $, for increasing resolution $ N $ using B-splines
of different order $ P $.}
\label{fig:rkT8error}
\end{figure}

\begin{figure}
\centering
\rotatebox{270}{\includegraphics[height=10cm]{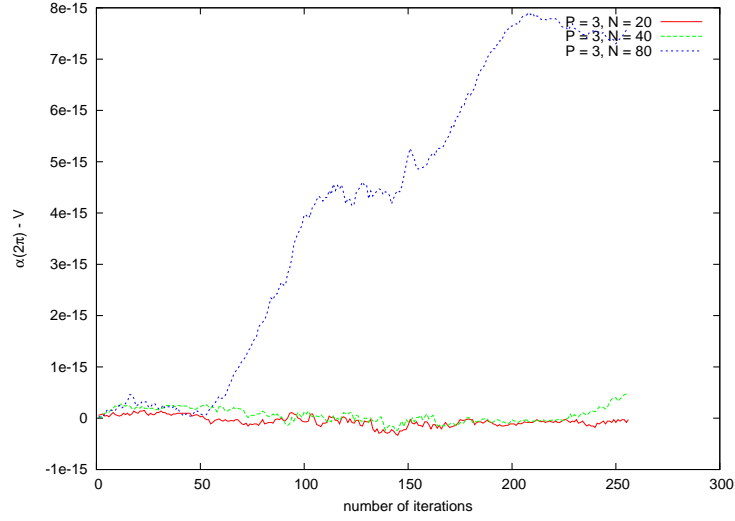}}
\caption{The difference of the exact area $ V $ of the drop 
  to the area of the drop given by $ \alpha(2\pi) $ during the 
  Rider-Kothe single vortex benchmark test
  for $ T = 8 $. 
}
\label{fig:massErrorAlpha}
\end{figure}

\begin{figure}
\centering
\rotatebox{270}{\includegraphics[height=10cm]{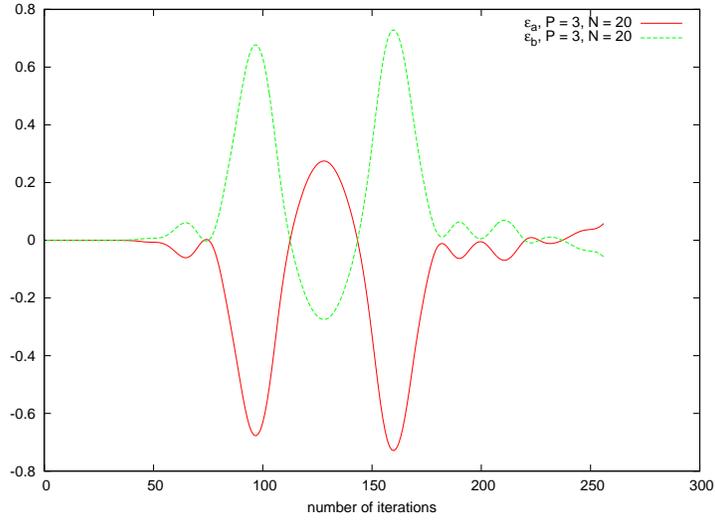}}
\caption{The quadrature errors $ \epsilon_a $, resp. $ \epsilon_b $, 
  defined in equations (\ref{eq:epsilonA}), resp. (\ref{eq:epsilonB}) 
  during the Rider-Kothe single vortex benchmark test
  for $ T = 8 $. The corresponding position of the interface is 
  displayed in figure \ref{fig:rkT8miss} at some instances in time.
  For this low resolution ($ N = 20 $, $ P = 3 $) the method breaks down.
  This is also indicated by large quadrature errors meaning that 
  the actual position of the interface, computed by means of equations 
  (\ref{eq:a2}), resp. (\ref{eq:b2}), includes a large deviation from 
  the exact position given by $ \alpha $ and $ \beta$. 
}
\label{fig:massErrorQuadratureFailed}
\end{figure}

\begin{figure}
\centering
\rotatebox{270}{\includegraphics[height=10cm]{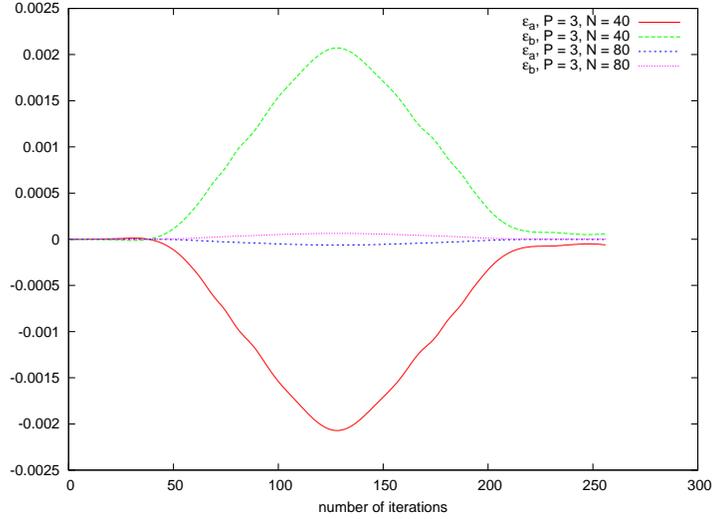}}
\caption{The quadrature errors $ \epsilon_a $, resp. $ \epsilon_b $, 
  defined in equations (\ref{eq:epsilonA}), resp. (\ref{eq:epsilonB}) 
  during the Rider-Kothe single vortex benchmark test
  for $ T = 8 $. This time the resolution is sufficient 
  ($N=40,80$, $ P = 3$) which is also indicated by a smaller quadrature error 
  compared to the one in figure \ref{fig:massErrorQuadratureFailed}.
  For maximum deformation of the drop we are confronted with a
  maximum quadrature error. 
}
\label{fig:massErrorQuadrature}
\end{figure}

\begin{figure}
\centering
\rotatebox{270}{\includegraphics[height=10cm]{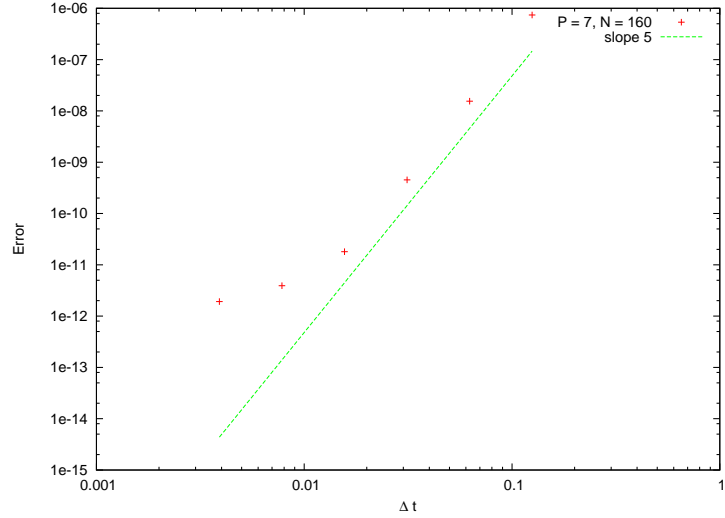}}
\caption{Error increase for the Rider-Kothe single vortex benchmark test
for $ T = 8 $, for increasing time step $ \Delta t $ using B-splines
of order $ P = 7 $ and a resolution of $ N = 160 $.}
\label{fig:rkT8errortemporal}
\end{figure}

The setup of the reversed single-vortex test of Rider and Kothe 
\cite{RiderKothe1998} consists of
a circular drop of radius $ r_0 =  0.15 $ placed at position $ ( x_0 , y_0 ) = (0.5,0.75) $ 
in a unit square box. The velocity field 
$ \vec{u} = ( - \partial_y \psi, \partial_x \psi )^T $ 
is obtained by means of the following stream function $ \psi $:
\be
\psi(x,y,t) = \frac{1}{\pi} \cos \left( \frac{ \pi t}{T} \right) 
\sin^2 \left( \pi x \right) \sin^2 \left( \pi y \right),
\ee
where $ T $ is the period at which the drop has
returned to its initial position. Thereby its interface at time $ T $ should
match the initial circle.
The discrepancy between the numerical interface at time $ T $ and the
exact circle serves as a measure of the numerical error.
As initial condition we used
\begin{eqnarray}
\alpha & = & 
\frac{1}{2} r_0^2 s + r_0\left( x_0 \sin s - y_0 \cos s + y_0 \right) 
\label{eq:initialConditionA}\\
\beta & = & \left( r_0 \cos s + x_0 \right) \left( r_0 \sin s + y_0 \right),
\label{eq:initialConditionB}
\end{eqnarray}
where $ s \in [0,2\pi] $. As mentioned in section \ref{sec:interpolation}, we 
divided $ [0,2\pi] $ into $ N $ equidistant sections. We performed three
series of tests for $ T = 1/2 $, $ T = 2 $ and $ T = 8 $. 
The results of these tests are shown in figures 
\ref{fig:rkT05}, \ref{fig:rkT05error}, \ref{fig:rkT2},\ref{fig:rkT2error},
\ref{fig:rkT8miss},\ref{fig:rkT8},\ref{fig:rkT8error},
\ref{fig:massErrorAlpha},\ref{fig:massErrorQuadratureFailed},
\ref{fig:massErrorQuadrature},\ref{fig:rkT8errortemporal} 
and tables \ref{tab:rkT05error}, \ref{tab:rkT2error},
\ref{tab:rkT8error}. For all simulations we chose $ n = 5 $ for the 
approximation of the trajectories for the advection step, cf. section 
\ref{sec:advection}. Depending on the order $ P $ of the B-spline interpolation
we chose a different value for the time steps $ \Delta t $ in order
to make the error contribution due to the advection step subdominant compared to
the error contribution of the interface Reconstruction. The advection step
itself can handle quite large time steps without displaying any sign of 
instability. All simulations
were performed using B-splines of order $ P = 3,5,7 $. For $ P = 3 $ the
time step was chosen $ \Delta t = 1/32 $, for $ P = 5 $, $ \Delta t = 1/128 $, 
and for $ P = 7 $, $ \Delta t = 1/256 $. The time step was kept fixed 
at these values even when going over to finer resolutions. In figure
\ref{fig:rkT05} the resulting position of the interface is shown for 
$ t = T/2 $ and $ t = T $ in the case $ T = 1/2 $, $ P = 3 $ and
a resolution of $ N = 10 $. The interface is well resolved and does not 
display any visual disturbances such as bumps or oscillations at maximum 
deformation and when it has returned to its initial position at $ t = T $. 
The same observation can be made for the 
case $ T = 2 $, cf. figure \ref{fig:rkT2}. Concerning the convergence of the
method for these two cases, e.g. $ T = 1/2 $ and $ T = 2 $, cf. figures
\ref{fig:rkT05error}, resp. \ref{fig:rkT2error} and 
tables \ref{tab:rkT05error}, resp. \ref{tab:rkT2error}, we observe that the
order of convergence corresponds approximately to the theoretical value of
$ P + 1 $, cf. equation (\ref{eq:bound}). For $ P = 7 $ and fine 
resolutions the error does not further decrease because of the round off limit. 
In addition, for the case $ T = 2 $ when going from $ N = 20 $ to $ N = 40 $, we
observe a sudden jump in the convergence, cf. figure \ref{fig:rkT2error} and
table \ref{tab:rkT2error}, for $ P = 5 $ and $ P =7 $. This phenomenon of 
accelerated convergence is even more pronounced in the case $ T = 8 $, cf. 
figure \ref{fig:rkT8error} and table \ref{tab:rkT8error}. An explanation 
for this sudden increase in convergence 
might lie in a underresolution of the problem for coarse 
resolutions $ N $, meaning that when increasing $ P $, keeping $ N $ fixed at
a small value no important gain in accuracy is observed. However, if the 
resolution is finer, i.e. large values of $ N $, increasing $ P $ will 
almost lead to spectral convergence, i.e. faster than algebraic. This points
to the eventuality of having insufficient sampling of the signal, 
i.e. an aliasing error. 
This is different to classical spectral methods such as methods based on 
Chebyshev or Legendre polynomials for which increasing the approximation 
order introduces an increase of spatial resolution by 
increasing the number of Gauss points. Periodic 
B-splines on the contrary offer the possibility of increasing $ P$ and $ N$ 
independently with the consequence of having eventually a 
persisting aliasing error when only increasing $ P $. 
Even worse a 
underresolved simulation can lead to an entirely wrong solution 
by the present method, 
as can be seen in figure \ref{fig:rkT8miss}. Here $ T = 8 $ and
the resolution $ N $ was fixed at 20 ($ P = 3 $). For $ t = T/2 $ still a 
few structures of the correct solution are recognizable, however for $ t = T $ 
the picture has entirely deteriorated. If underresolved, the numerical solution
can display self intersections. This indicates that the present method is less
robust to underresolution. However, for well resolved cases, which for 
the case $ T = 8 $ start at only 40 knots, cf. figure 
\ref{fig:rkT8}, the present numerical scheme produces a very accurate 
result. \\
Concerning area conservation, we observe 
from figure \ref{fig:massErrorAlpha} that the value $ \alpha(2\pi) $ 
is, apart from round off contributions, equal to the initial area of the 
drop. However, the quadrature errors $ \epsilon_a $, resp. $ \epsilon_b $, 
equations (\ref{eq:epsilonA}), resp. (\ref{eq:epsilonB}), can become rather 
important during simulation, as for instance for the 
underresolved case $ T = 8 $, $ N = 20 $ and $ P = 3 $, cf. figure 
\ref{fig:massErrorQuadratureFailed}, indicating a possible 
discrepancy between the actual area of the drop and $ \alpha(2\pi) $. 
For well resolved cases the quadrature errors are smaller, cf. 
figure \ref{fig:massErrorQuadrature}, but seem to increase with 
increasing deformation of the drop. \\
As a last numerical experiment we investigated the accuracy of the 
advection scheme derived in section \ref{sec:advection}. 
By fixing the number of Gauss Labatto Legendre nodes to $ n = 5 $, 
the interpolating polynomial has order 4 for which reason we 
expect the advection scheme to converge with $ 5^{th} $ order
accuracy with respect to the time step $ \Delta t $. 
In order to observe the error contribution by the advection 
scheme, we chose a B-spline of order $ P = 7 $ and a spatial 
resolution of $ N = 160 $, such that the error contribution by 
the interface representation is subdominant. Decreasing
the time step $ \Delta t $ leads indeed to a fifth order 
convergence of the numerical error up to the point at which 
the error contribution by the interface representation 
becomes dominant, cf. figure \ref{fig:rkT8errortemporal}. 

\begin{table}
\centering
\[
\begin{array}{l|l|l|l}
P & N  & \mbox{Error} & O \\
\hline
3 & 10 &3.66583\times 10^{-5} & -  \\
  & 20 &1.91077\times 10^{-6} & 4.26\\
  & 40 &1.14024\times 10^{-7} & 4.07\\
  & 80 &6.96072\times 10^{-9} & 4.03\\
\hline
5 & 10 &1.74018\times 10^{-6} & -\\
  & 20 &1.47129\times 10^{-8} & 6.87\\
  & 40 &1.87246\times 10^{-10} &6.30\\
  & 80 &2.69292\times 10^{-12} &6.12\\
\hline
7 & 10 &1.18328\times 10^{-7} & -\\
  & 20 &1.79321\times 10^{-10} & 9.37\\
  & 40 &5.09673\times 10^{-13} & 8.46\\
  & 80 &6.38136\times 10^{-15} & 6.32
\end{array}
\]
\caption{Results for the Rider-Kothe single vortex benchmark test
for $ T = 1/2 $}
\label{tab:rkT05error}
\end{table}

\begin{table}
\centering
\[
\begin{array}{l|l|l|l}
P & N  & \mbox{Error} & O \\
\hline
3 & 10 &2.96591\times 10^{-4} & -\\
  & 20 &3.54562\times 10^{-6} &6.39\\
  & 40 &2.13069\times 10^{-7} &4.06\\
  & 80 &1.29146\times 10^{-8} &4.04\\
\hline
5 & 10 &5.50018\times 10^{-5} & -\\
  & 20 &3.90107\times 10^{-7} & 7.14\\
  & 40 &2.51014\times 10^{-10} &10.60\\
  & 80 &3.72075\times 10^{-12} &6.08\\
\hline
7 & 10 &2.25578\times 10^{-5} & - \\
  & 20 &1.91445\times 10^{-7} & 6.88 \\
  & 40 &1.49644\times 10^{-11} &13.64\\
  & 80 &4.62276\times 10^{-14} &8.33
\end{array}
\]
\caption{Results for the Rider-Kothe single vortex benchmark test
for $ T = 2 $}
\label{tab:rkT2error}
\end{table}

\begin{table}
\centering
\[
\begin{array}{l|l|l|l}
P & N  & \mbox{Error} & O \\
\hline
3 & 40 &1.28241\times 10^{-4} & - \\
  & 80 &3.57649\times 10^{-6} &5.16\\
  & 160&1.84163\times 10^{-7} &4.30\\
  & 320&1.08844\times 10^{-8} &4.08\\
\hline
5 & 40 &2.34006\times 10^{-5} & - \\
  & 80 &4.21891\times 10^{-8} &9.12\\
  &160 &3.82007\times 10^{-10} &6.79\\
  &320 &5.19595\times 10^{-12} &6.20\\
\hline
7 & 40 &1.16624\times 10^{-5} &- \\
  & 80 &7.93665\times 10^{-9} &10.52\\
  &160 &1.92373\times 10^{-12}&12.01\\
  &320 &3.63856\times 10^{-13}&2.40
\end{array}
\]
\caption{Results for the Rider-Kothe single vortex benchmark test
for $ T = 8 $}
\label{tab:rkT8error}
\end{table}
\clearpage
\subsection{Numerical verification part 2}

\begin{figure}
\centering
\begin{minipage}[t][0.23\textheight][t]{0.5\linewidth}
\centering
\rotatebox{270}{
\includegraphics[height=0.2\textheight]{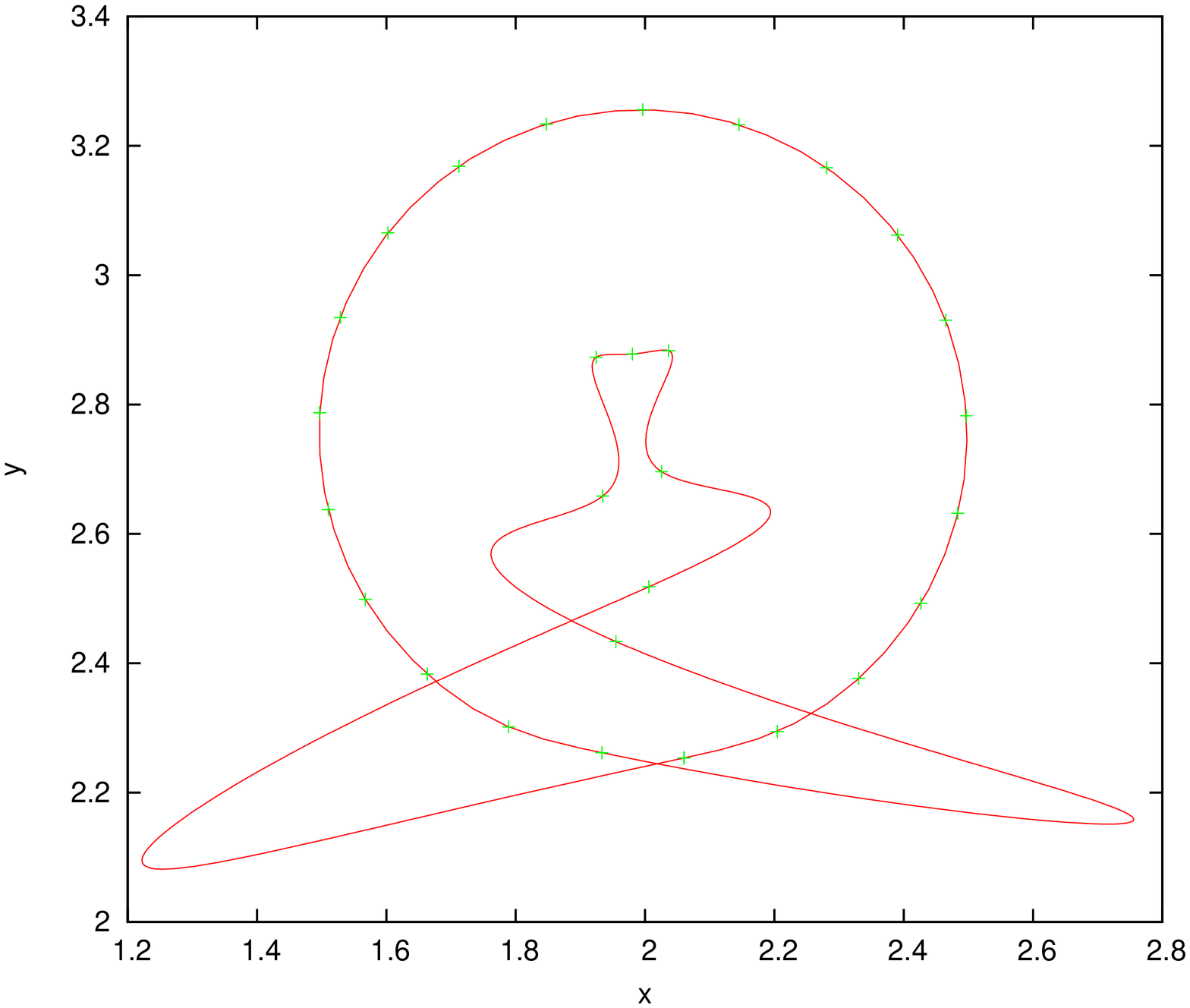}
}\\
{$ N = 28, t = 0 $}
\end{minipage} \nolinebreak
\begin{minipage}[t][0.23\textheight][t]{0.5\linewidth}
\centering
\rotatebox{270}{
\includegraphics[height=0.2\textheight]{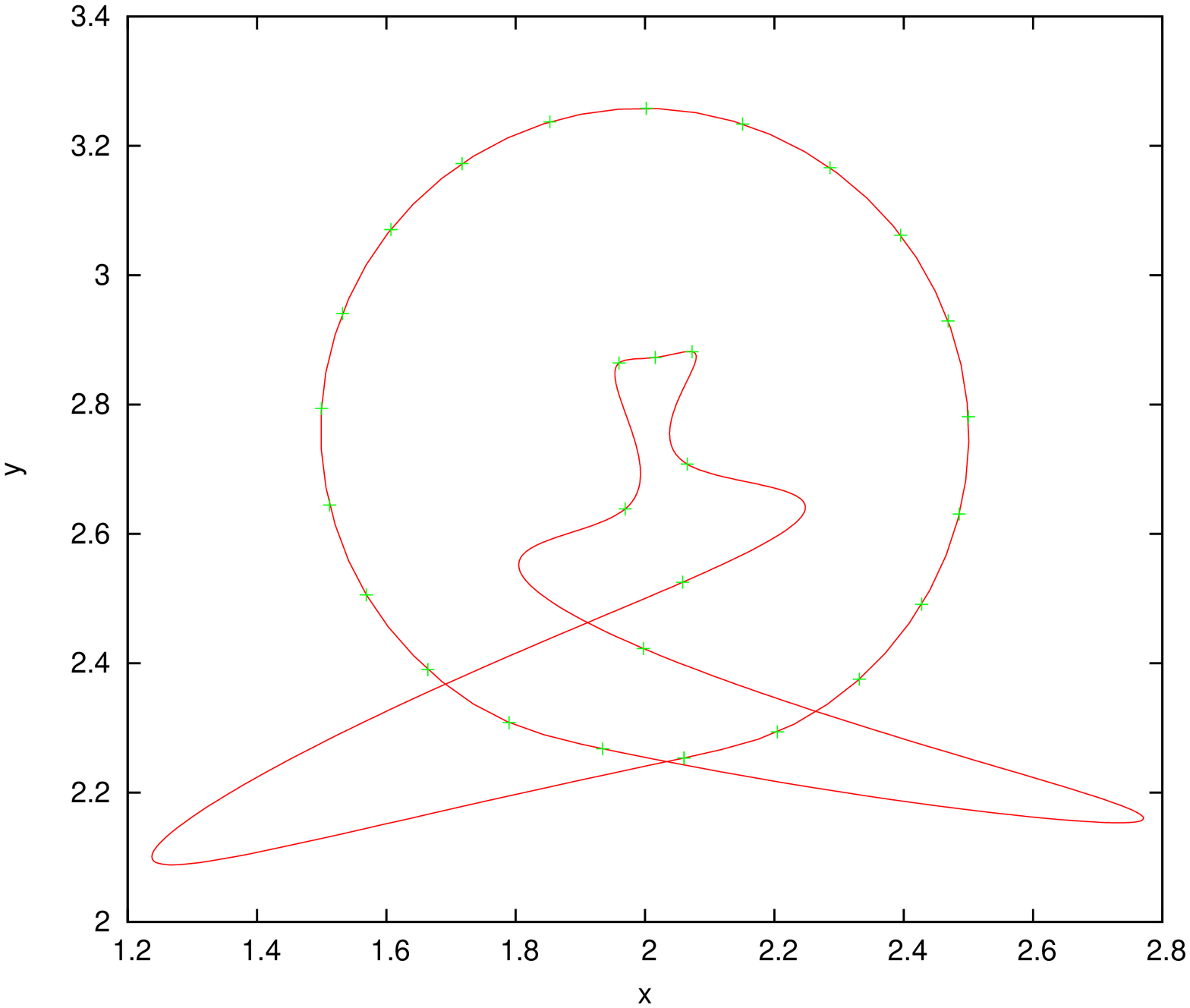}
}\\
$ N =28, t = T $
\end{minipage}\\
\begin{minipage}[t][0.23\textheight][t]{0.5\linewidth}
\centering
\rotatebox{270}{
\includegraphics[height=0.2\textheight]{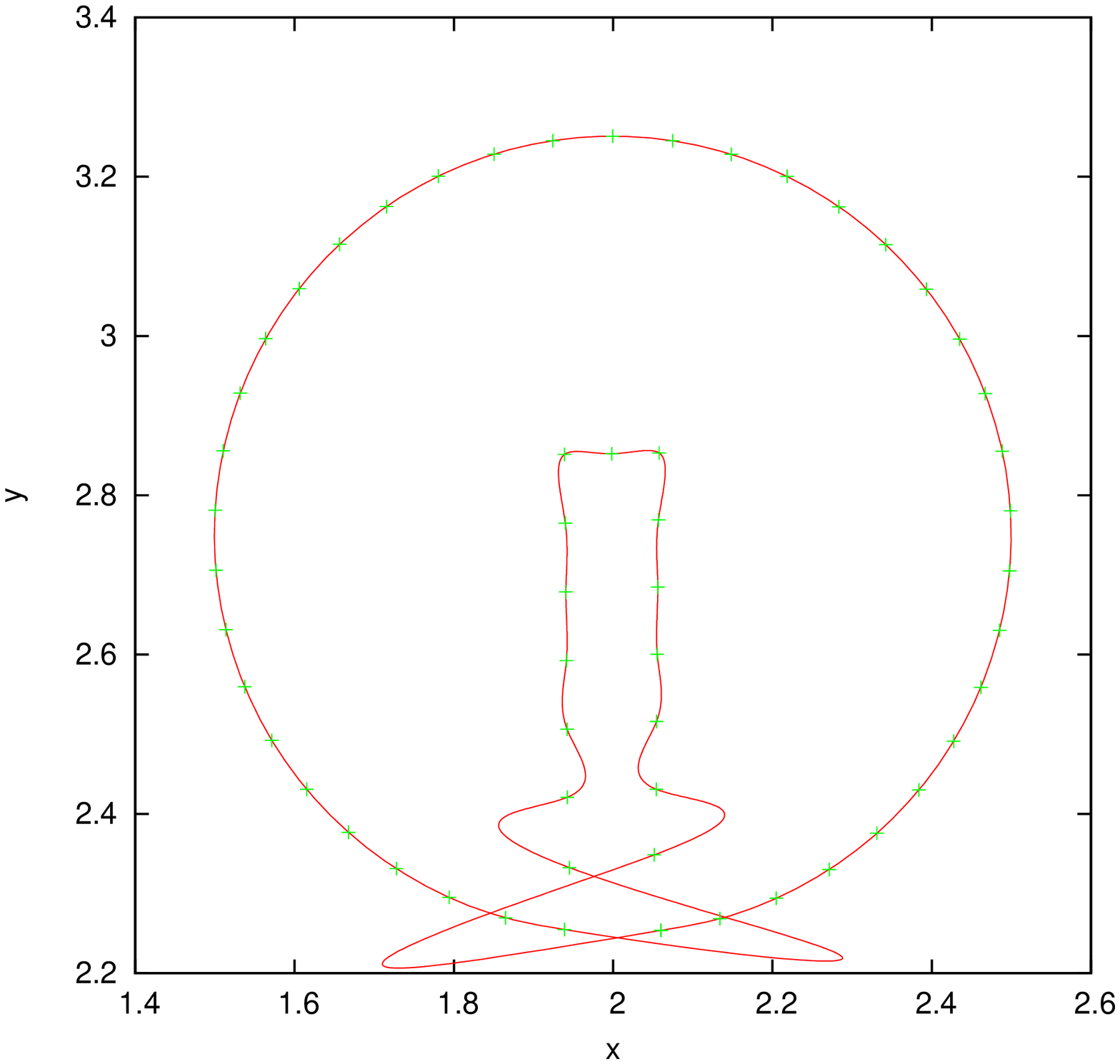}
}\\
$ N =56, t = 0 $
\end{minipage} \nolinebreak
\begin{minipage}[t][0.23\textheight][t]{0.5\linewidth}
\centering
\rotatebox{270}{
\includegraphics[height=0.2\textheight]{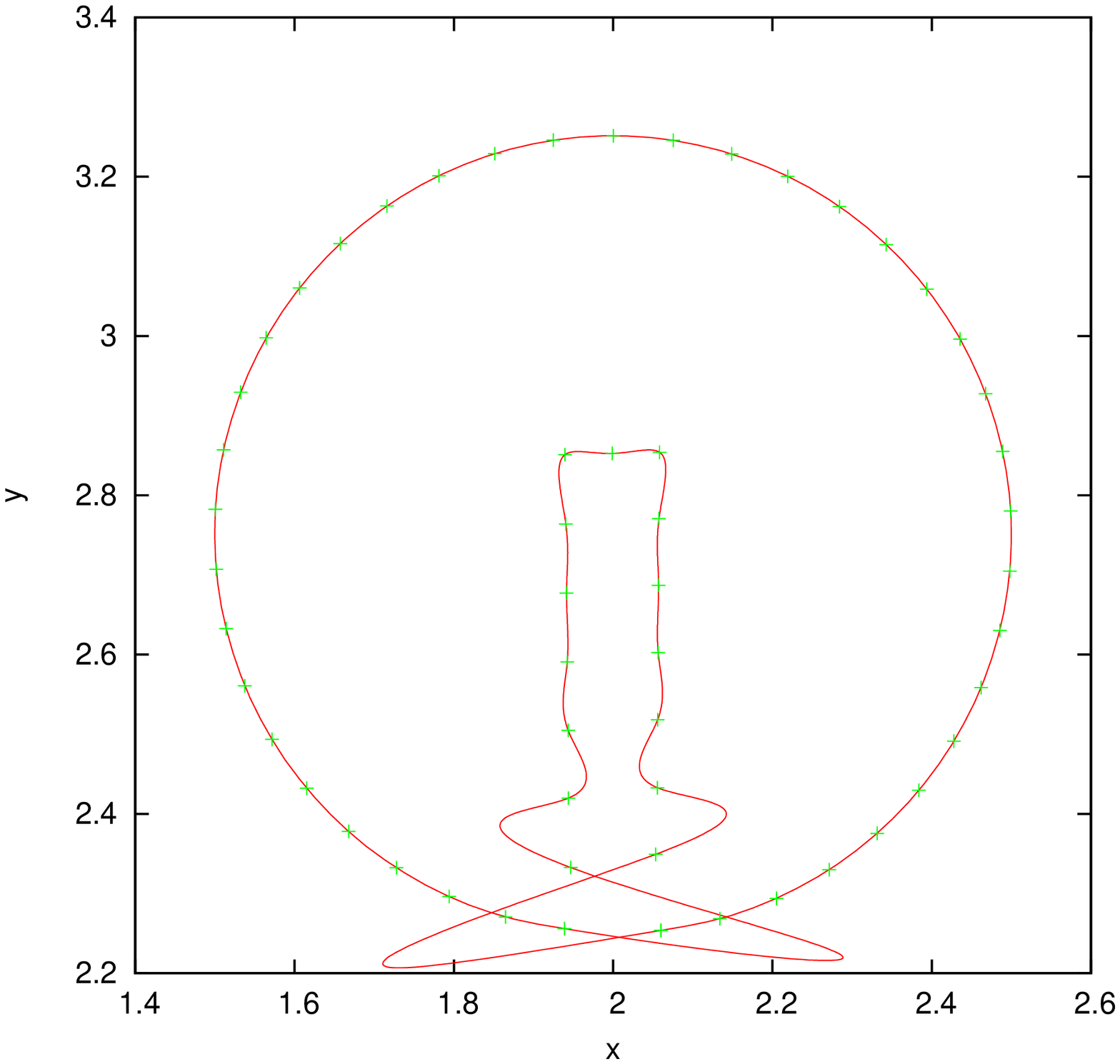}
}\\
$ N = 56, t = T $
\end{minipage}\\
\begin{minipage}[t][0.23\textheight][t]{0.5\linewidth}
\centering
\rotatebox{270}{
\includegraphics[height=0.2\textheight]{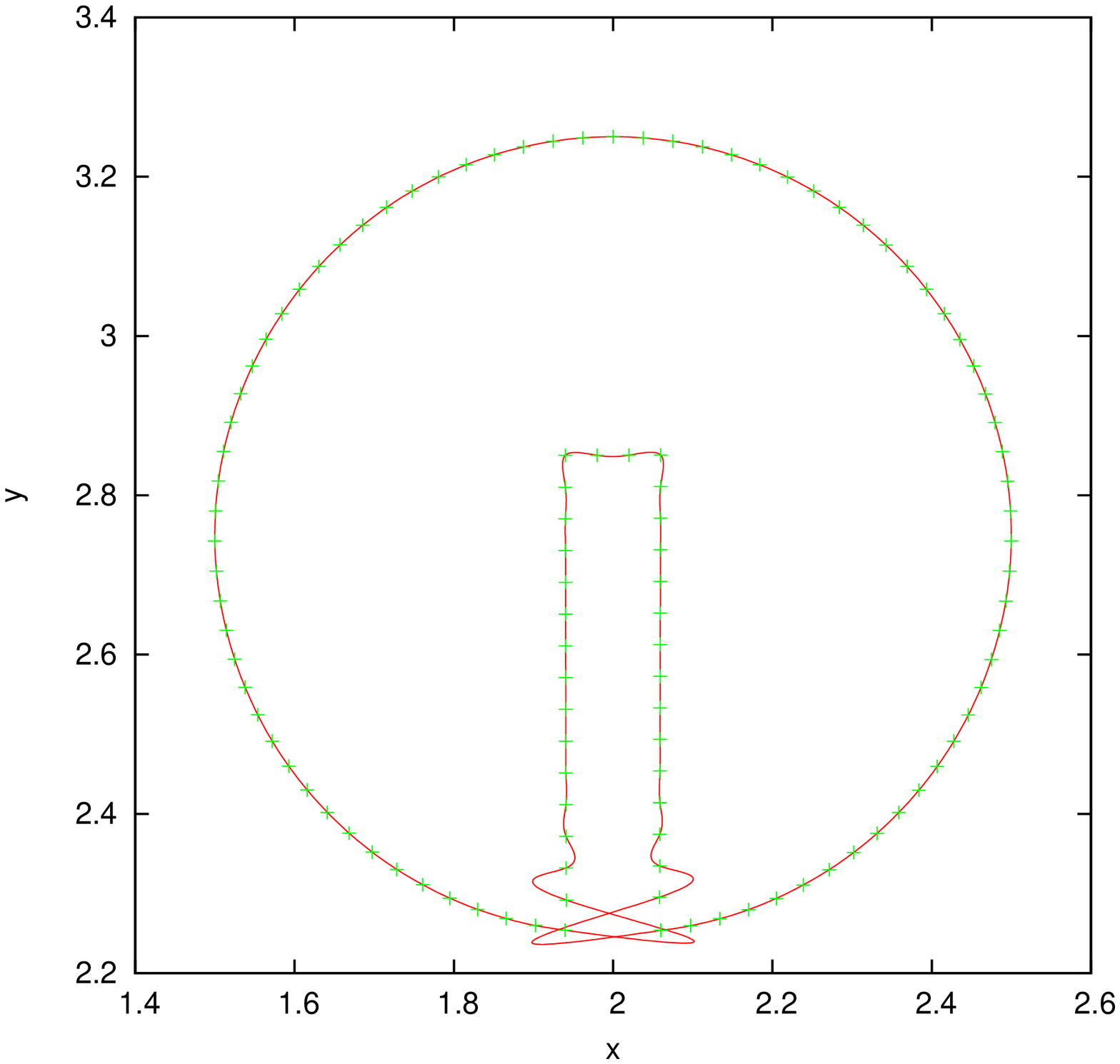}
}\\
{$ N =112, t = 0$}
\end{minipage} \nolinebreak
\begin{minipage}[t][0.23\textheight][t]{0.5\linewidth}
\centering
\rotatebox{270}{
\includegraphics[height=0.2\textheight]{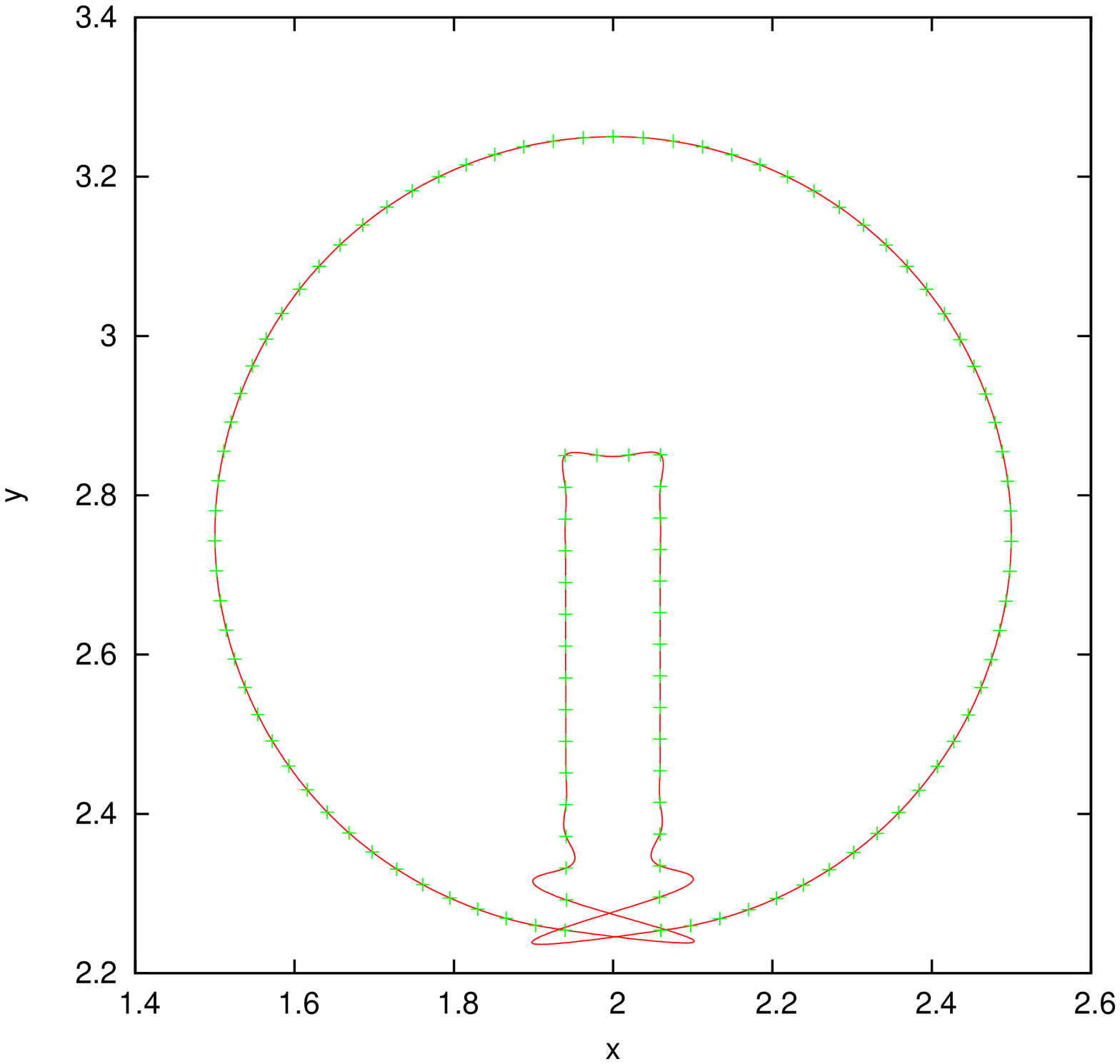}
}\\
$ N = 112, t = T $
\end{minipage}\\
\begin{minipage}[t][0.23\textheight][t]{0.5\linewidth}
\centering
\rotatebox{270}{
\includegraphics[height=0.2\textheight]{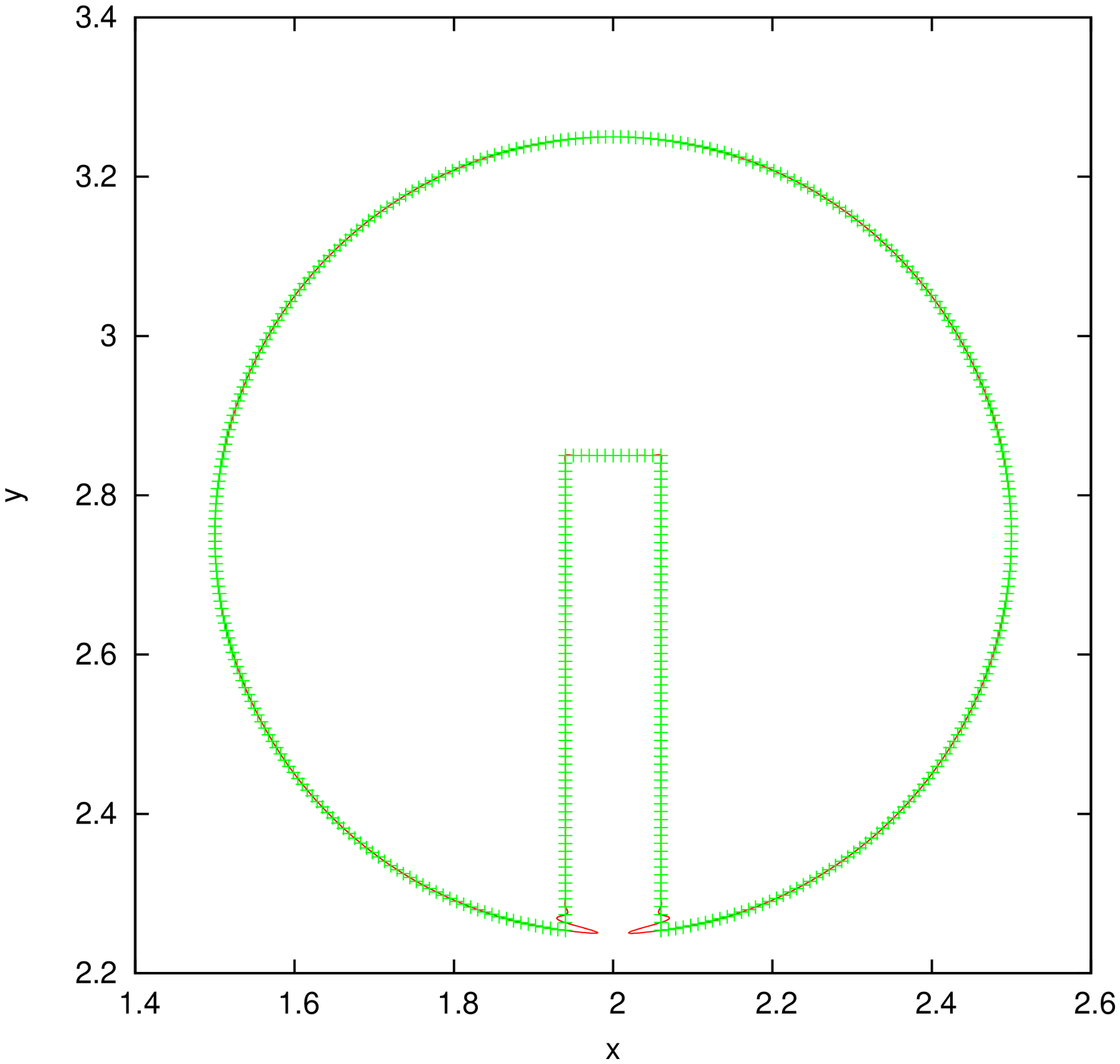}
}\\
$ N = 448, t = 0 $
\end{minipage} \nolinebreak
\begin{minipage}[t][0.23\textheight][t]{0.5\linewidth}
\centering
\rotatebox{270}{
\includegraphics[height=0.2\textheight]{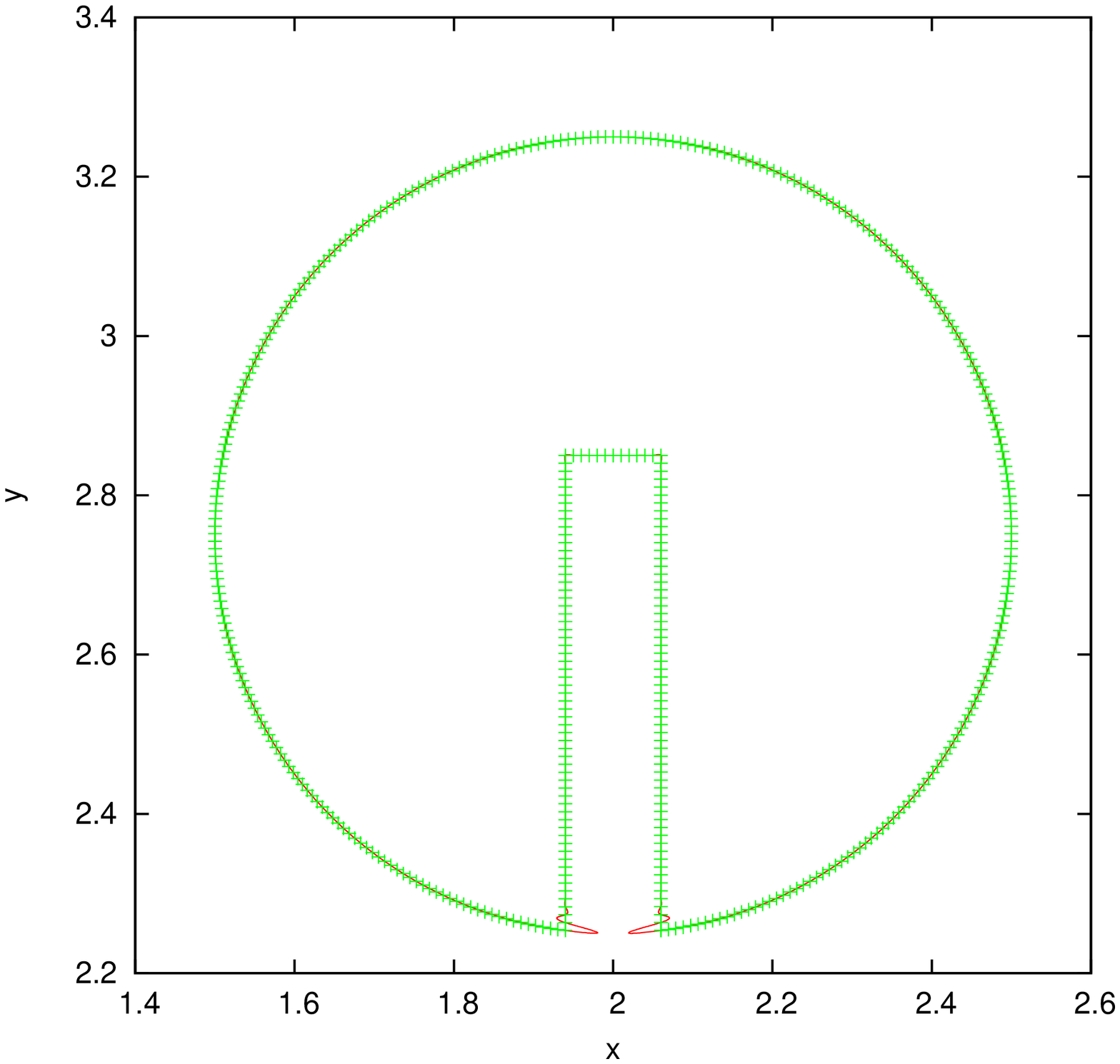}
}\\
$ N = 448, t = T $
\end{minipage}\\
\caption{Result of Zalesak's slotted disk rotation test for 
different resolutions $N $. The order of B-spline interpolation is $ P = 3$. 
Shown are the graphs at the initial position, $ t = 0 $,
and the position after a 
full rotation, $ t = T $. }
\label{fig:slottedP3}
\end{figure}

\begin{figure}
\centering
\begin{minipage}[t][0.23\textheight][t]{0.5\linewidth}
\centering
\rotatebox{270}{
\includegraphics[height=0.2\textheight]{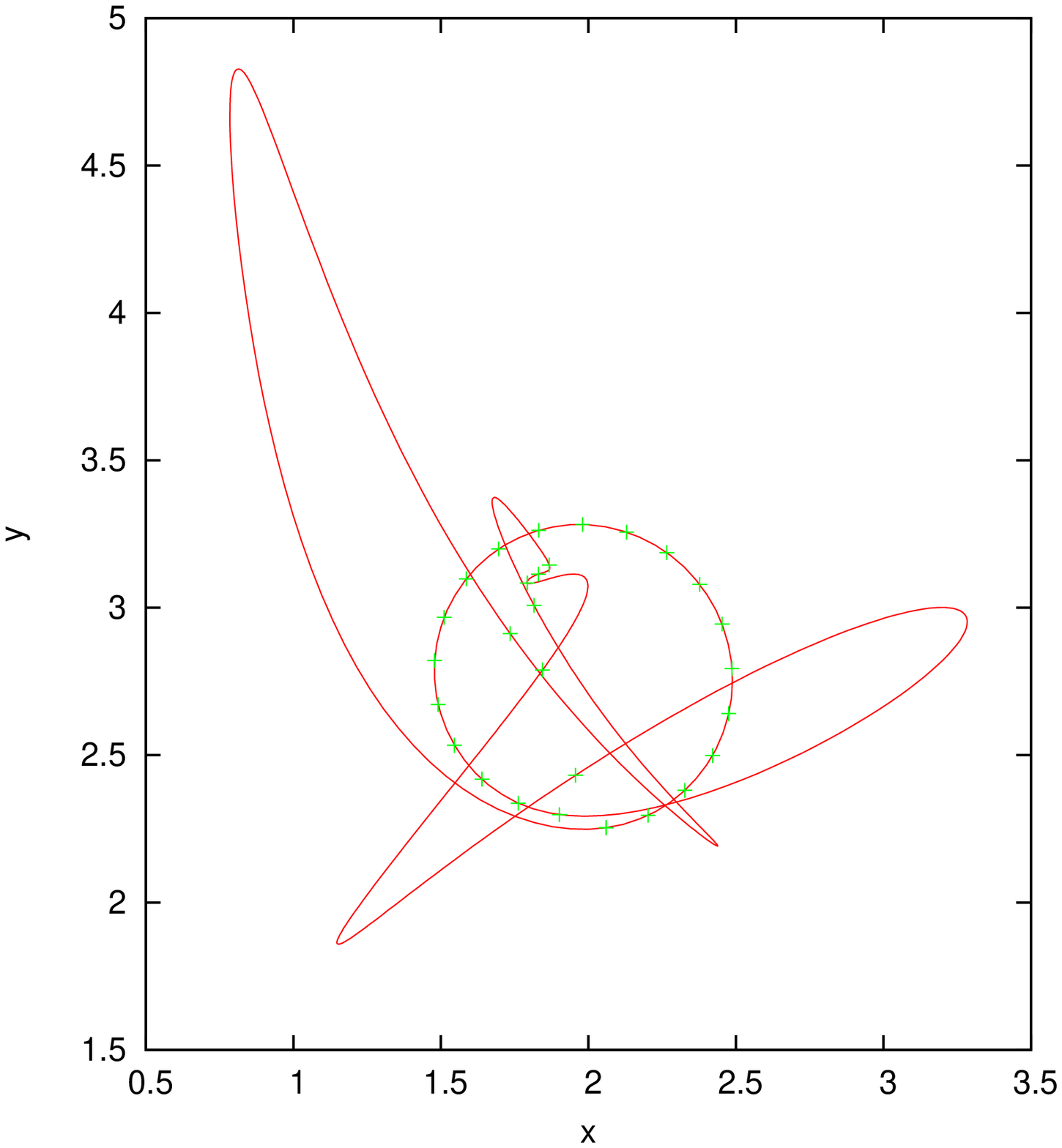}
}\\
{$ N = 28, t = 0 $}
\end{minipage} \nolinebreak
\begin{minipage}[t][0.23\textheight][t]{0.5\linewidth}
\centering
\rotatebox{270}{
\includegraphics[height=0.2\textheight]{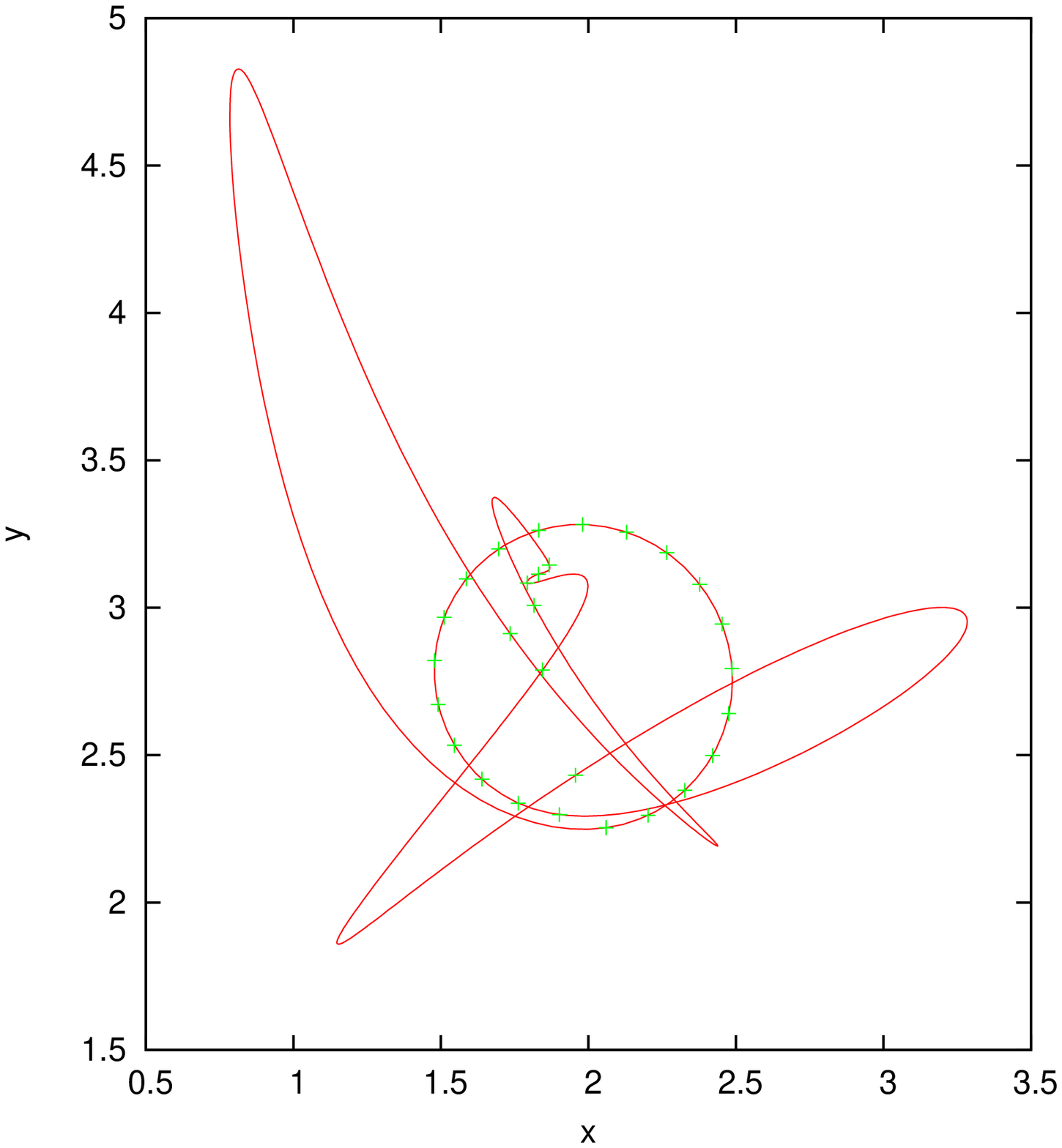}
}\\
$ N =28, t = T $
\end{minipage}\\
\begin{minipage}[t][0.23\textheight][t]{0.5\linewidth}
\centering
\rotatebox{270}{
\includegraphics[height=0.2\textheight]{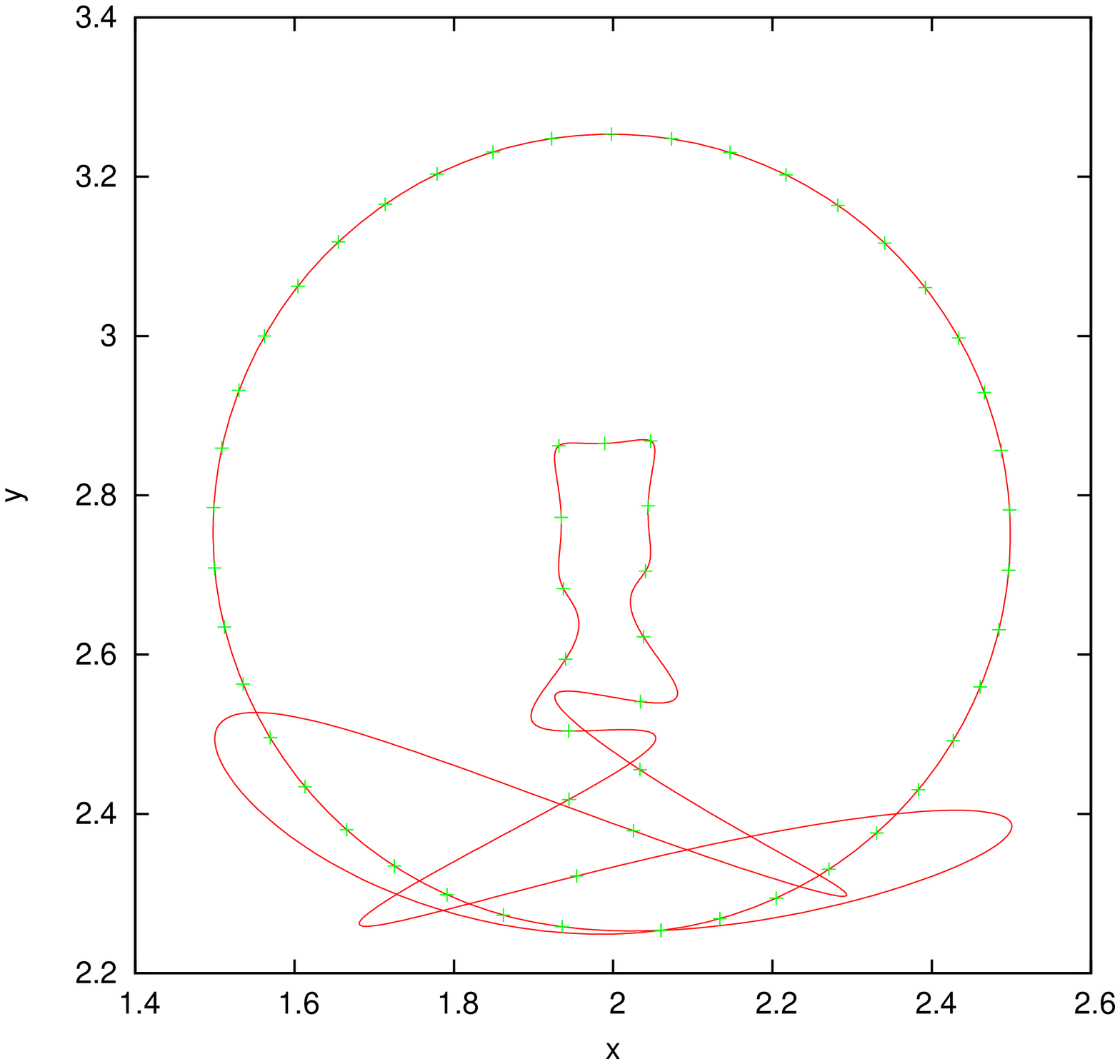}
}\\
$ N =56, t = 0 $
\end{minipage} \nolinebreak
\begin{minipage}[t][0.23\textheight][t]{0.5\linewidth}
\centering
\rotatebox{270}{
\includegraphics[height=0.2\textheight]{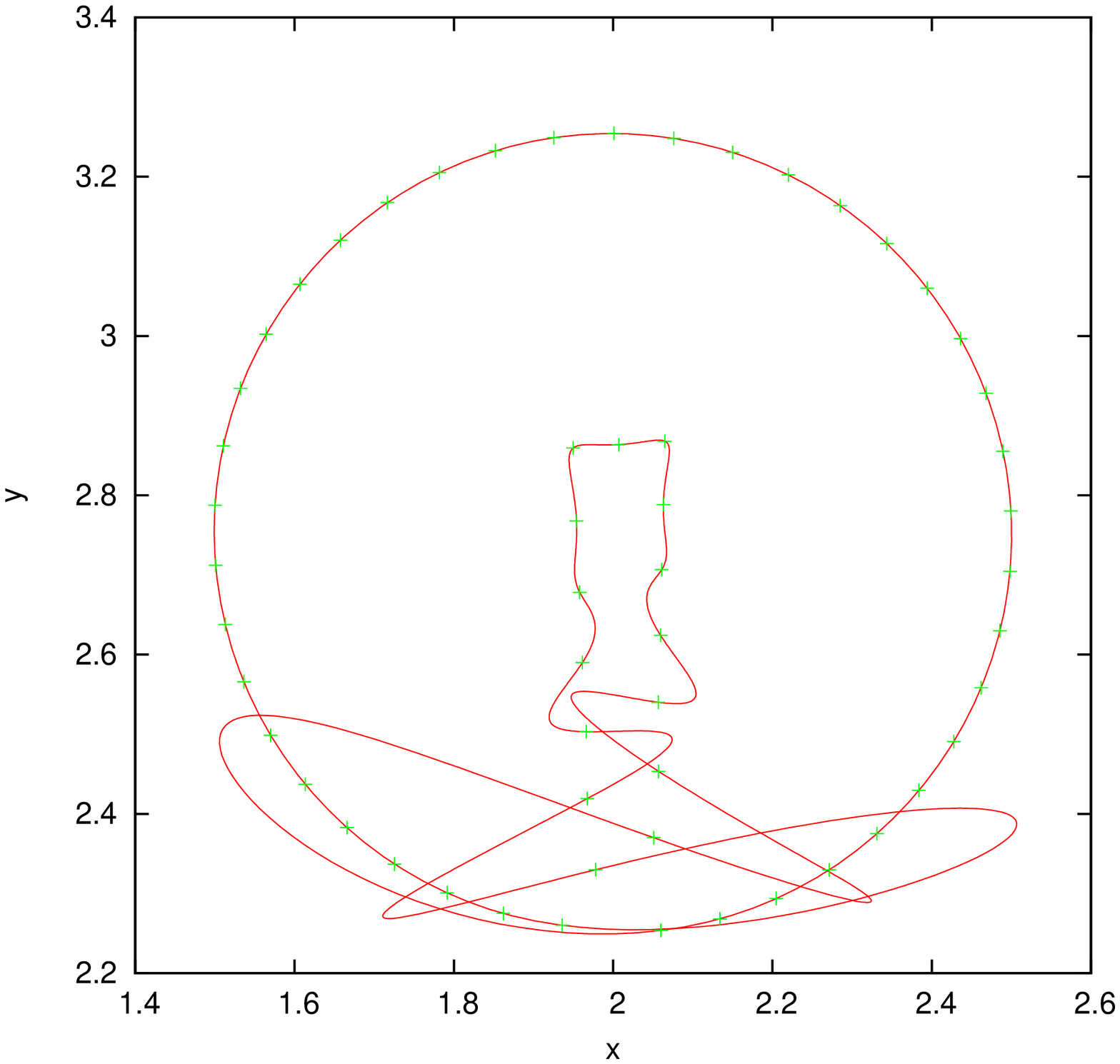}
}\\
$ N = 56, t = T $
\end{minipage}\\
\begin{minipage}[t][0.23\textheight][t]{0.5\linewidth}
\centering
\rotatebox{270}{
\includegraphics[height=0.2\textheight]{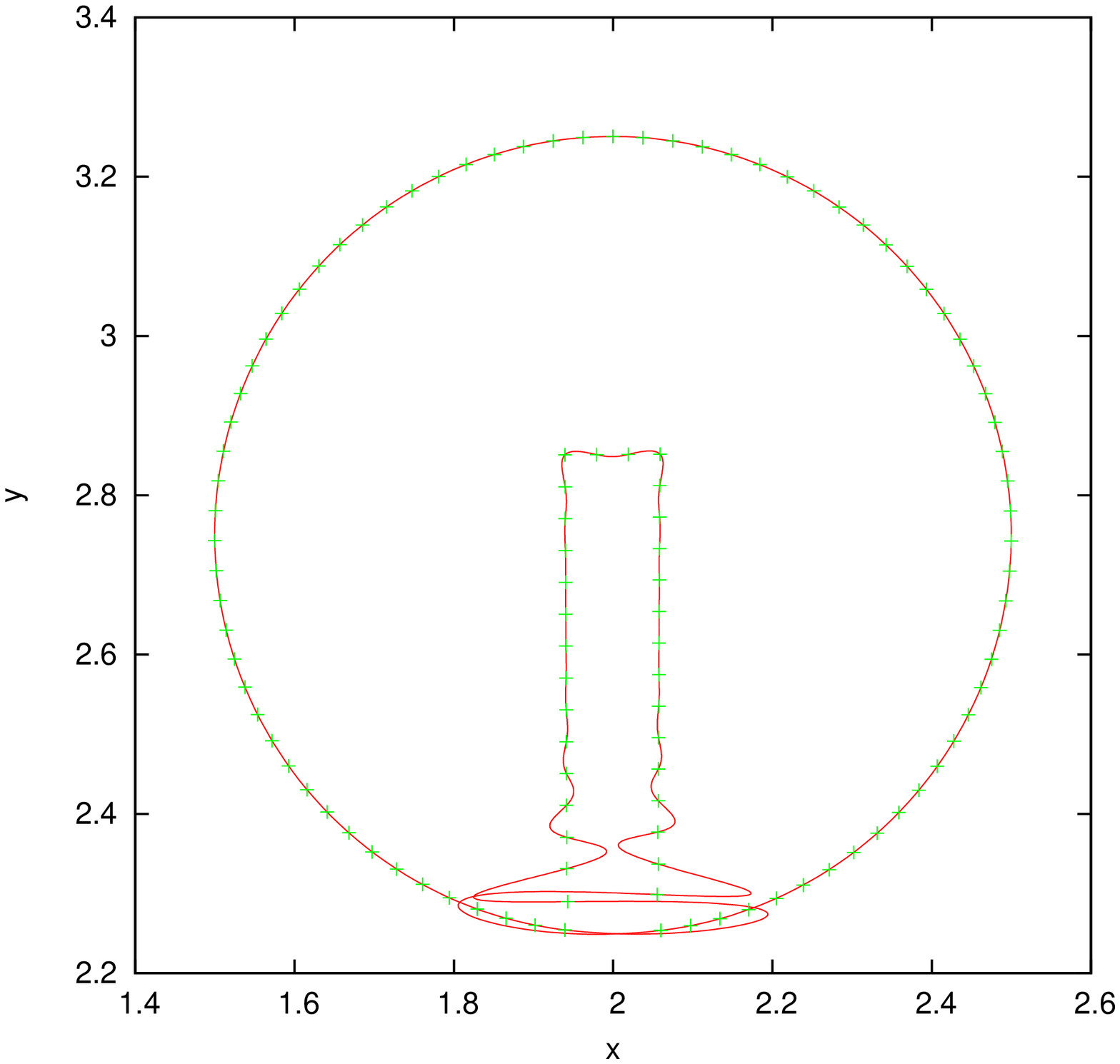}
}\\
{$ N =112, t = 0$}
\end{minipage} \nolinebreak
\begin{minipage}[t][0.23\textheight][t]{0.5\linewidth}
\centering
\rotatebox{270}{
\includegraphics[height=0.2\textheight]{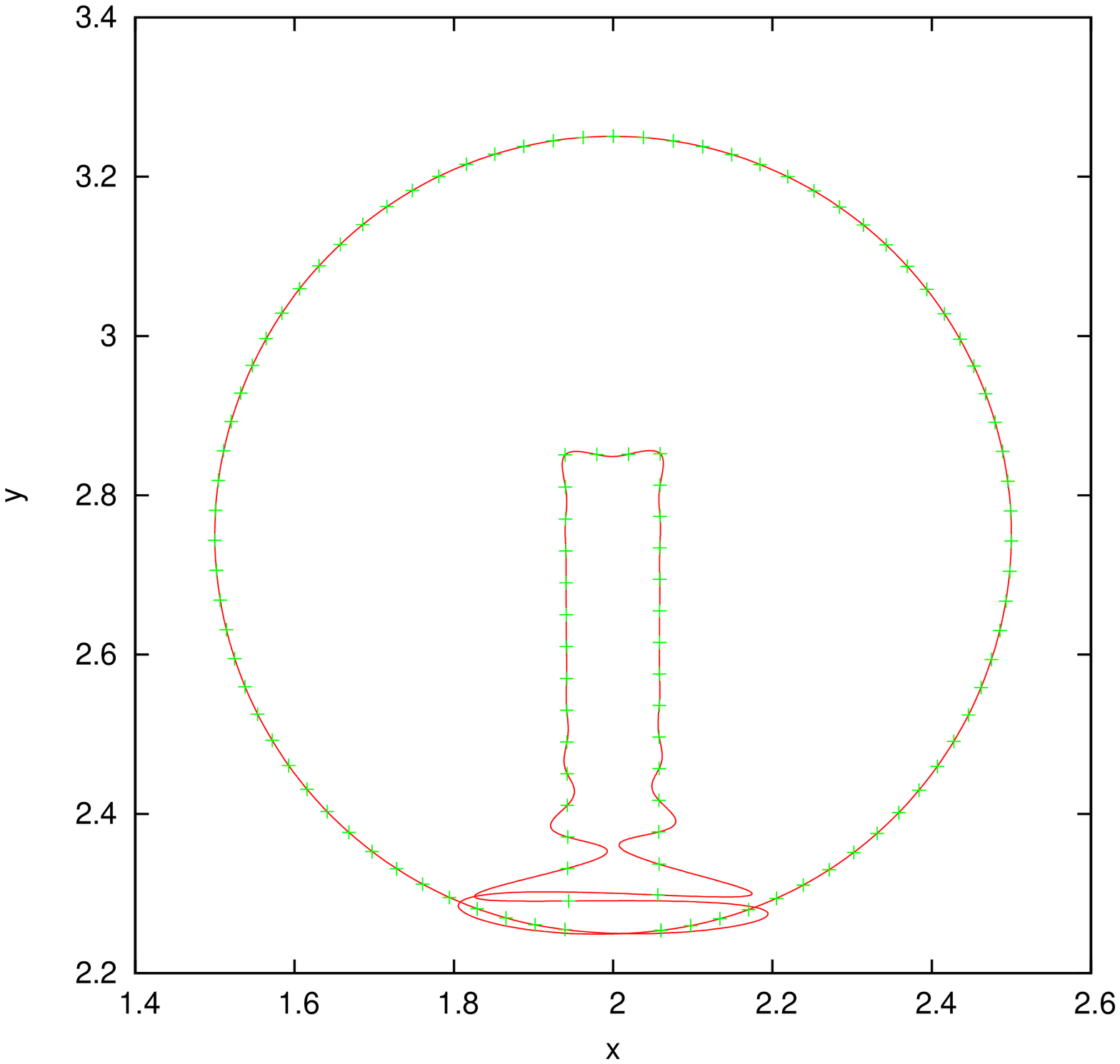}
}\\
$ N = 112, t = T $
\end{minipage}\\
\begin{minipage}[t][0.23\textheight][t]{0.5\linewidth}
\centering
\rotatebox{270}{
\includegraphics[height=0.2\textheight]{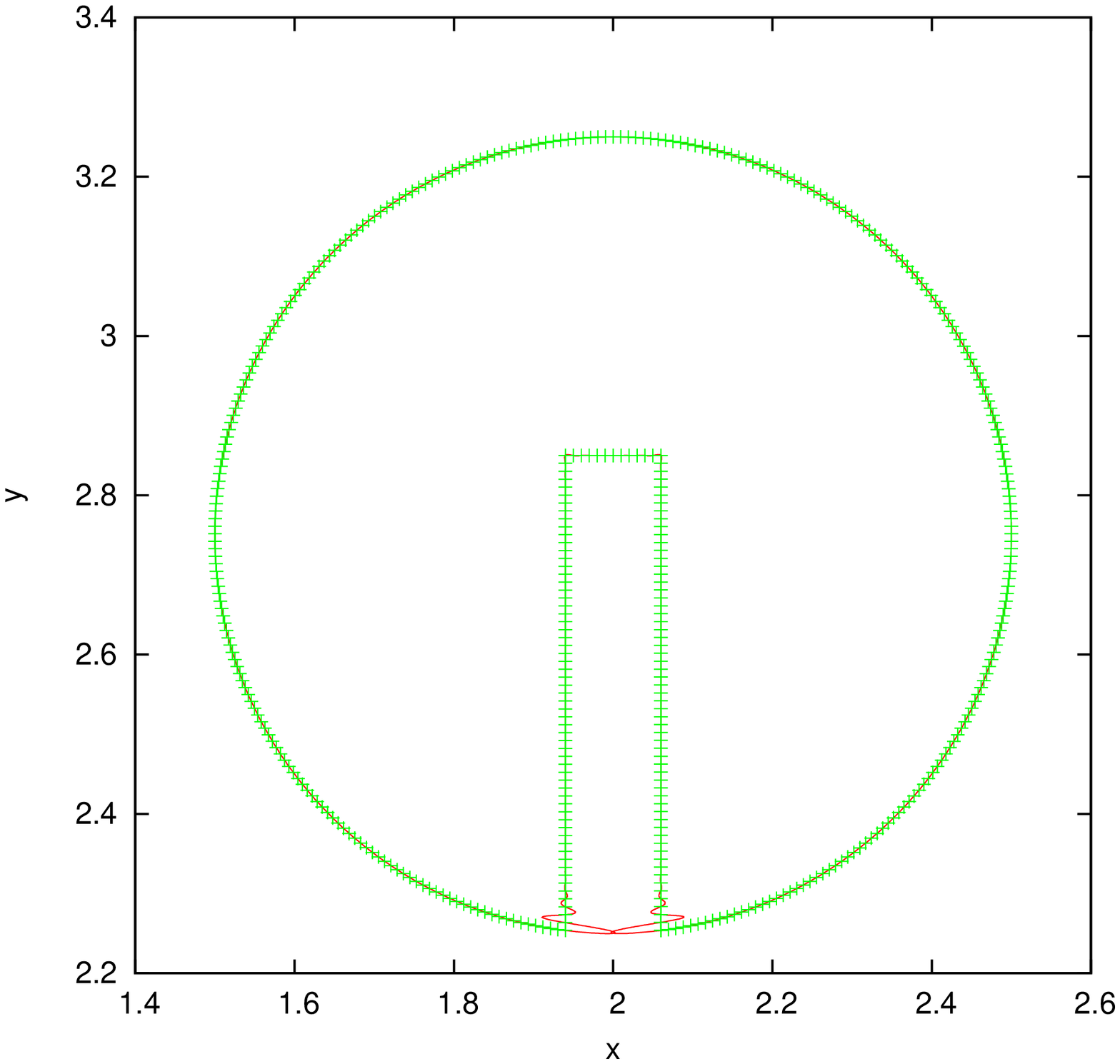}
}\\
$ N = 448, t = 0 $
\end{minipage} \nolinebreak
\begin{minipage}[t][0.23\textheight][t]{0.5\linewidth}
\centering
\rotatebox{270}{
\includegraphics[height=0.2\textheight]{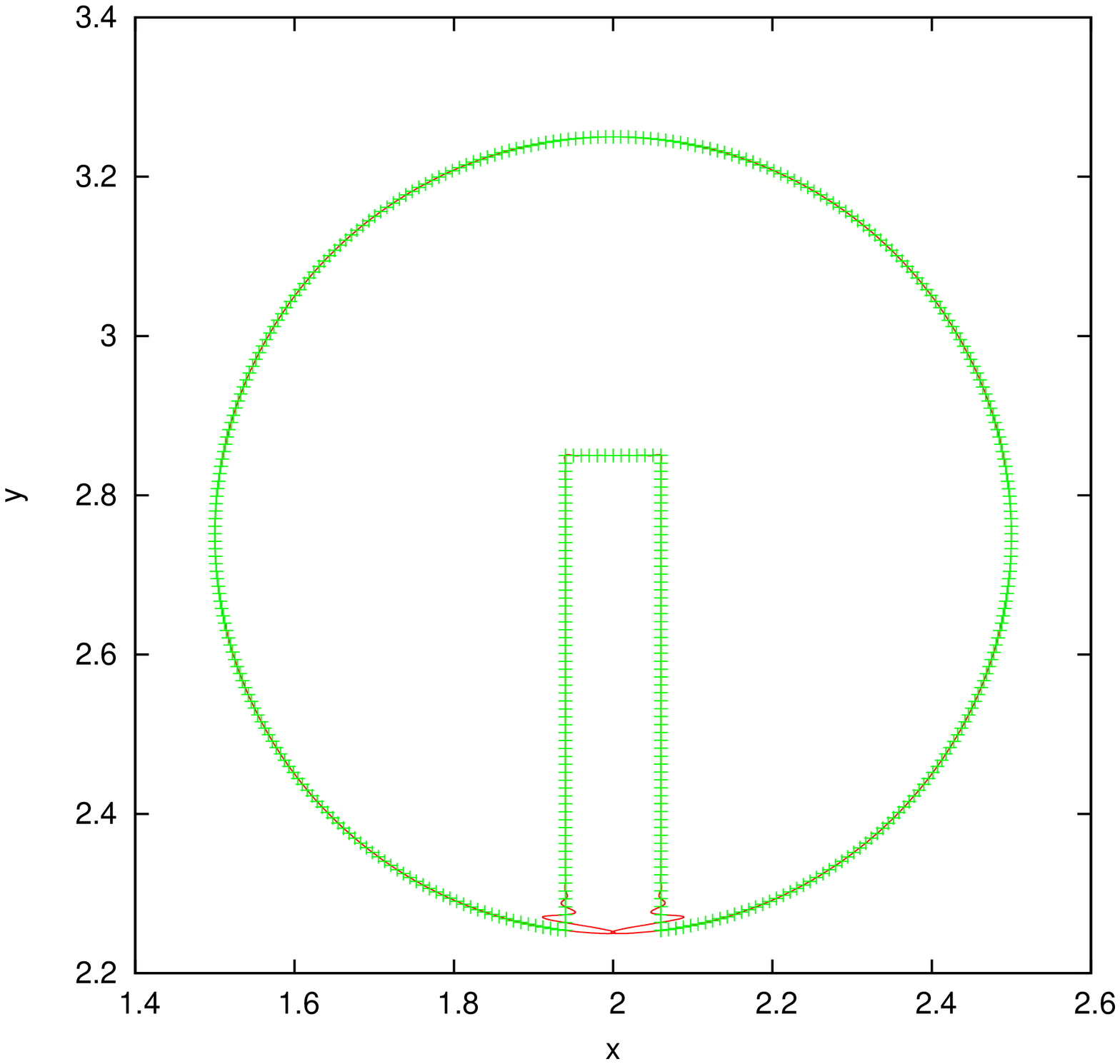}
}\\
$ N = 448, t = T $
\end{minipage}\\
\caption{Result of Zalesak's slotted disk rotation test for 
different resolutions $N $. The order of B-spline interpolation is $ P = 5$. 
Shown are the graphs at the initial position, $ t = 0 $,
and the position after a 
full rotation, $ t = T $. }
\label{fig:slottedP5}
\end{figure}

\begin{figure}
\centering
\begin{minipage}[t][0.23\textheight][t]{0.5\linewidth}
\centering
\rotatebox{270}{
\includegraphics[height=0.2\textheight]{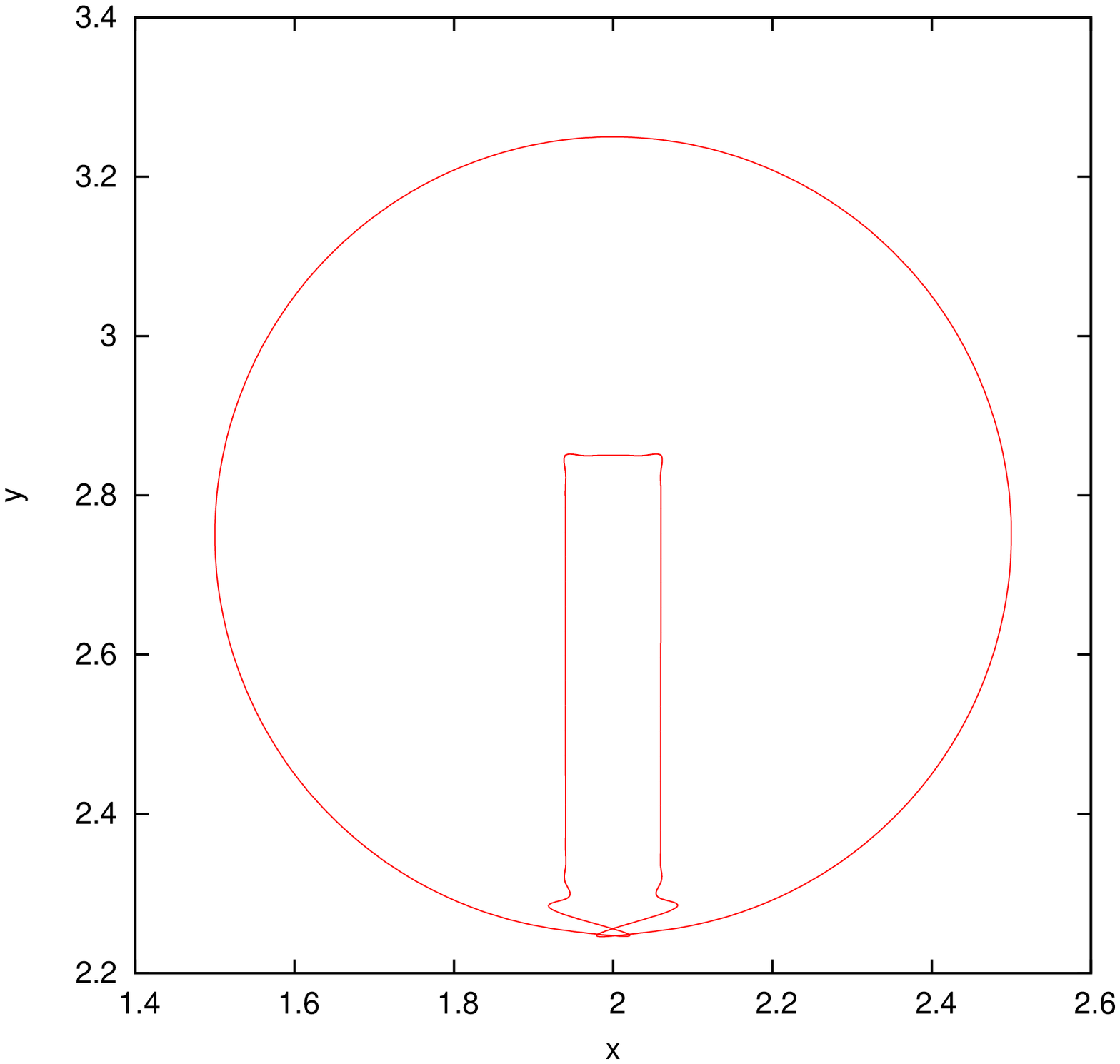}
}\\
{$ t = 0 $}
\end{minipage} \nolinebreak
\begin{minipage}[t][0.23\textheight][t]{0.5\linewidth}
\centering
\rotatebox{270}{
\includegraphics[height=0.2\textheight]{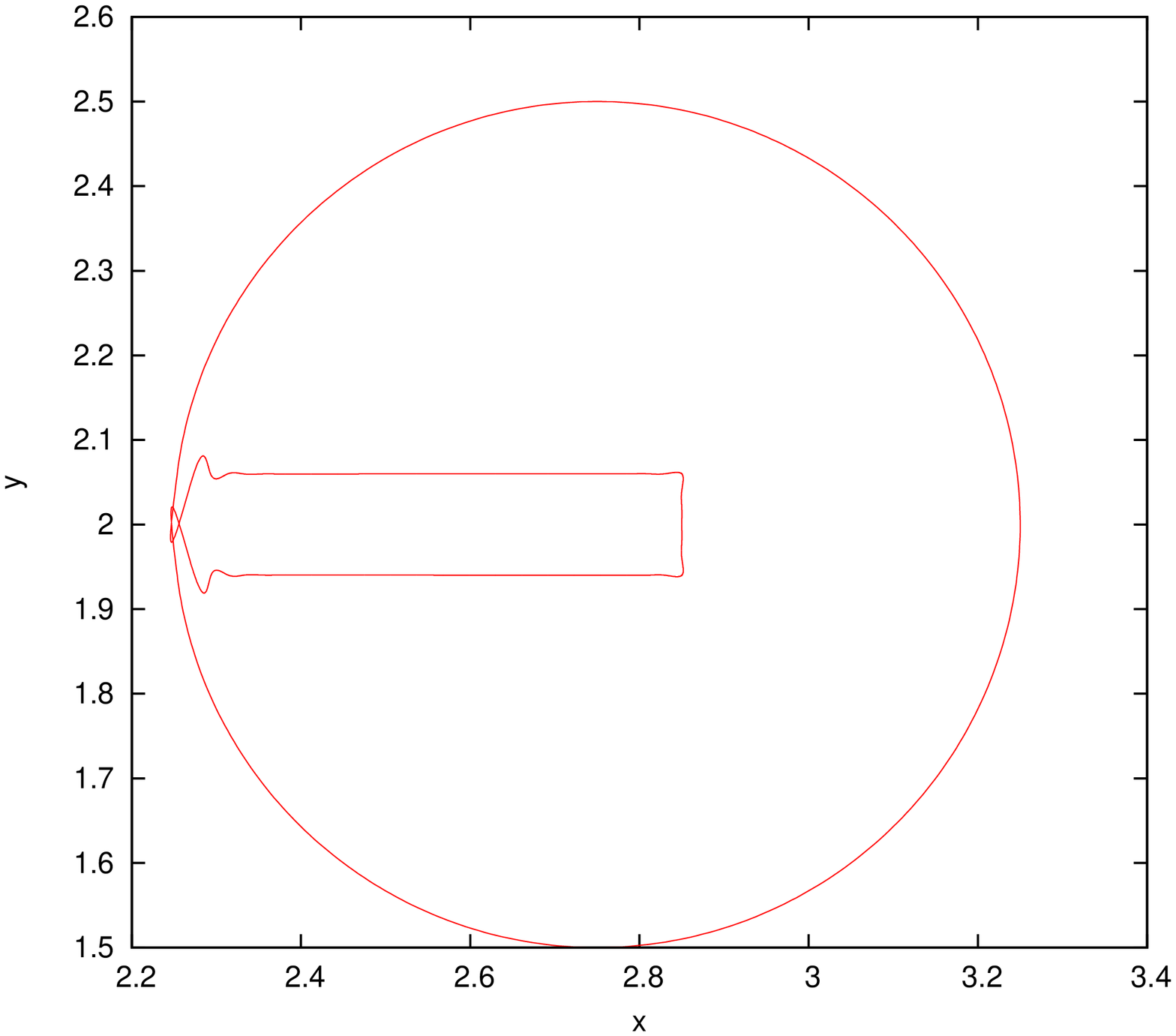}
}\\
$ t = T/4 $
\end{minipage}\\
\begin{minipage}[t][0.23\textheight][t]{0.5\linewidth}
\centering
\rotatebox{270}{
\includegraphics[height=0.2\textheight]{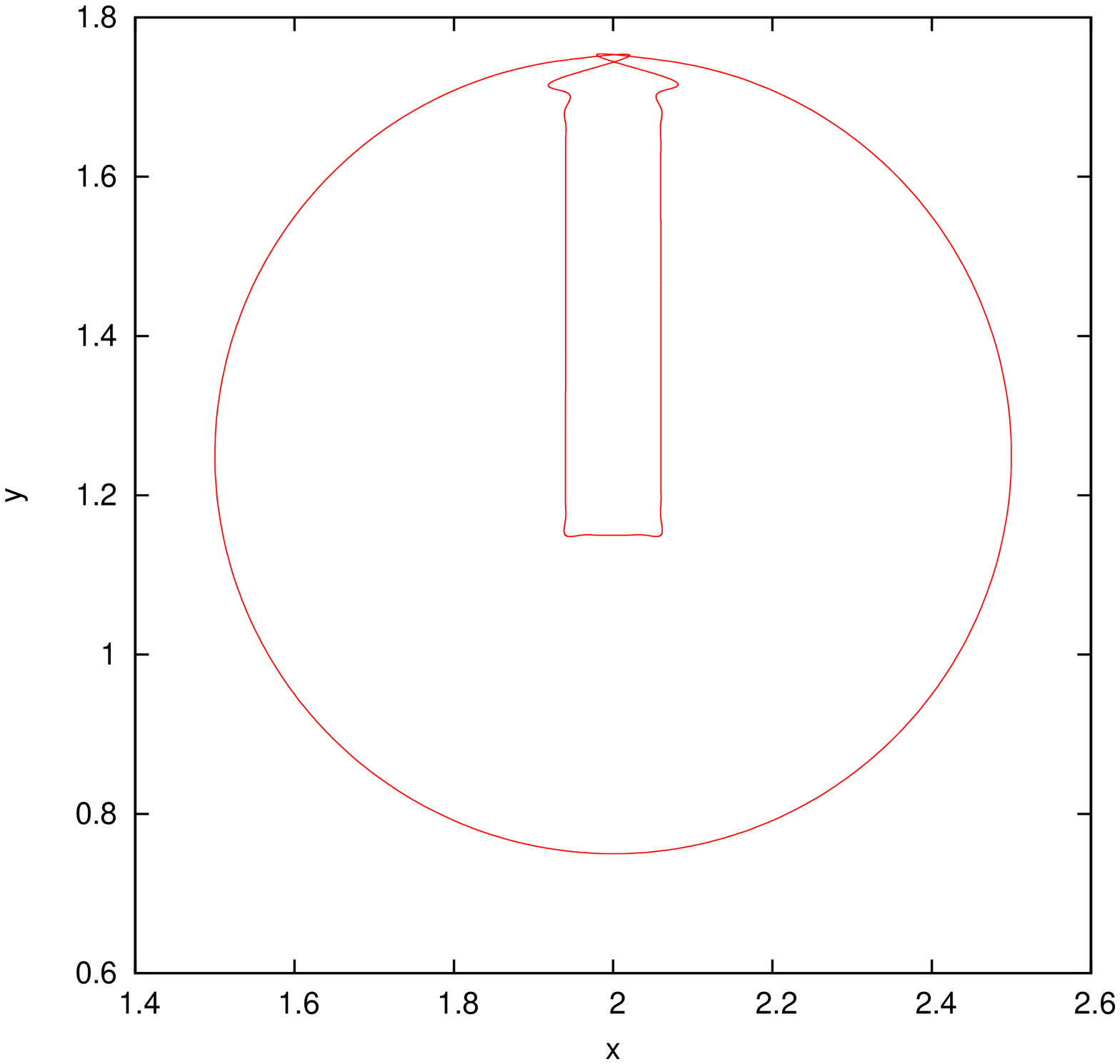}
}\\
$ t = T/2 $
\end{minipage} \nolinebreak
\begin{minipage}[t][0.23\textheight][t]{0.5\linewidth}
\centering
\rotatebox{270}{
\includegraphics[height=0.2\textheight]{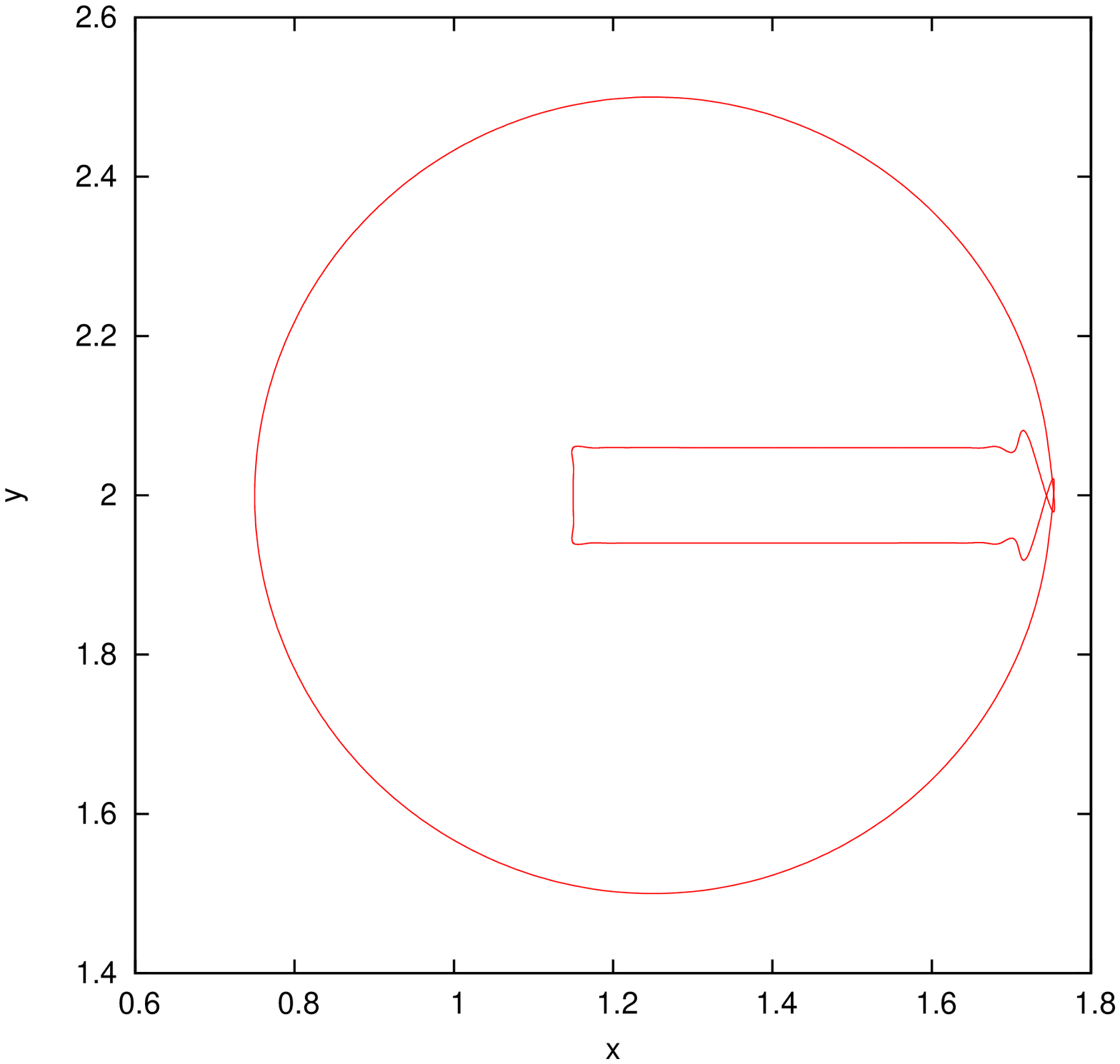}
}\\
$ t = 3T/4 $
\end{minipage}\\
\begin{minipage}[t][0.23\textheight][t]{0.5\linewidth}
\centering
\rotatebox{270}{
\includegraphics[height=0.2\textheight]{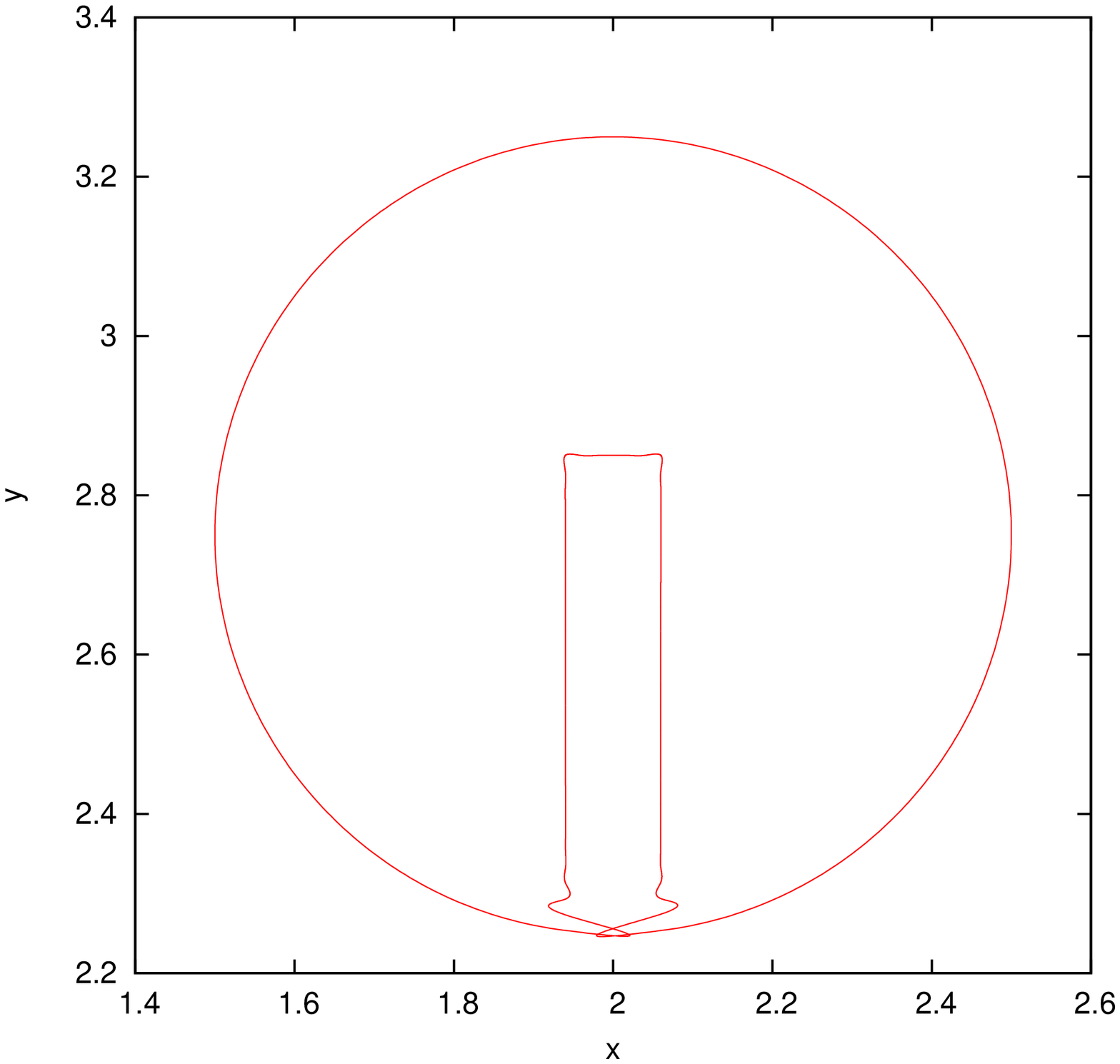}
}\\
{$ t = T$}
\end{minipage} \nolinebreak
\begin{minipage}[t][0.23\textheight][t]{0.5\linewidth}
\centering
\rotatebox{270}{
}\\
$  $
\end{minipage}\\
\caption{Result of Zalesak's slotted disk rotation test for 
$N = 224 $ and $ P = 3 $. 
The graphs are shown at different points in time. }
\label{fig:slottedP3Rotation}
\end{figure}

\begin{figure}
\centering
\rotatebox{270}{\includegraphics[height=10cm]{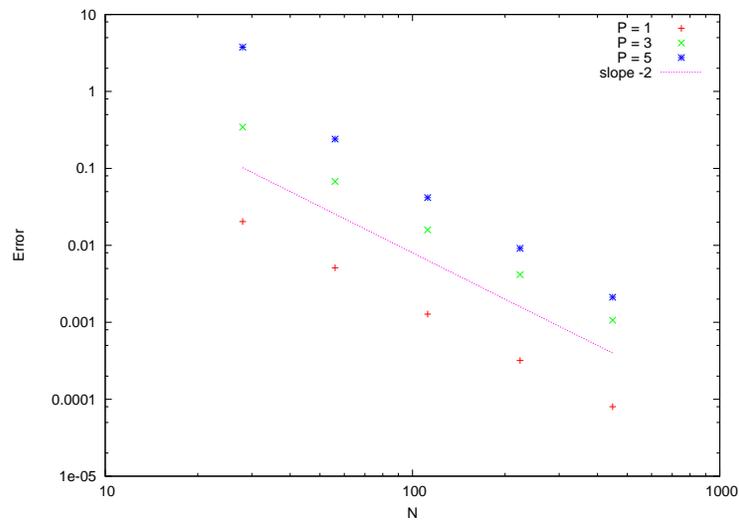}}
\caption{Error decrease for Zalesak's slotted disk rotation test. 
Due to the discontinuity in the first derivative of the solution,
the present method exhibits only second order accuracy with respect to
the resolution $ N $. In addition, the fifth order B-spline interpolation
is less accurate than the third order one, since the
spurious oscillations are larger in the former case. }
\label{fig:slottedError}
\end{figure}

\begin{figure}
\centering
\begin{minipage}[t][0.23\textheight][t]{0.5\linewidth}
\centering
\rotatebox{270}{
\includegraphics[height=0.2\textheight]{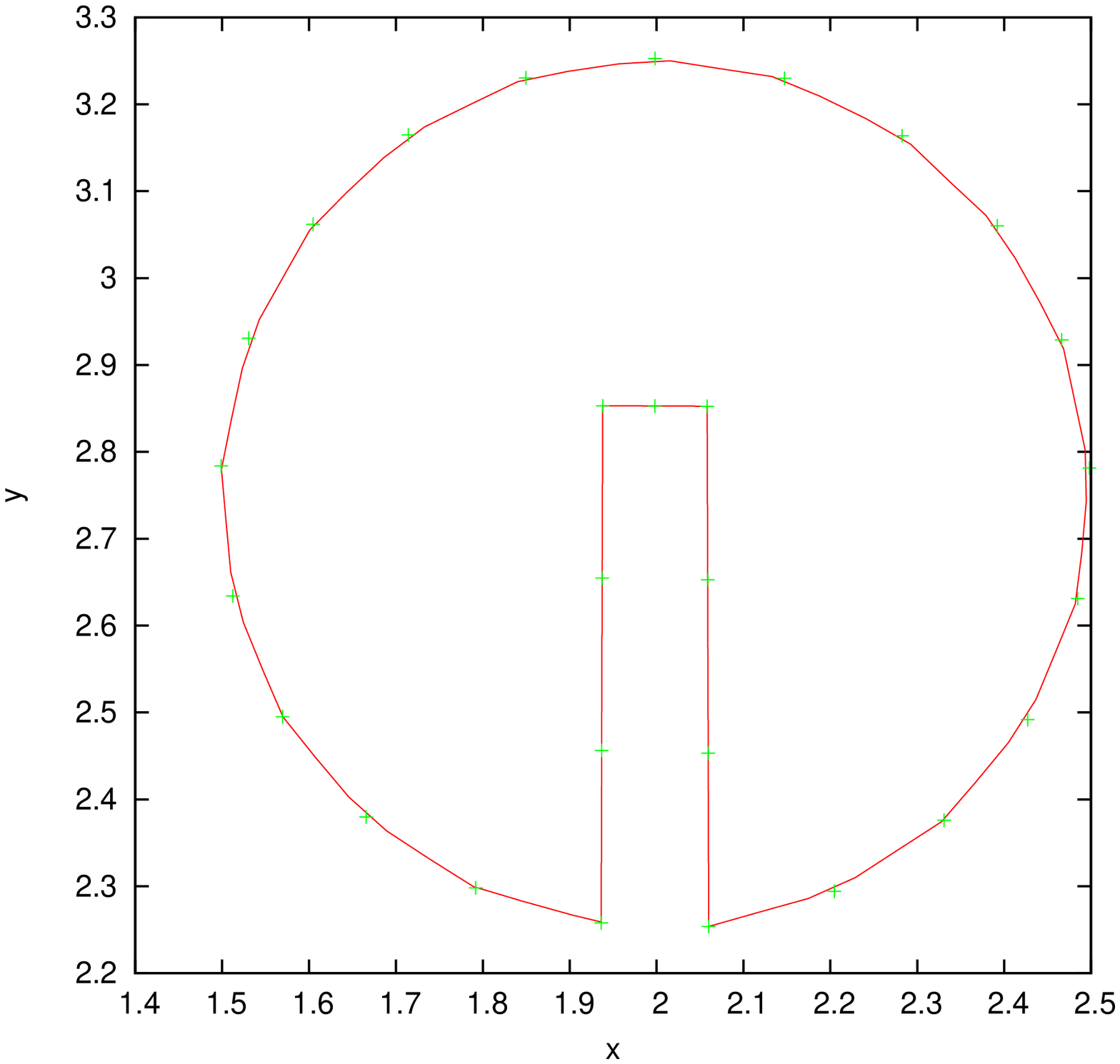}
}\\
{$ t = 0 $}
\end{minipage} \nolinebreak
\begin{minipage}[t][0.23\textheight][t]{0.5\linewidth}
\centering
\rotatebox{270}{
\includegraphics[height=0.2\textheight]{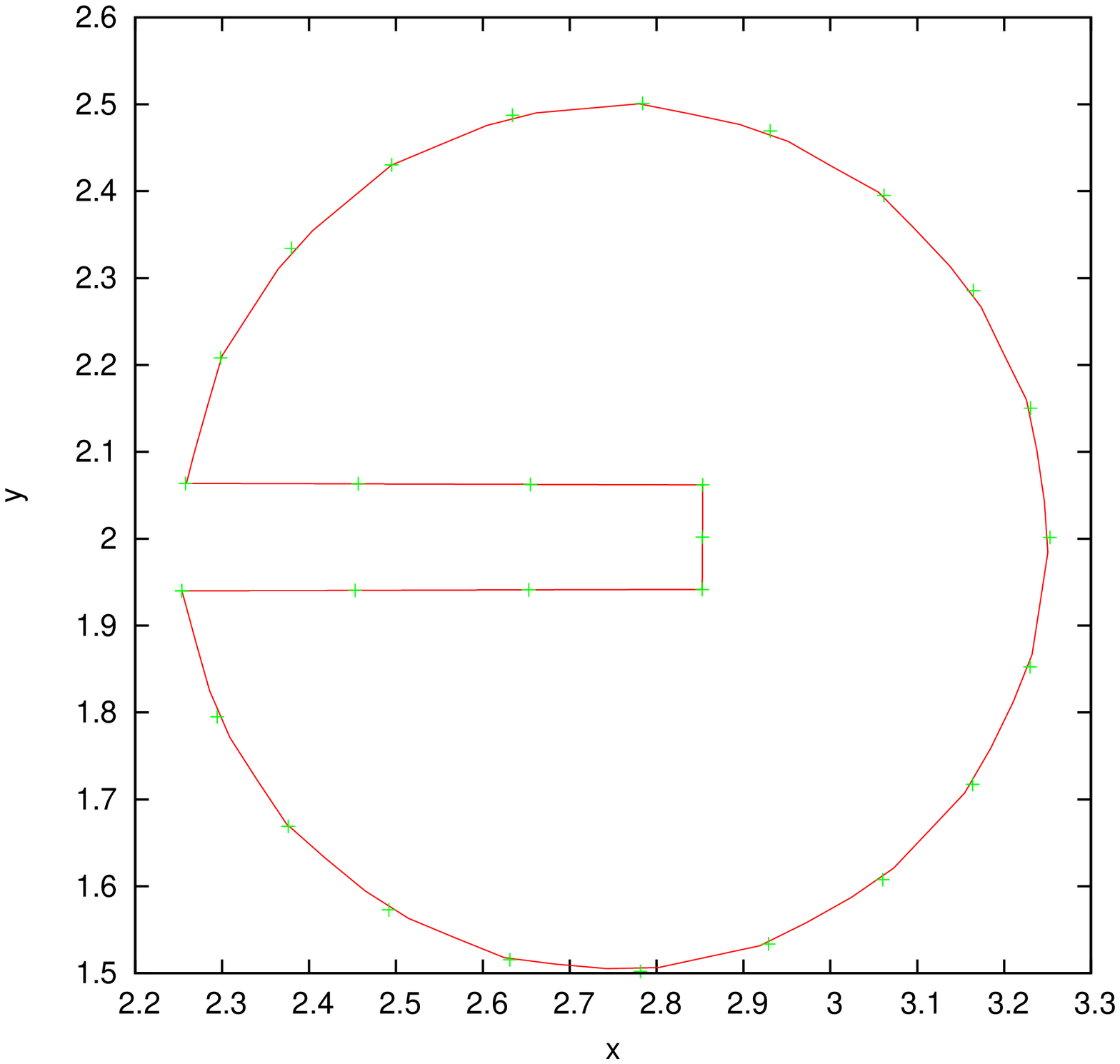}
}\\
$ t = T/4 $
\end{minipage}\\
\begin{minipage}[t][0.23\textheight][t]{0.5\linewidth}
\centering
\rotatebox{270}{
\includegraphics[height=0.2\textheight]{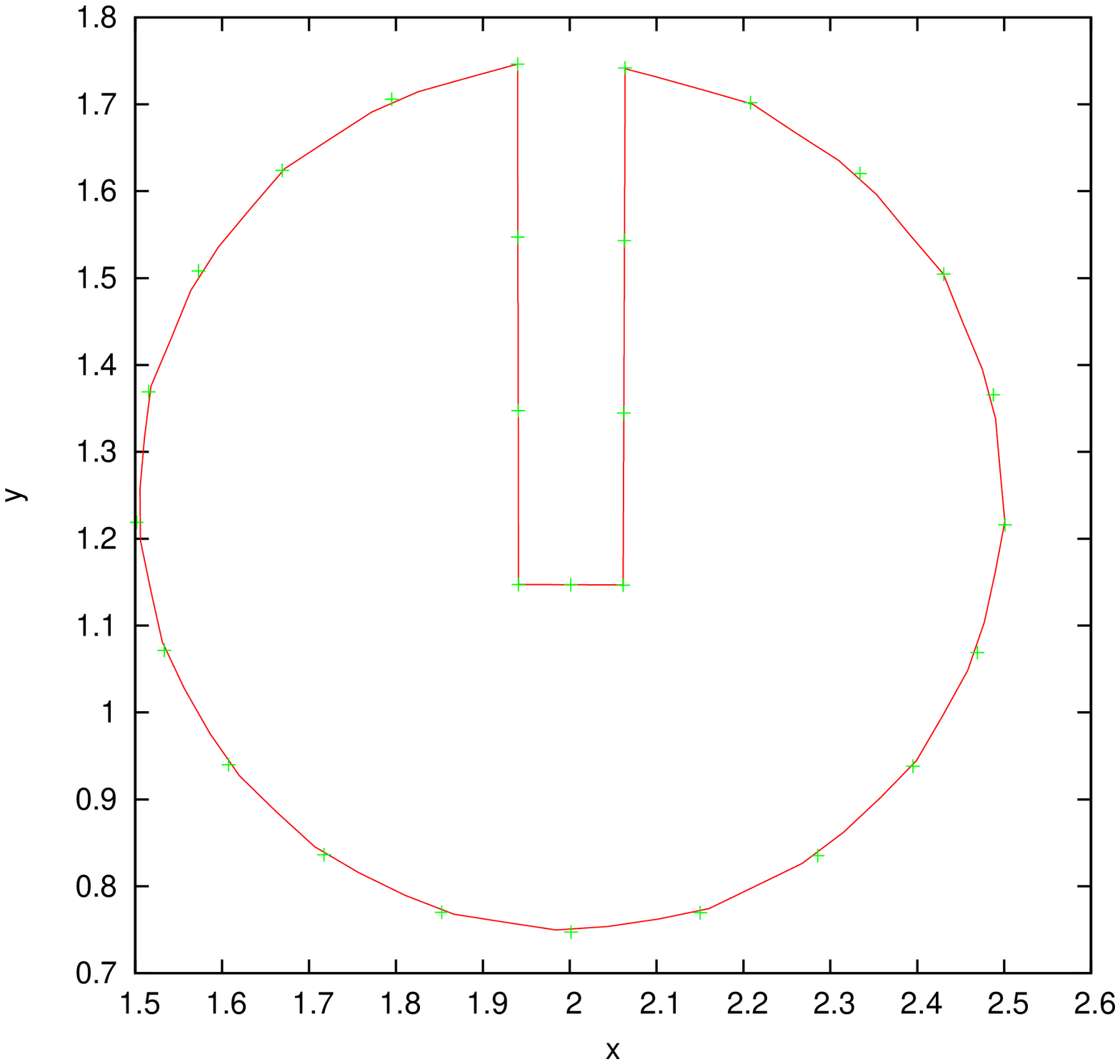}
}\\
$ t = T/2 $
\end{minipage} \nolinebreak
\begin{minipage}[t][0.23\textheight][t]{0.5\linewidth}
\centering
\rotatebox{270}{
\includegraphics[height=0.2\textheight]{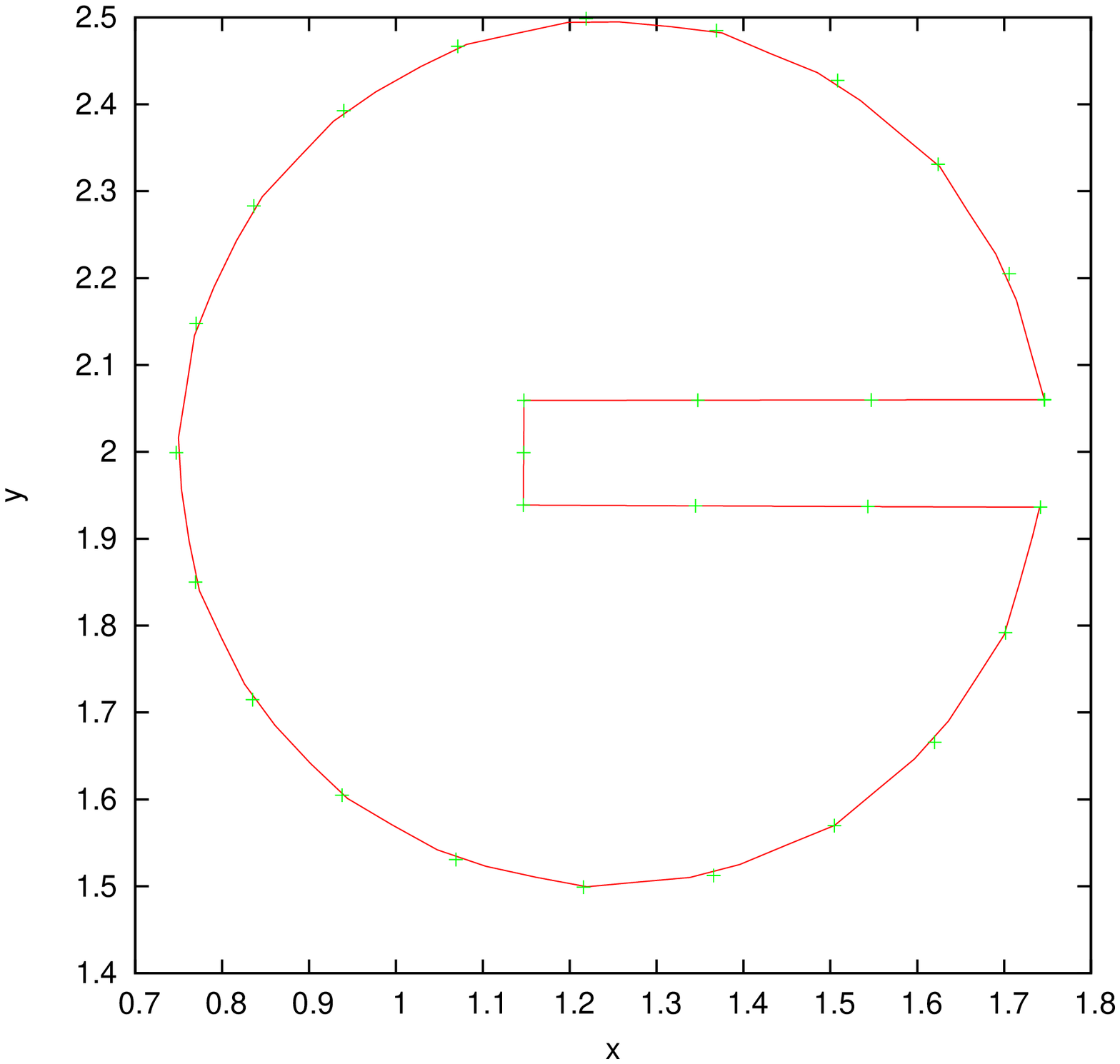}
}\\
$ t = 3T/4 $
\end{minipage}\\
\begin{minipage}[t][0.23\textheight][t]{0.5\linewidth}
\centering
\rotatebox{270}{
\includegraphics[height=0.2\textheight]{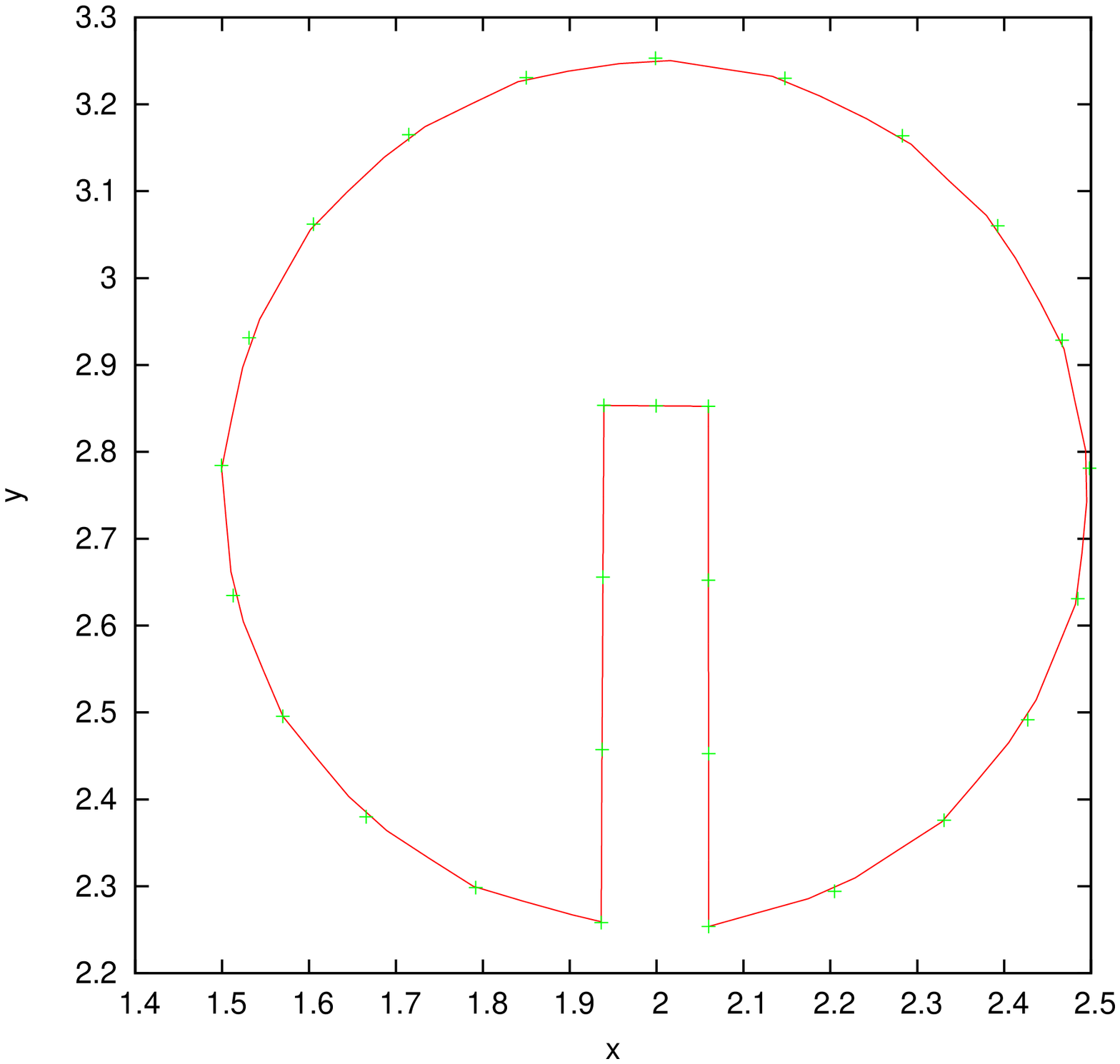}
}\\
{$ t = T$}
\end{minipage} \nolinebreak
\begin{minipage}[t][0.23\textheight][t]{0.5\linewidth}
\centering
\rotatebox{270}{
}\\
$  $
\end{minipage}\\
\caption{Result of Zalesak's slotted disk rotation test for 
$N = 28 $ and $ P = 1 $. 
The graphs are shown at different points in time. }
\label{fig:slottedP1Rotation}
\end{figure}

The slotted disk rotation test of Zalesak \cite{Zalesak1979} 
uses a solid body rotation 
to advect a slotted disk. The stream function $ \psi $ is given by 
\be
\psi(x,y) =  \frac{\omega}{2} \left\{ ( x - x_0 )^2 + ( y- y_0)^2 \right\},
\ee
where $ \omega $ is chosen in such a way as to allow a complete rotation of
the drop in $ 2524 $ iterations. The computational box is four by four and the 
drop has a radius of $ r_0 = 1/2 $. It is situated at 
$ ( x_0 , y_0 ) = ( 2, 2.75) $. 
The coordinates of the four corners of the slot are given by:
\begin{eqnarray}
(x_a , y_a ) &=&( x_0 + \frac{3}{50} , y_0 - \sqrt{ r_0^2 - (x_a - x_0 )^2 } )\\
(x_b , y_b ) &=&( x_a , y_0 + \frac{5}{50} ) \\
(x_c , y_c ) &=&( x_0 - \frac{3}{50} , y_b ) \\
(x_d , y_d ) &=&( x_c , y_a ).
\end{eqnarray}
As an initial condition we 
chose a description of the initial interface by means of four functions
$ \alpha_1, \alpha_2, \alpha_3, \alpha_4 $, defined the following way:
\begin{eqnarray}
\alpha_1(s) & = & 
\frac{1}{2} \left( r_0^2 s - 
r_0 \left( x_0 \left( \cos \left(s+s_a \right) - \cos \left( s_a \right) \right)+
y_0 \left( \sin \left( s + s_a \right) - \sin \left( s_a \right) \right) \right)
\right),\\
\alpha_2(s) & = & \frac{1}{2}x_c s,\\
\alpha_3(s) & = & -\frac{1}{2}y_b s,\\
\alpha_4(s) & = & \frac{1}{2}x_a s, \\
\end{eqnarray}
where $ s_a = \arcsin \left( \frac{x_a - x_0 }{r_0 } \right) $ and
$ s_b = \arcsin \left( \frac{x_0 - x_c }{r_0 } \right) $.
The function $ \alpha $ is composed by means of these four functions.
\be
\alpha(s) = \left\{ \begin{array}{ll} \alpha_1(s) & 0 < s \le 2 \pi - s_b - s_a \\
  \begin{array}{l}
    \! \! \alpha_2(s - 2 \pi + s_b + s_a ) \\
    \, \, + \alpha_1(2 \pi - s_b - s_a) 
    \end{array}
  & \begin{array}{l} 
      \! \! 2 \pi - s_b - s_a \\
      \, \, < s \le 2 \pi - s_b - s_a + y_b - y_a 
      \end{array} \\
  \begin{array}{l}
\! \! \alpha_3(s- 2 \pi + s_b + s_a - y_b + y_a ) \\
\, \, + \alpha_2(y_b - y_a )\\
\, \, + \alpha_1(2 \pi - s_b - s_a) 
\end{array}
& \begin{array}{l} 
\! \! 2 \pi - s_b - s_a + y_b - y_a \\
\, \, < s \\
\, \, \le 2 \pi - s_b - s_a + y_b - y_a + x_a - x_c 
\end{array}
\\
 \begin{array}{l}
\! \! \alpha_4(2 \pi - s_b - s_a + y_b - y_a + x_a - x_c - s ) \\
\, \, +\alpha_3(x_a - x_c ) \\
\, \, + \alpha_2(y_b - y_a ) \\
\, \, + \alpha_1(2 \pi - s_b - s_a) \end{array}
& \begin{array}{l} 
\! \! 2 \pi - s_b - s_a + y_b - y_a + x_a - x_c \\
\, \, < s \\
\, \, \le 2 \pi - s_b - s_a + 2(y_b - y_a) + x_a - x_c 
\end{array}
\end{array}
\right.
\ee
The parameter $ s $ takes values in the interval
$ [ 0 , 2 \pi - s_b - s_a + 2(y_b - y_a) + x_a - x_c  ] $ this time. 
We discretized this interval in such a way that the corner positions of the 
slot are at knots of the discretization. The function $ \alpha $ is then 
interpolated at these knots. The function $ \beta $ is handled likewise, 
with $ \beta $ given by:
\be
\beta(s) = \left\{ \begin{array}{ll} 
\begin{array}{l} 
\! \! \left( r_0 \sin \left( s + s_a \right) + x_0 \right)\\
\, \, \left( -r_0 \cos \left( s + s_a \right) + y_0 \right) 
\end{array} &
0 < s \le 2 \pi - s_b - s_a \\
x_d(y_a + s -  2 \pi + s_b + s_a ) &
\begin{array}{l} 
  \! \! 2 \pi - s_b - s_a \\
  \, \, < s \le 2 \pi - s_b - s_a + y_b - y_a 
\end{array} \\
\begin{array}{l} 
\! \! y_b(x_d + s - 2 \pi + s_b + \\
\,\, s_a - y_b + y_a + x_d ) 
\end{array}&
 \begin{array}{l} 
\! \! 2 \pi - s_b - s_a + y_b - y_a \\
\, \, < s \\
\, \, \le 2 \pi - s_b - s_a + y_b - y_a + x_a - x_c 
\end{array}
\\
 \begin{array}{l} 
\!\! x_a(y_b - s + 2 \pi - s_b - \\
\, \, s_a + y_b - y_a + x_a - x_c )
\end{array}
 & \begin{array}{l} 
\! \! 2 \pi - s_b - s_a + y_b - y_a + x_a - x_c \\
\, \, < s \\
\, \, \le 2 \pi - s_b - s_a + 2(y_b - y_a) + x_a - x_c. 
   \end{array}
\end{array}
\right.
\ee
The initial condition is only $ \mathcal{C}^0 $ because of the
kinks at the corners of the slot. Although using high order periodic B-splines,
we expect the convergence rate therefore to be of only second order at most. 
In addition, discontinuities can give rise to the development of 
spurious oscillations of the interpolant at these discontinuities,
the so called Gibbs phenomenon \cite{CourantHilbert1924}. 
The higher the order of the periodic B-splines the further these spurious
oscillations spread along the interface, as can be seen when comparing
figures \ref{fig:slottedP3} and \ref{fig:slottedP5}, where we compare the
interface position at time $ t = 0 $ and $ t = T $, 
after one rotation, for
different resolutions. Intermediate steps are shown for $ N = 160 $ and 
$ P = 3$ in figure \ref{fig:slottedP3Rotation}, indicating that the
main contribution to the overall error does indeed not come from the
advection but from the interpolation of a function with 
discontinuities in its first derivative. The order of convergence is
reduced to second order no matter the order of the B-spline interpolation,
cf. figure \ref{fig:slottedError}, resp. table \ref{tab:slottedError}. 
The absolute error is even larger for
higher order B-spline interpolation, due to the larger spurious oscillations. 
We remark that choosing an order $ P = 1 $ for the B-spline interpolation
gives us a PLIC like description of the interface, which for the present
benchmark test produces more accurate results since it only requires 
$ \mathcal{C}^0 $ continuity of the function to be interpolated. This leads to 
the result that for $ P = 1 $ the slot stays sharp during the entire 
simulation as can be seen in figure \ref{fig:slottedP1Rotation}. 

\begin{table}
\centering
\[
\begin{array}{l|l|l|l}
P & N  & \mbox{Error} & O \\
\hline
1 & 28 &2.03638\times 10^{-2} & - \\
  & 56 &5.10861\times 10^{-3} & 1.99 \\
  & 112&1.27831\times 10^{-3} & 2.00  \\
  & 224&3.19646\times 10^{-4} & 2.00  \\
  & 448&7.99148\times 10^{-5} & 2.00  \\
\hline
3 & 28 &3.82303\times 10^{-1} & - \\
  & 56 &6.76916\times 10^{-2} &2.50\\
  & 112&1.58615\times 10^{-2} &2.09\\
  & 224&4.25359\times 10^{-3} &1.90\\
  & 448&1.08273\times 10^{-3} &1.94\\
\hline
5 & 28 &3.76212\times 10^{0}  & - \\
  & 56 &2.40403\times 10^{-1} &3.97\\
  &112 &4.16235\times 10^{-2} &2.53\\
  &224 &9.12356\times 10^{-3} &2.19\\
  &448 &2.12014\times 10^{-3} &2.11
\end{array}
\]
\caption{Results for Zalesak's slotted disk rotation test for $ P = 3 $ and
$ P = 5 $.}
\label{tab:slottedError}
\end{table}

\clearpage
\subsection{Numerical verification part 3}

\begin{figure}
\centering
\begin{minipage}[t][0.23\textheight][t]{0.5\linewidth}
\centering
\rotatebox{270}{
\includegraphics[height=0.2\textheight]{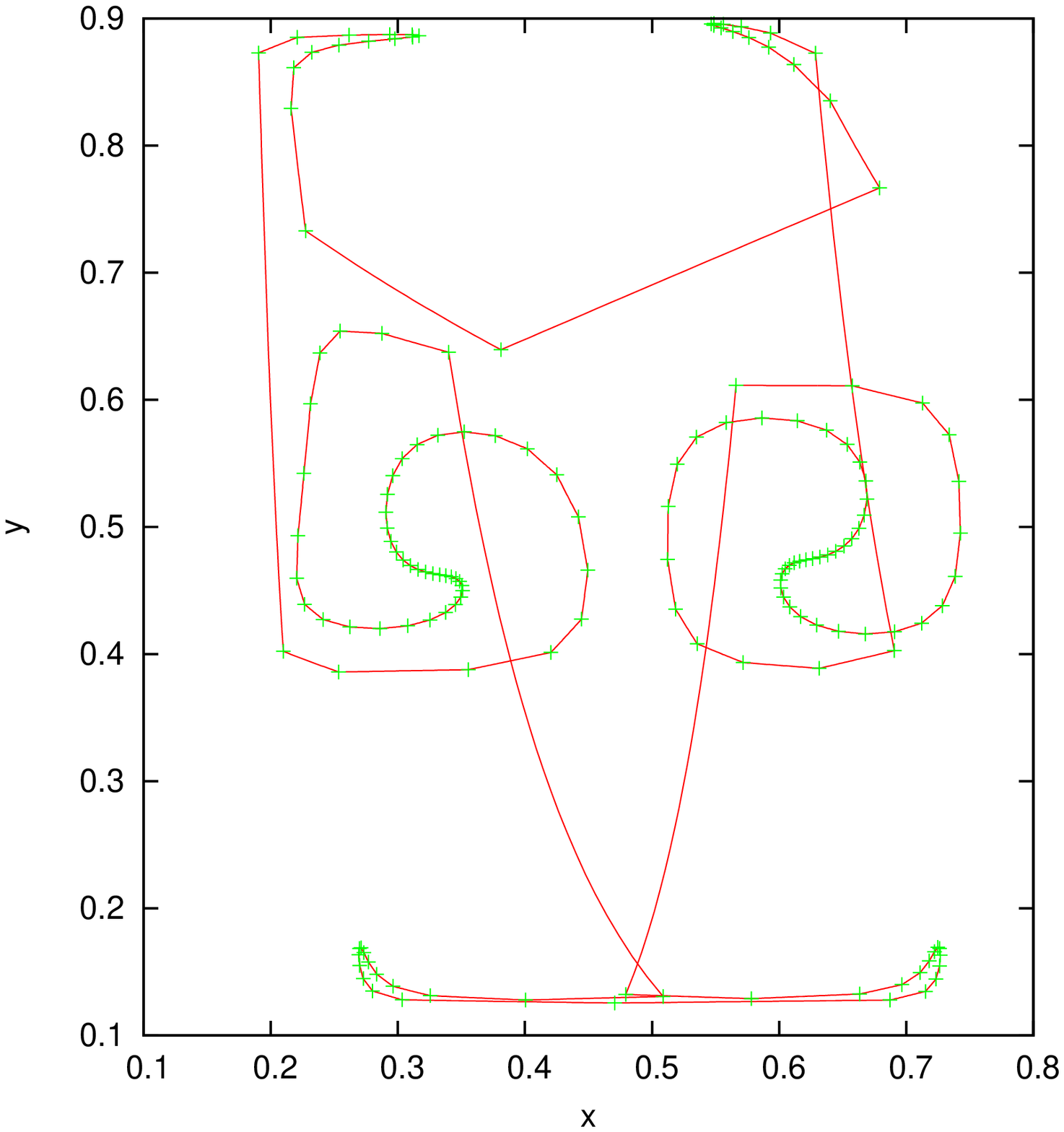}
}\\
{$ N =156, t = T/2 $}
\end{minipage} \nolinebreak
\begin{minipage}[t][0.23\textheight][t]{0.5\linewidth}
\centering
\rotatebox{270}{
\includegraphics[height=0.2\textheight]{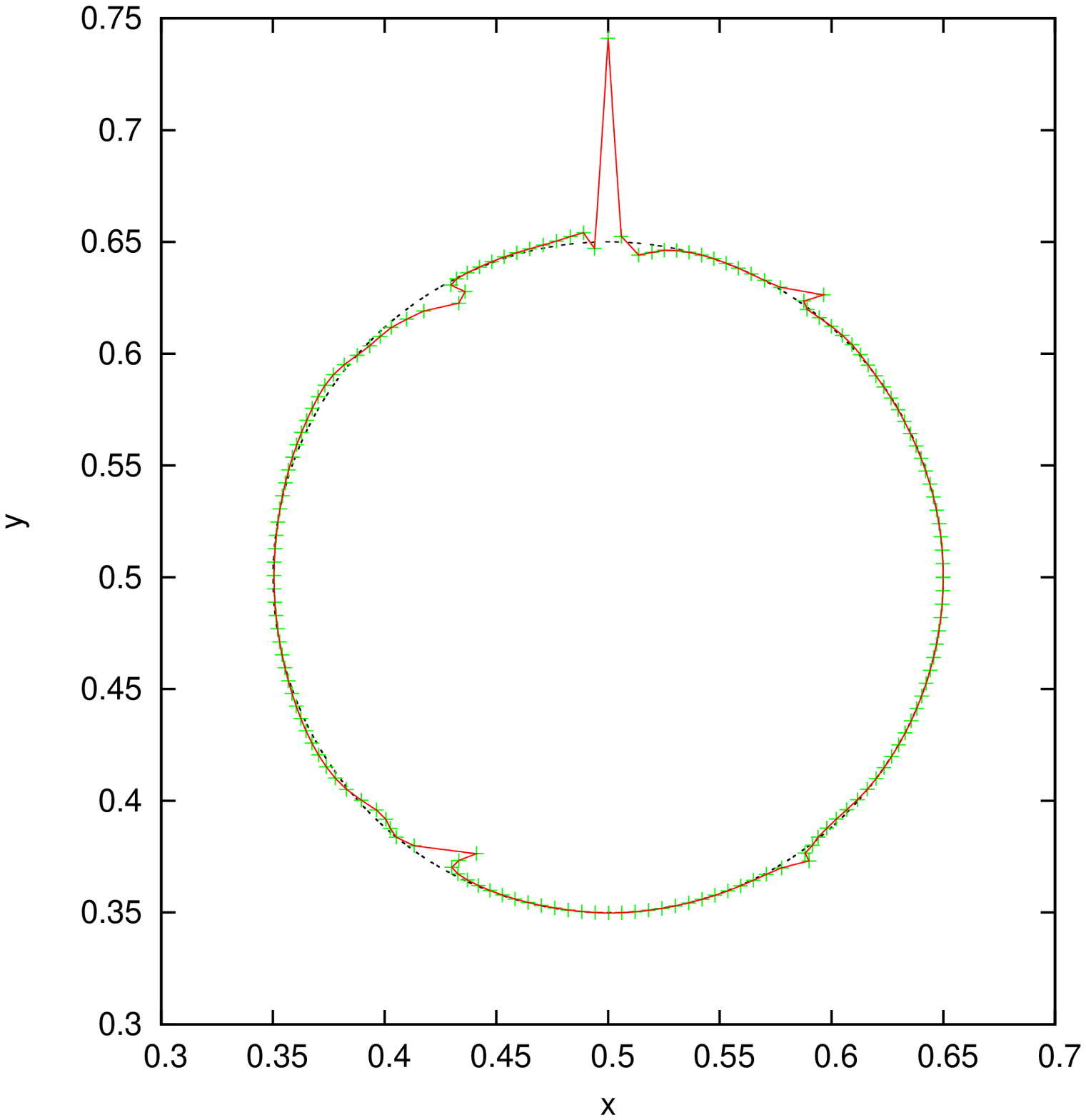}
}\\
$ N =156, t = T $
\end{minipage}\\
\begin{minipage}[t][0.23\textheight][t]{0.5\linewidth}
\centering
\rotatebox{270}{
\includegraphics[height=0.2\textheight]{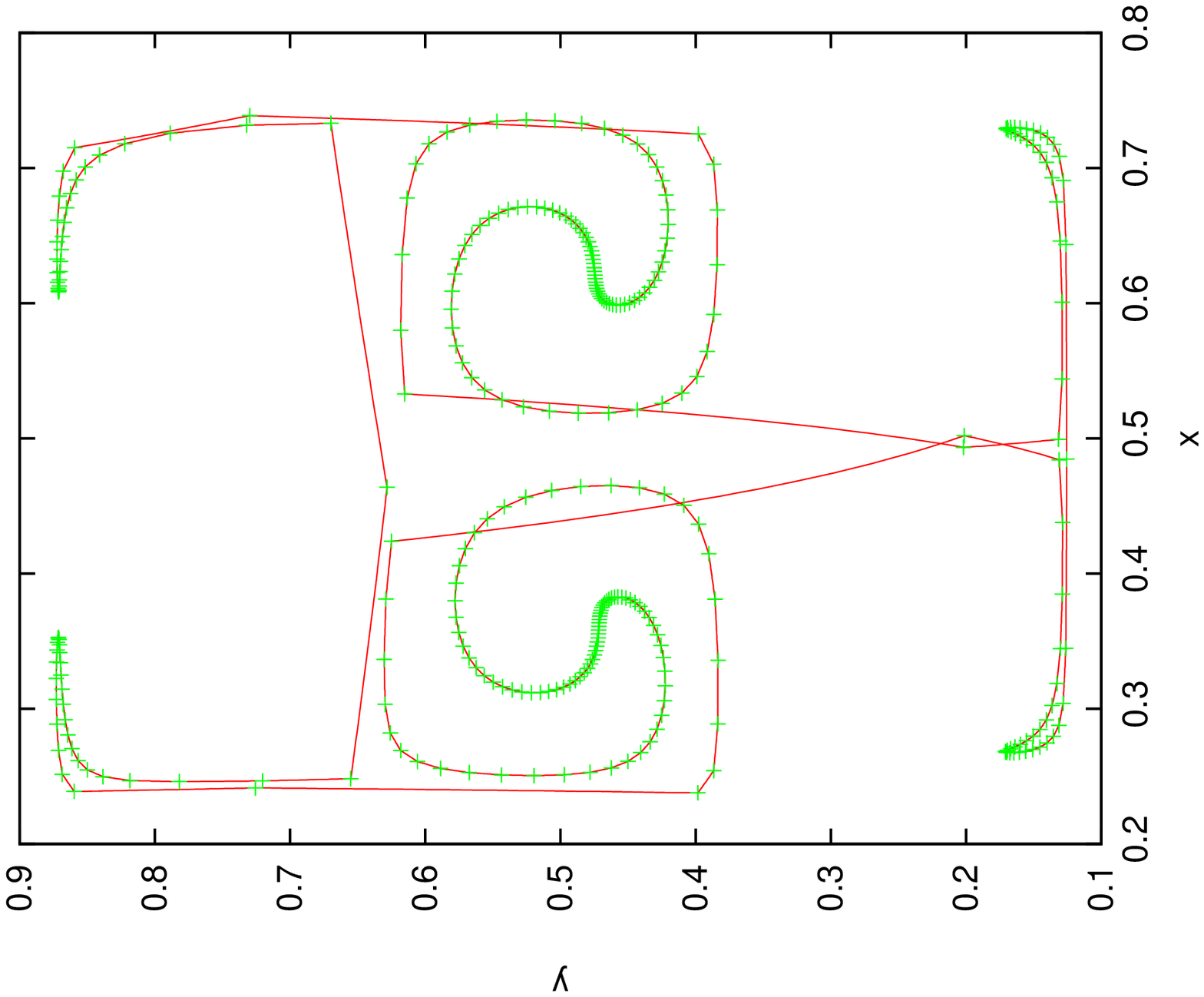}
}\\
$ N =312, t = T/2 $
\end{minipage} \nolinebreak
\begin{minipage}[t][0.23\textheight][t]{0.5\linewidth}
\centering
\rotatebox{270}{
\includegraphics[height=0.2\textheight]{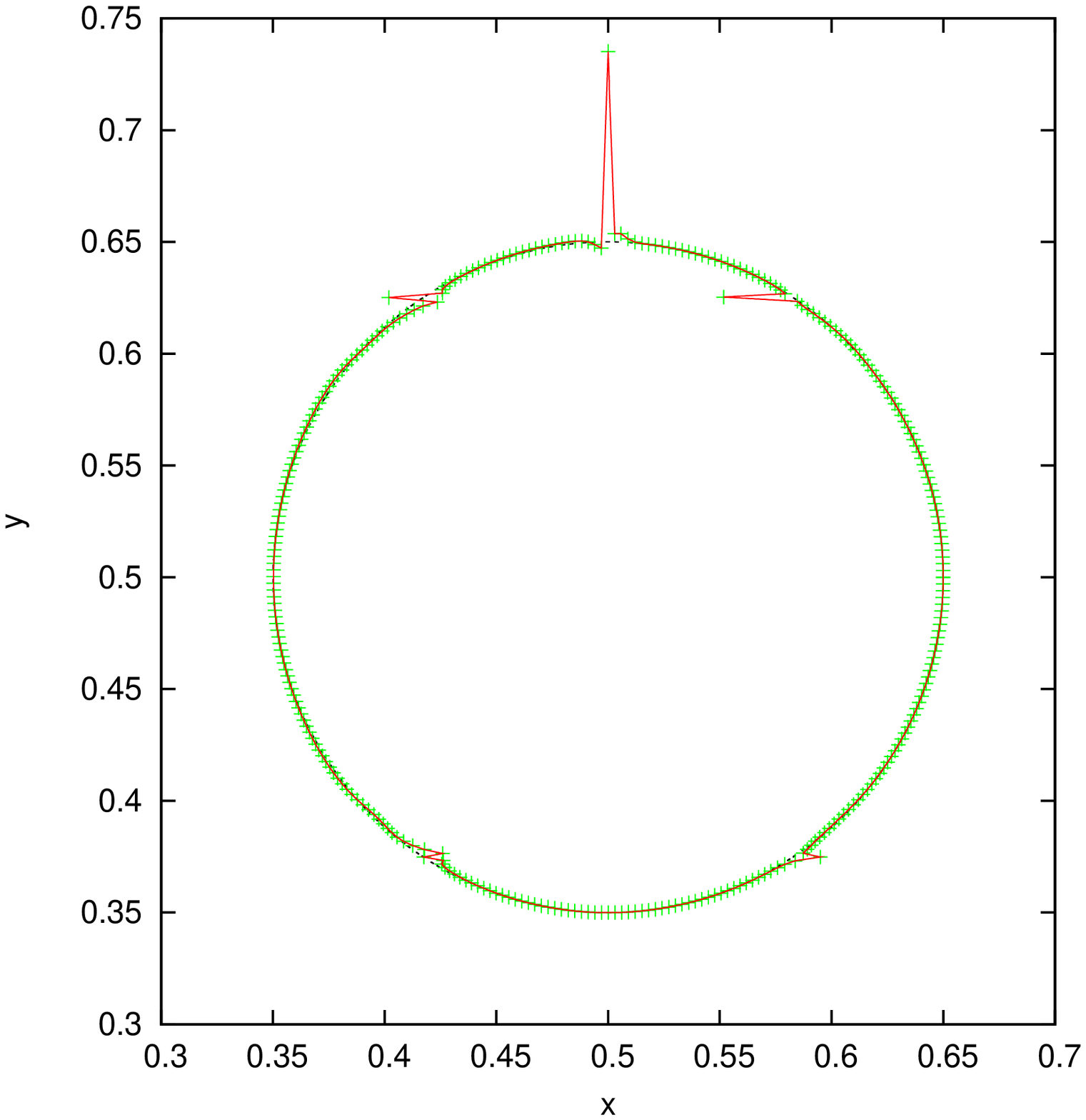}
}\\
$ N = 312, t = T $
\end{minipage}\\
\begin{minipage}[t][0.23\textheight][t]{0.5\linewidth}
\centering
\rotatebox{270}{
\includegraphics[height=0.2\textheight]{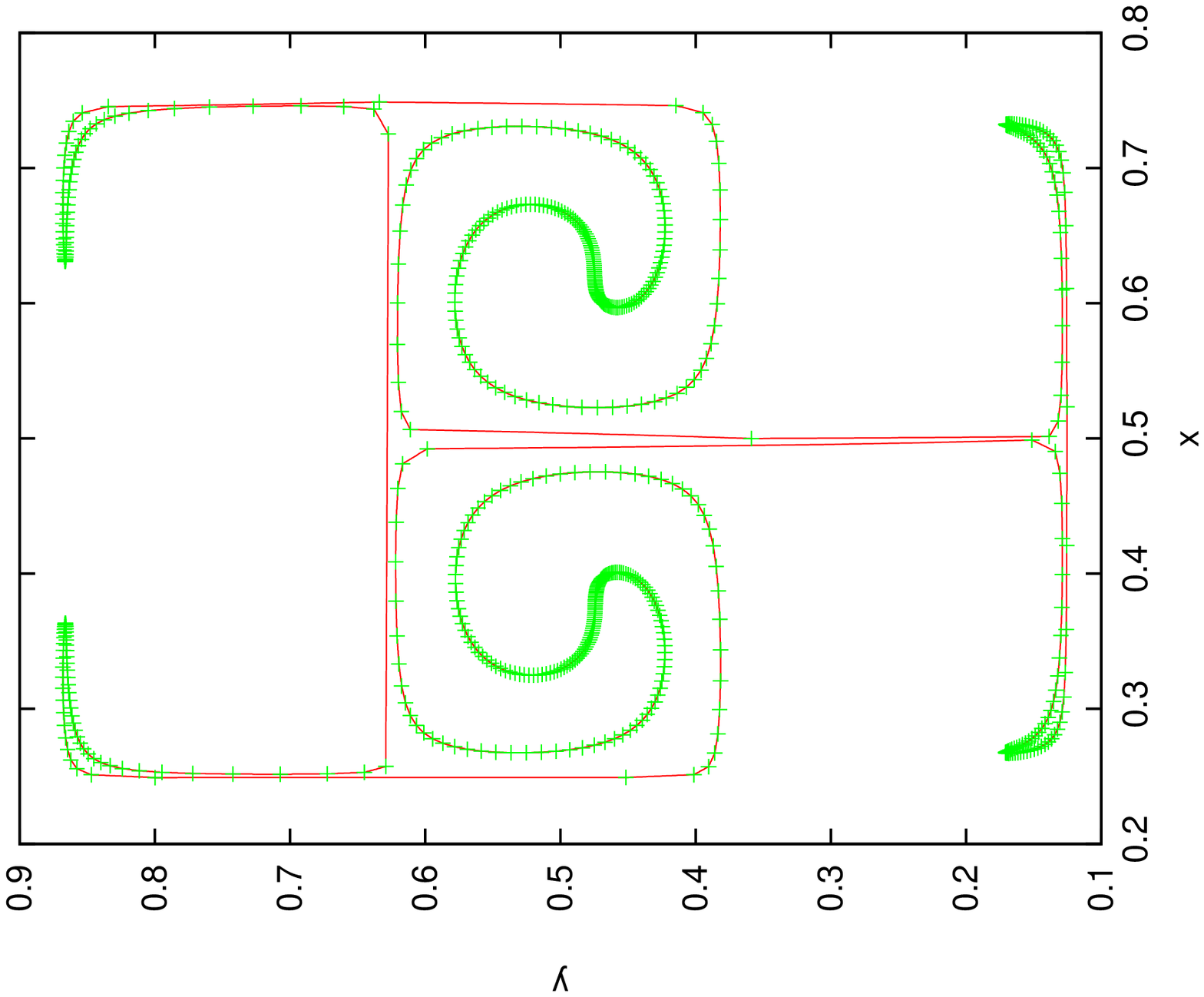}
}\\
{$ N =625, t = T/2$}
\end{minipage} \nolinebreak
\begin{minipage}[t][0.23\textheight][t]{0.5\linewidth}
\centering
\rotatebox{270}{
\includegraphics[height=0.2\textheight]{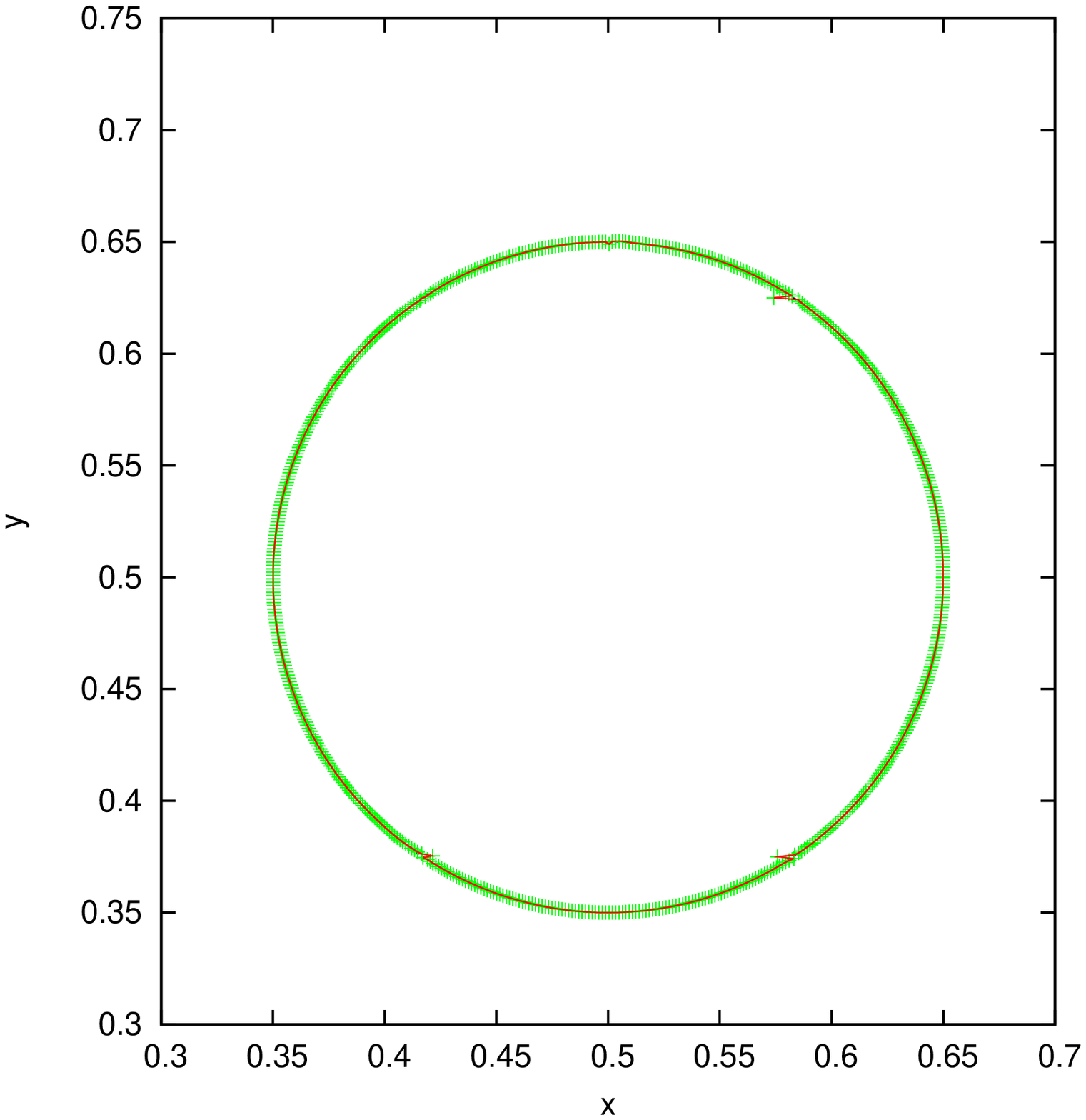}
}\\
$ N = 625, t = T $
\end{minipage}\\
\begin{minipage}[t][0.23\textheight][t]{0.5\linewidth}
\centering
\rotatebox{270}{
\includegraphics[height=0.2\textheight]{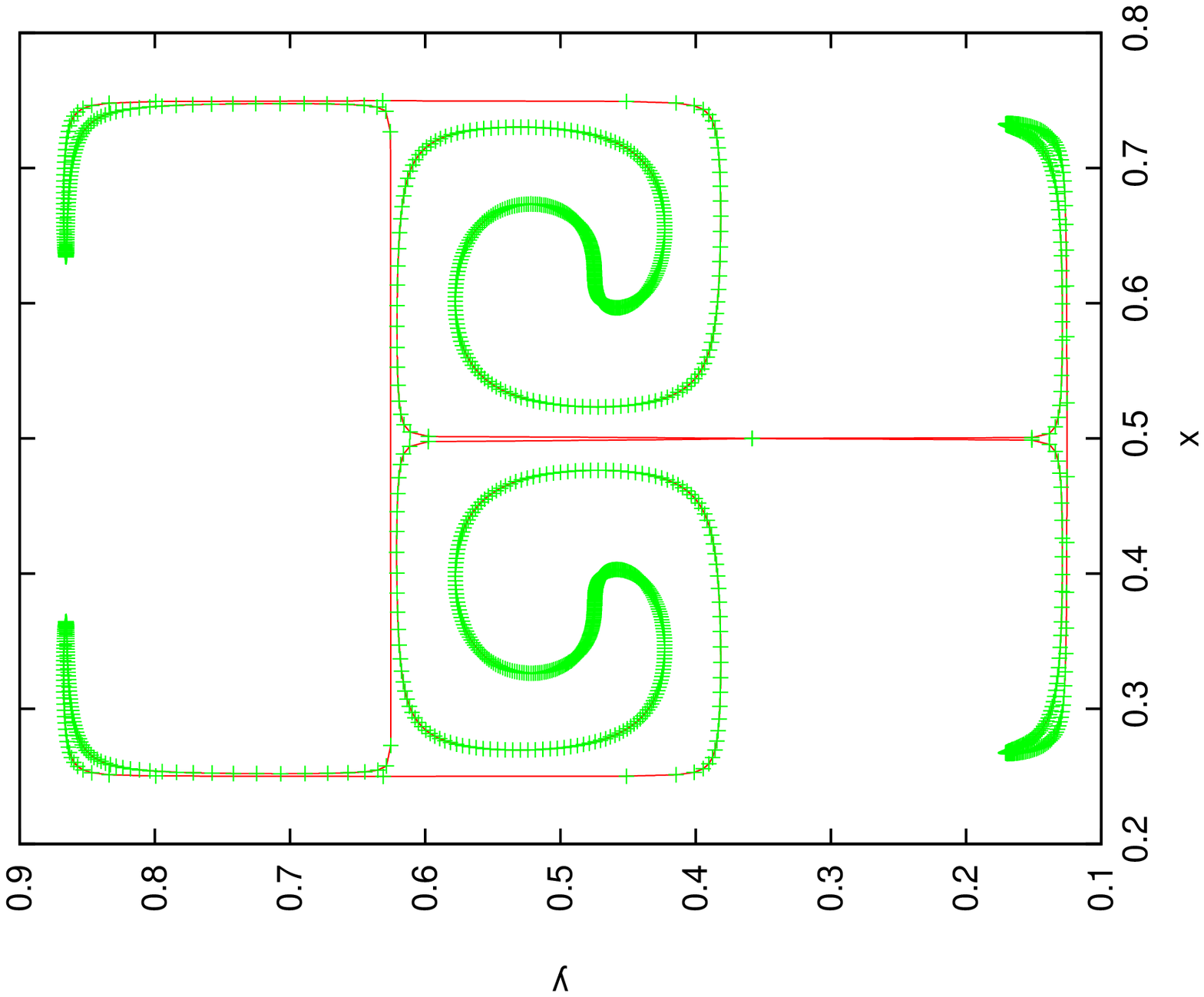}
}\\
$ N = 1250, t = T/2 $
\end{minipage} \nolinebreak
\begin{minipage}[t][0.23\textheight][t]{0.5\linewidth}
\centering
\rotatebox{270}{
\includegraphics[height=0.2\textheight]{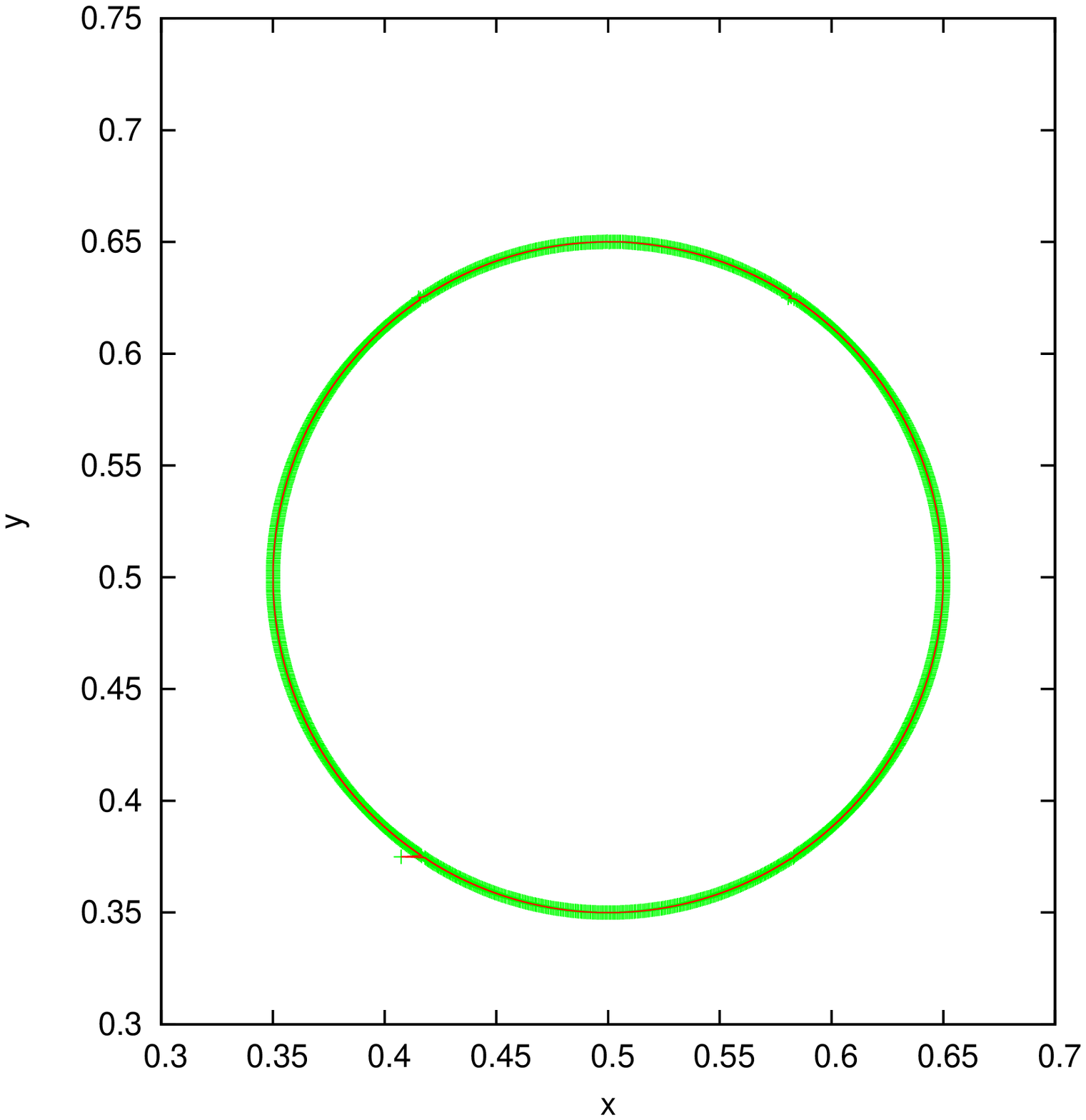}
}\\
$ N = 1250, t = T $
\end{minipage}\\
\caption{Result of the deformation field test for
different resolutions $N $. The order of B-spline interpolation is $ P = 1$. 
The graphs are shown at the maximal deformation, $ t = T/2 $ and
after returning to the initial position $ t = T $. The black dashed line
is the exact solution for $ t = T $. }
\label{fig:deformationP1}
\end{figure}

\begin{figure}
\centering
\begin{minipage}[t][0.23\textheight][t]{0.5\linewidth}
\centering
\rotatebox{270}{
\includegraphics[height=0.2\textheight]{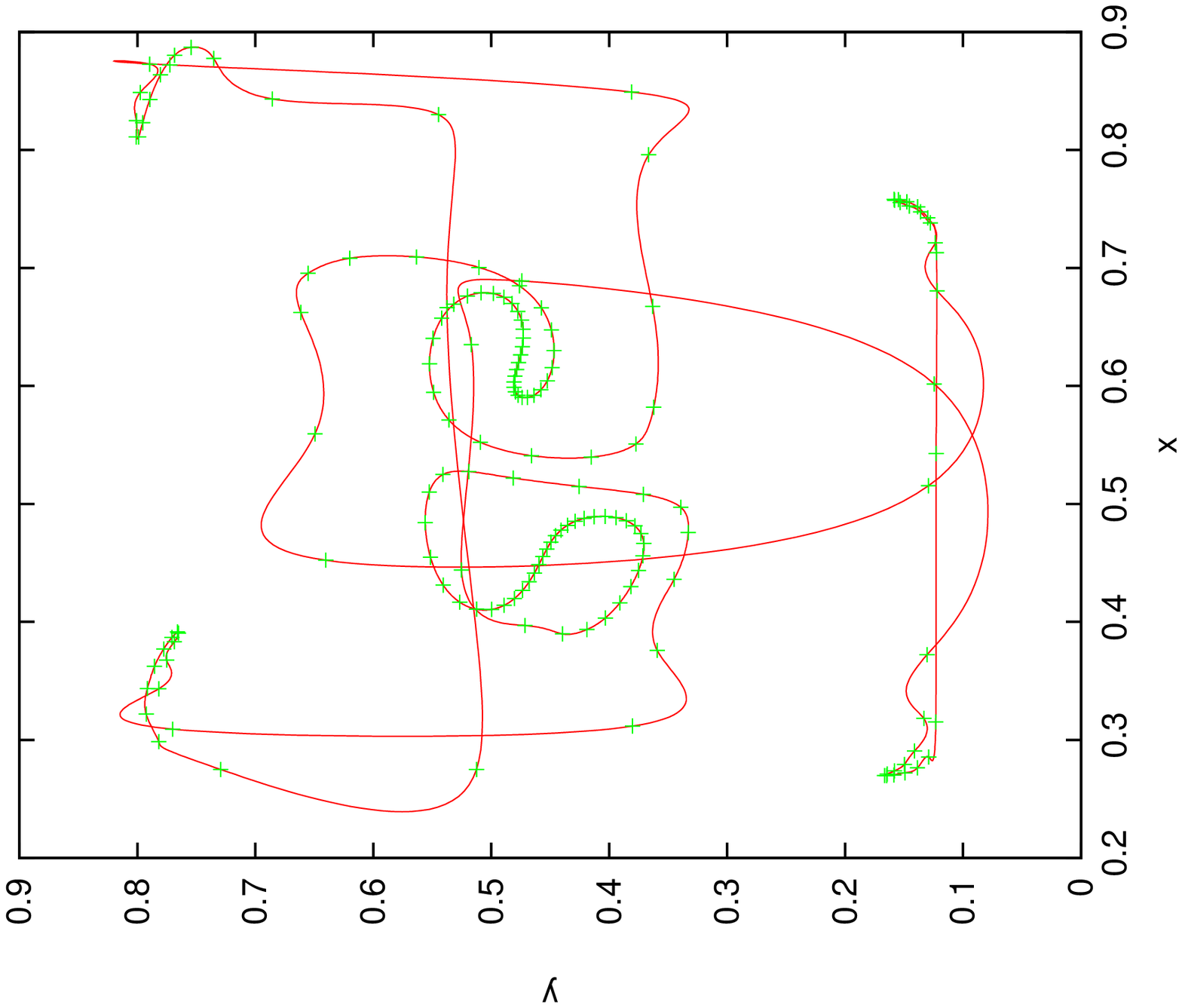}
}\\
{$ N =156, t = T/2 $}
\end{minipage} \nolinebreak
\begin{minipage}[t][0.23\textheight][t]{0.5\linewidth}
\centering
\rotatebox{270}{
\includegraphics[height=0.2\textheight]{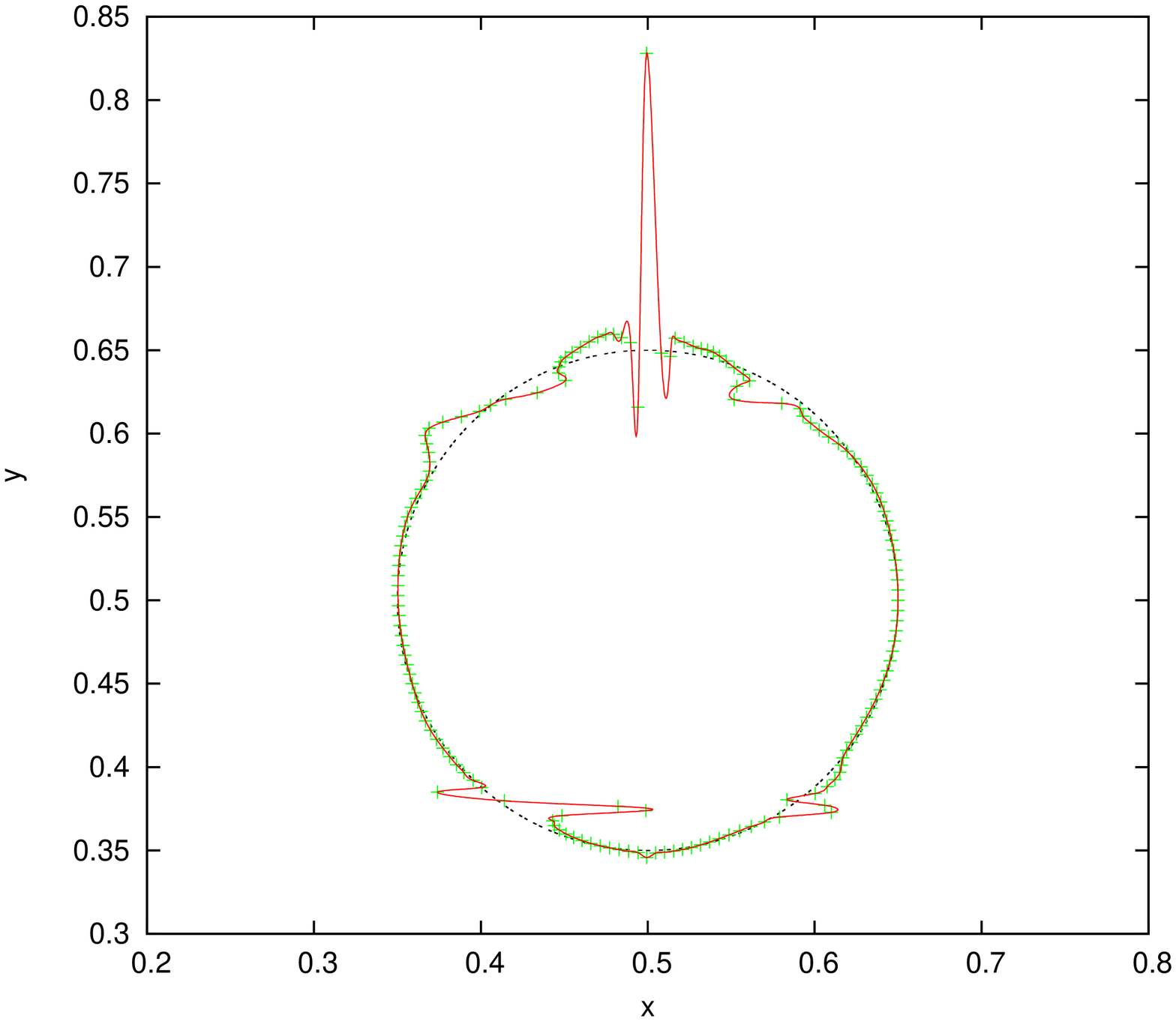}
}\\
$ N =156, t = T $
\end{minipage}\\
\begin{minipage}[t][0.23\textheight][t]{0.5\linewidth}
\centering
\rotatebox{270}{
\includegraphics[height=0.2\textheight]{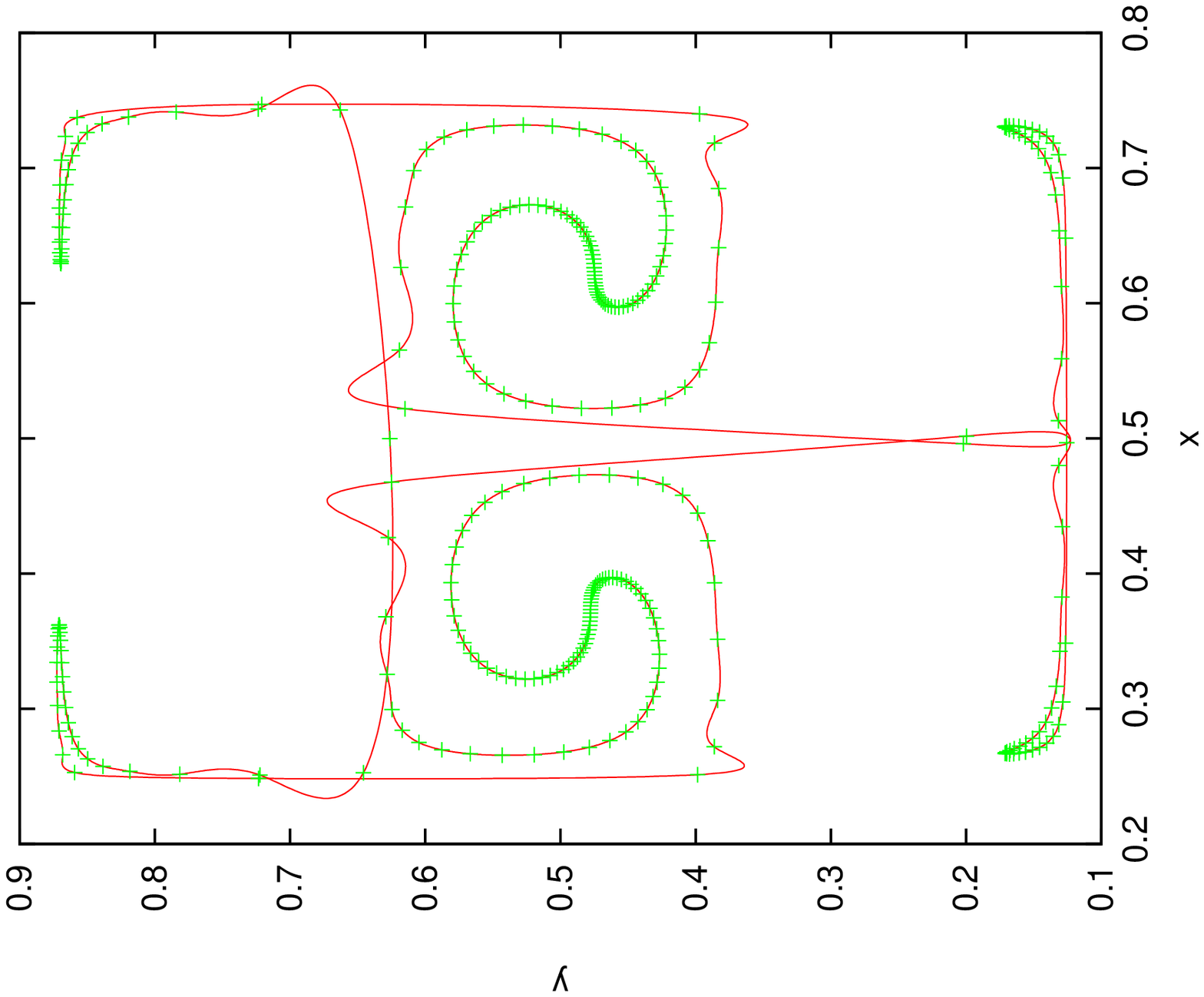}
}\\
$ N =312, t = T/2 $
\end{minipage} \nolinebreak
\begin{minipage}[t][0.23\textheight][t]{0.5\linewidth}
\centering
\rotatebox{270}{
\includegraphics[height=0.2\textheight]{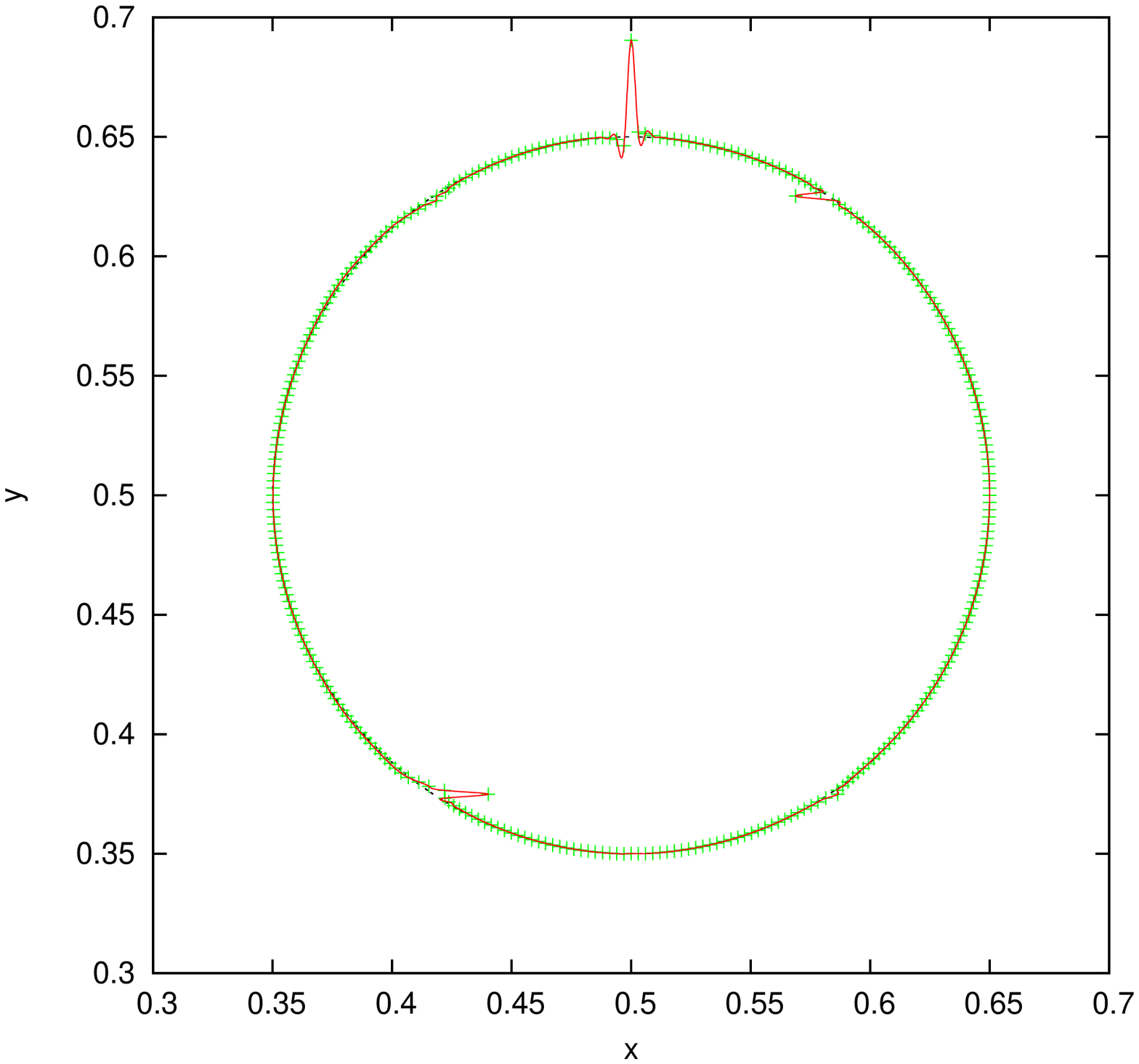}
}\\
$ N = 312, t = T $
\end{minipage}\\
\begin{minipage}[t][0.23\textheight][t]{0.5\linewidth}
\centering
\rotatebox{270}{
\includegraphics[height=0.2\textheight]{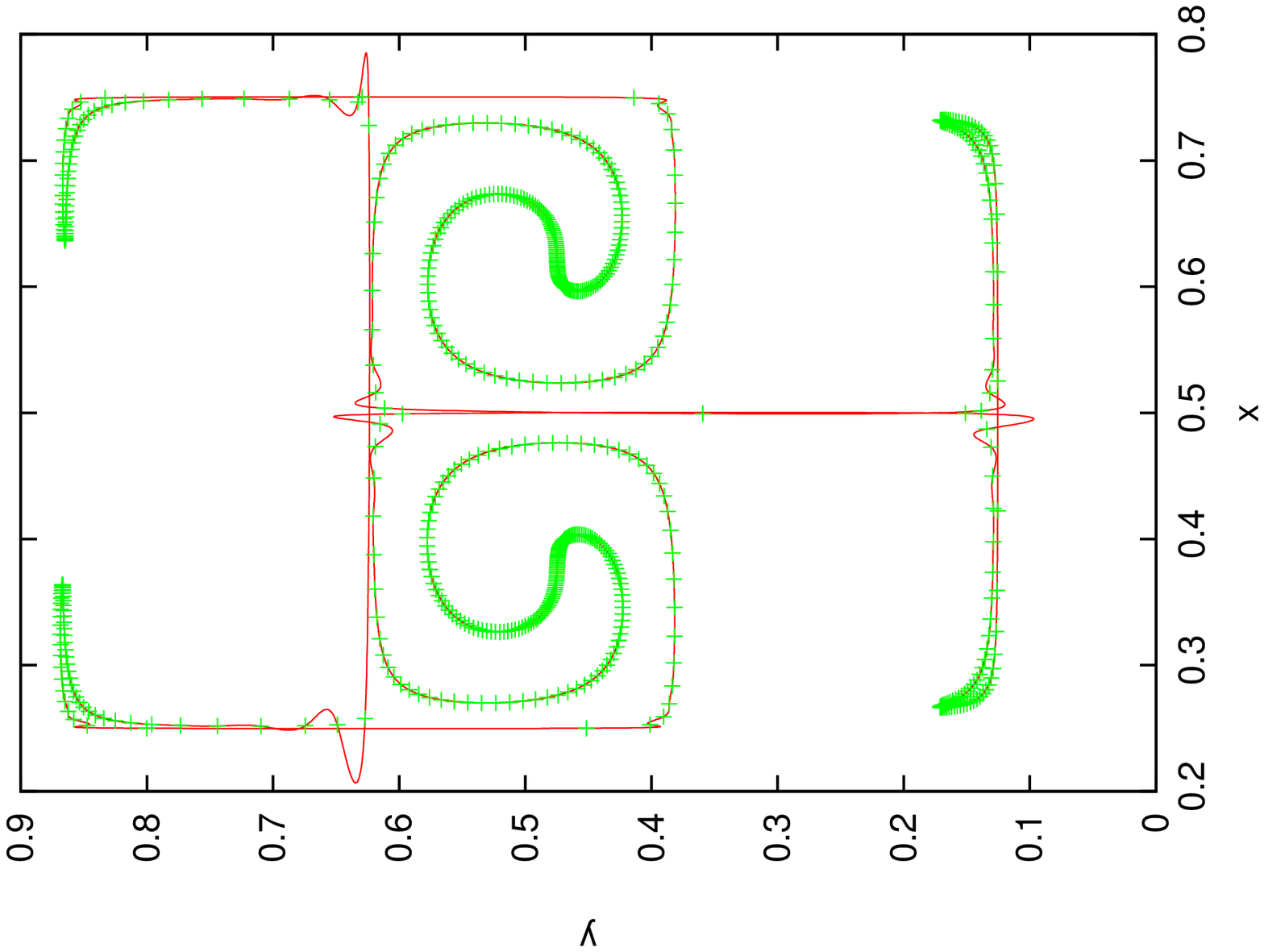}
}\\
{$ N =625, t = T/2$}
\end{minipage} \nolinebreak
\begin{minipage}[t][0.23\textheight][t]{0.5\linewidth}
\centering
\rotatebox{270}{
\includegraphics[height=0.2\textheight]{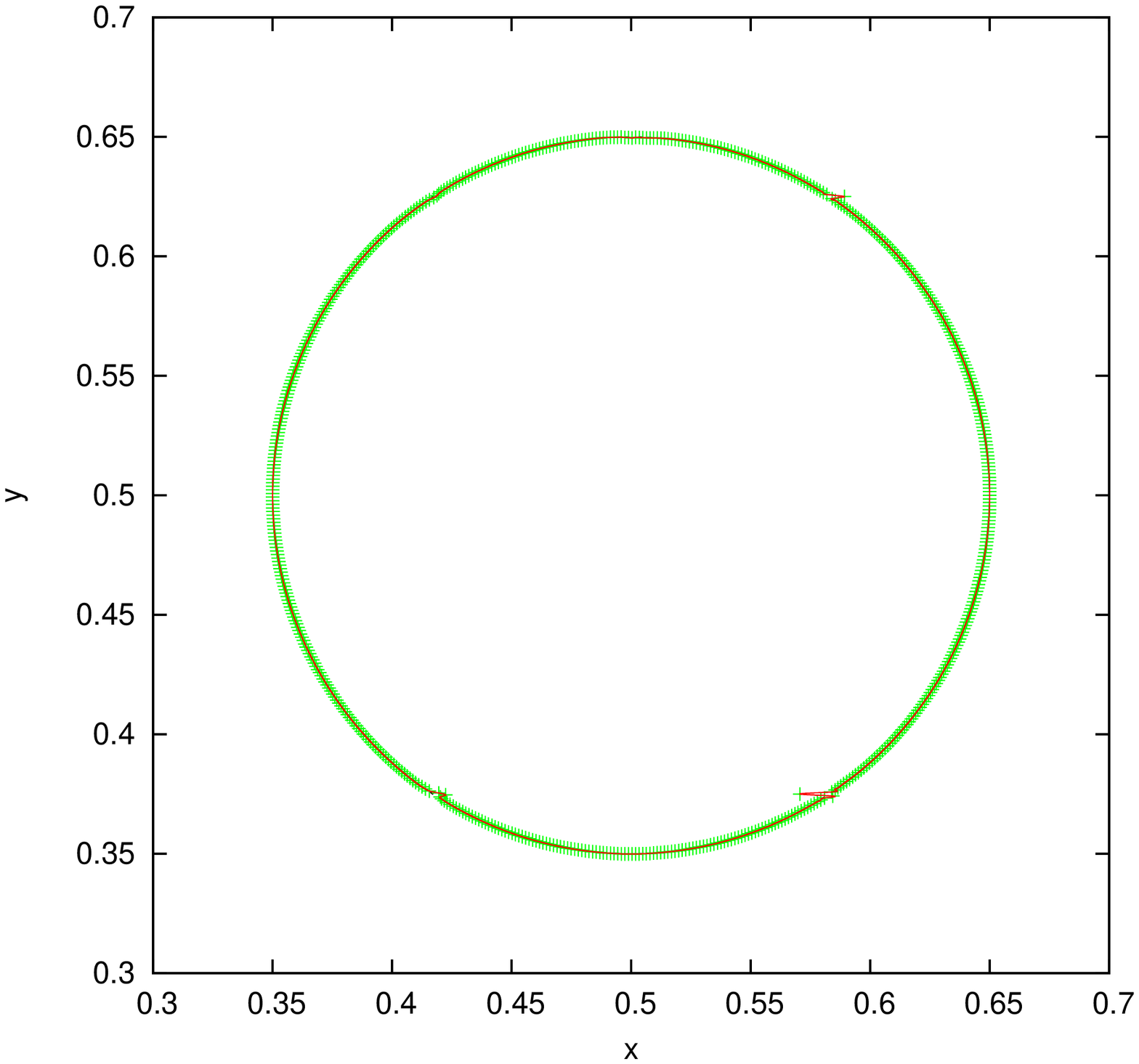}
}\\
$ N = 625, t = T $
\end{minipage}\\
\begin{minipage}[t][0.23\textheight][t]{0.5\linewidth}
\centering
\rotatebox{270}{
\includegraphics[height=0.2\textheight]{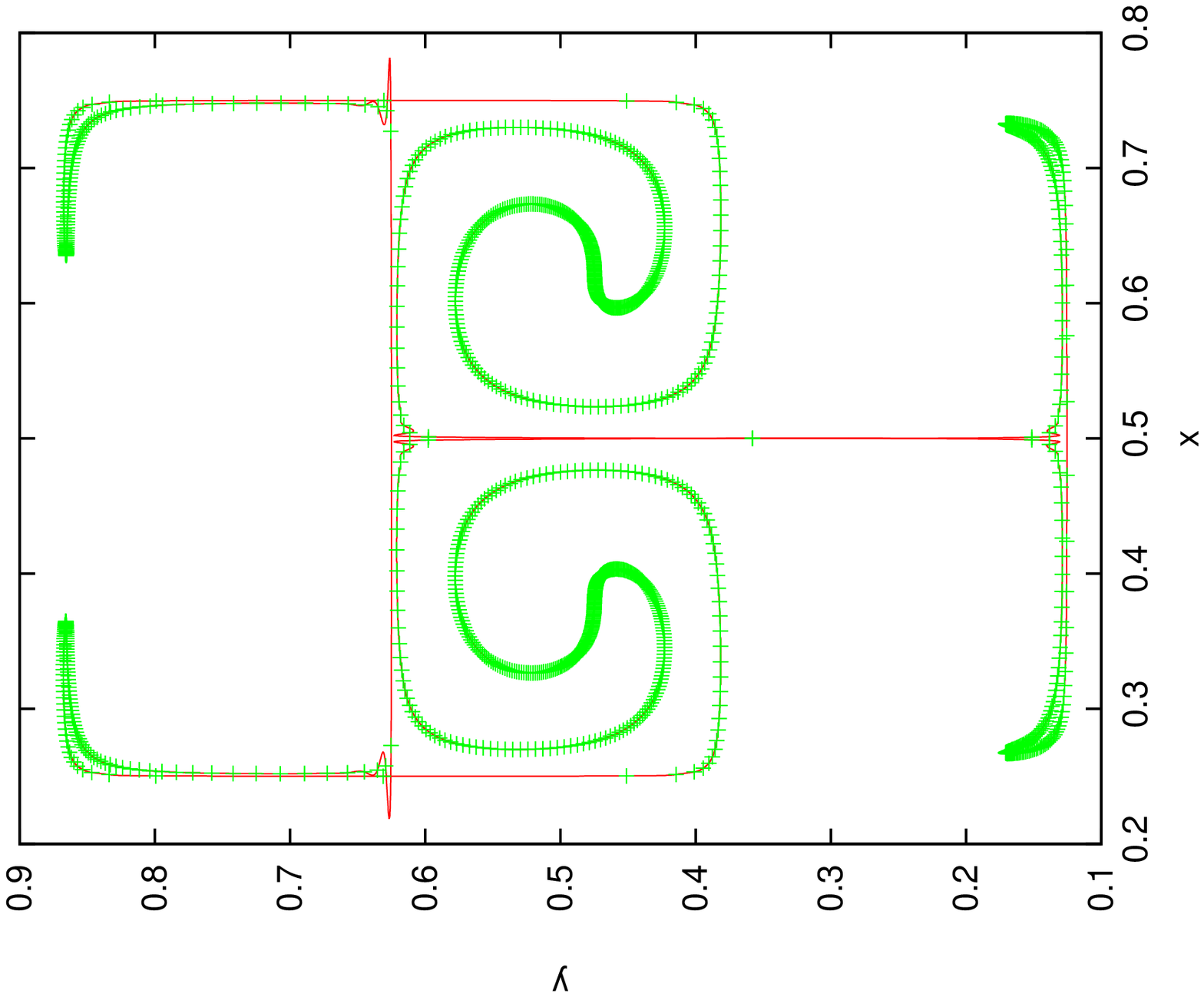}
}\\
$ N = 1250, t = T/2 $
\end{minipage} \nolinebreak
\begin{minipage}[t][0.23\textheight][t]{0.5\linewidth}
\centering
\rotatebox{270}{
\includegraphics[height=0.2\textheight]{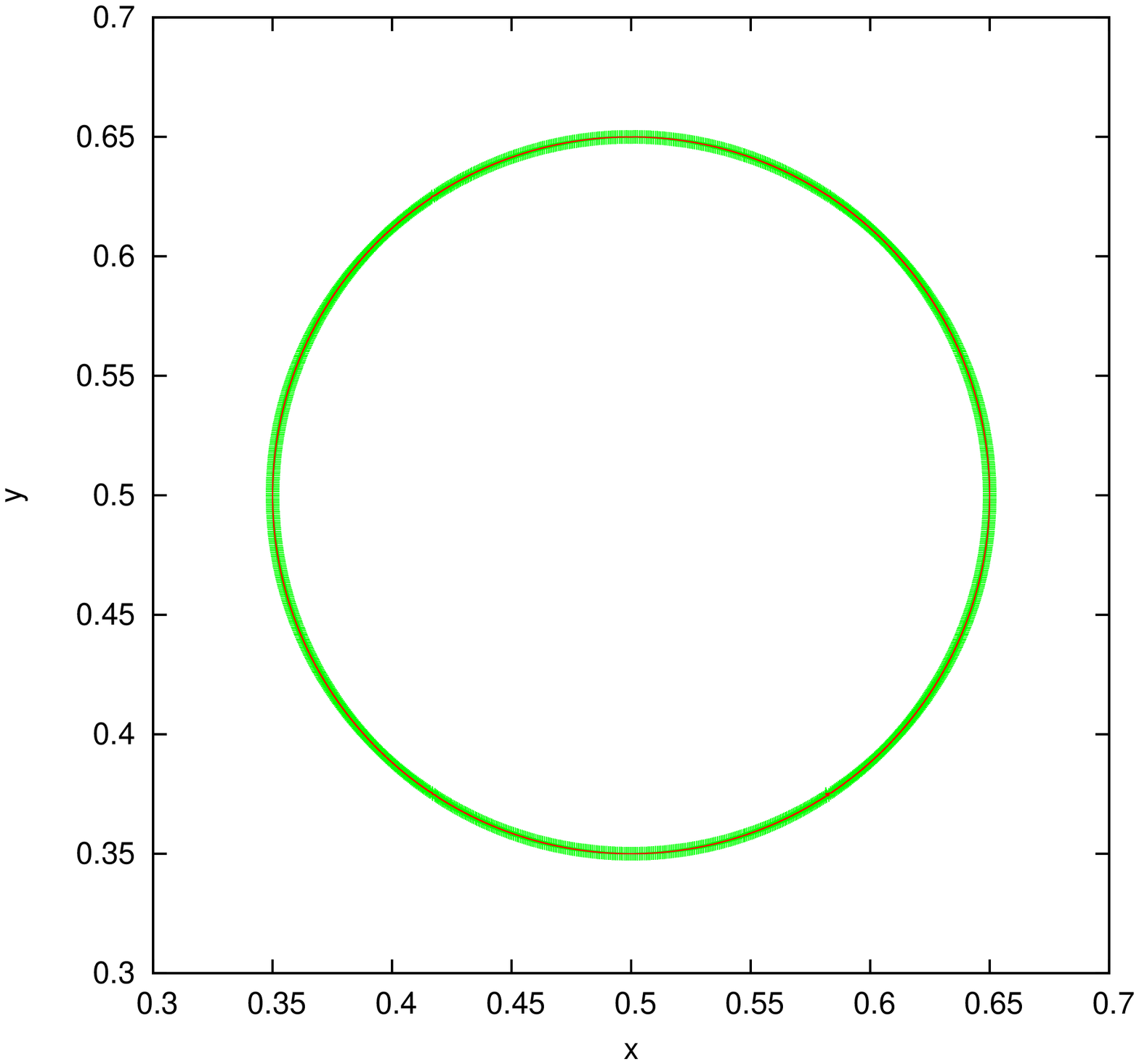}
}\\
$ N = 1250, t = T $
\end{minipage}\\
\caption{Result of the deformation field test for
different resolutions $N $. The order of B-spline interpolation is $ P = 3$. 
The graphs are shown at the maximal deformation, $ t = T/2 $ and
after returning to the initial position $ t = T $. The black dashed line
is the exact solution for $ t = T $. }
\label{fig:deformationP3}
\end{figure}

\begin{figure}
\centering
\begin{minipage}[t][0.23\textheight][t]{0.5\linewidth}
\centering
\rotatebox{270}{
\includegraphics[height=0.2\textheight]{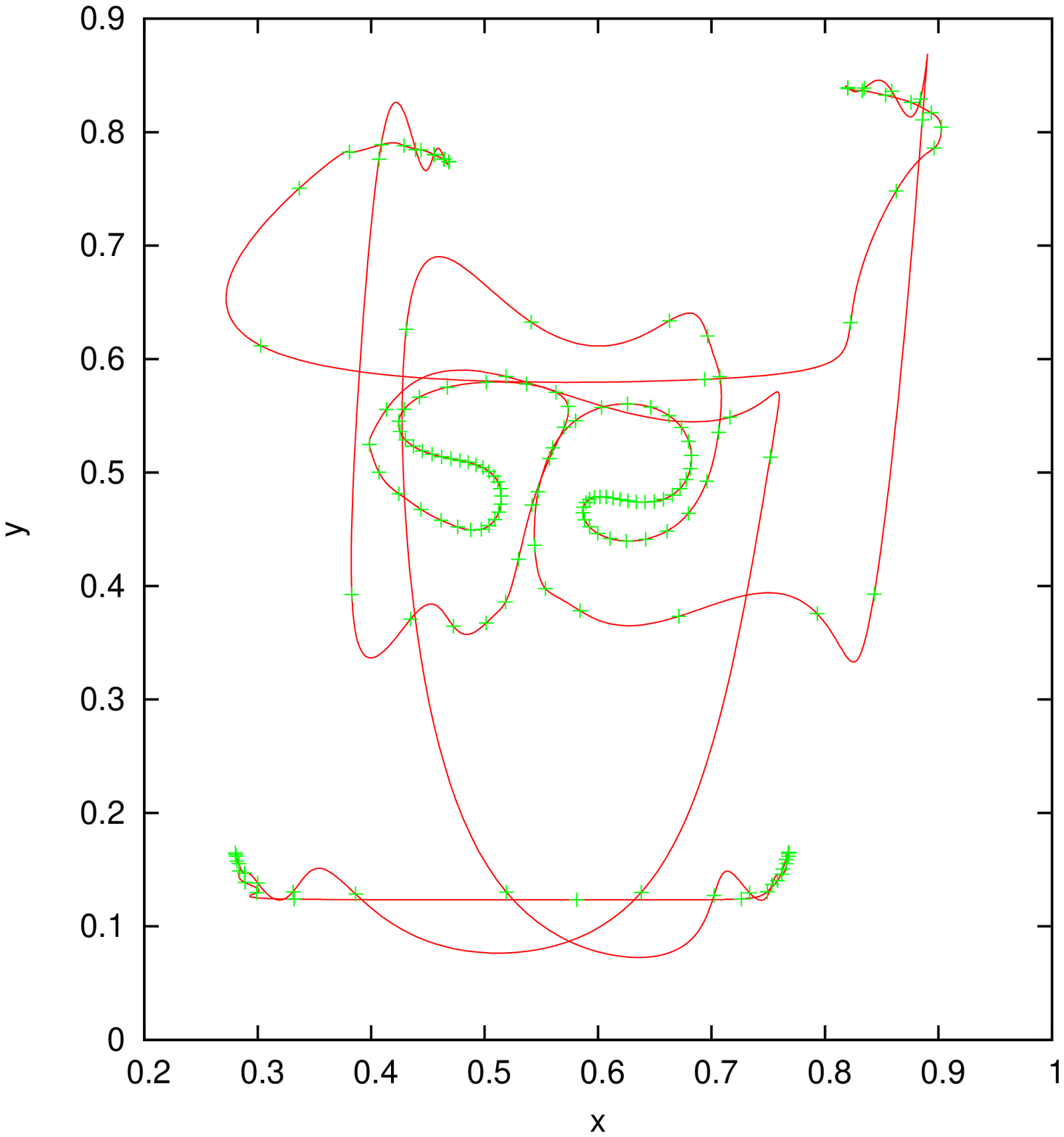}
}\\
{$ N =156, t = T/2 $}
\end{minipage} \nolinebreak
\begin{minipage}[t][0.23\textheight][t]{0.5\linewidth}
\centering
\rotatebox{270}{
\includegraphics[height=0.2\textheight]{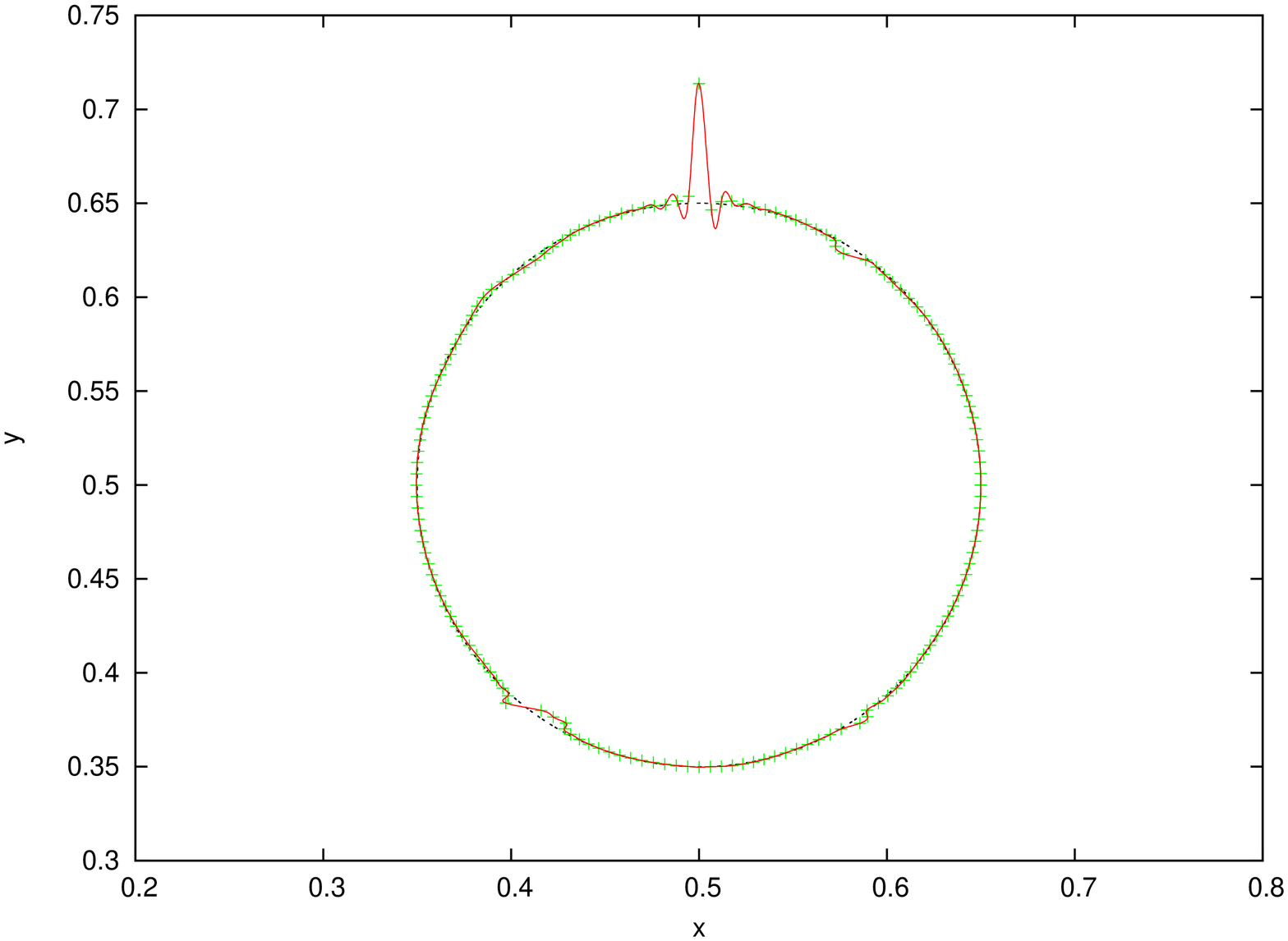}
}\\
$ N =156, t = T $
\end{minipage}\\
\begin{minipage}[t][0.23\textheight][t]{0.5\linewidth}
\centering
\rotatebox{270}{
\includegraphics[height=0.2\textheight]{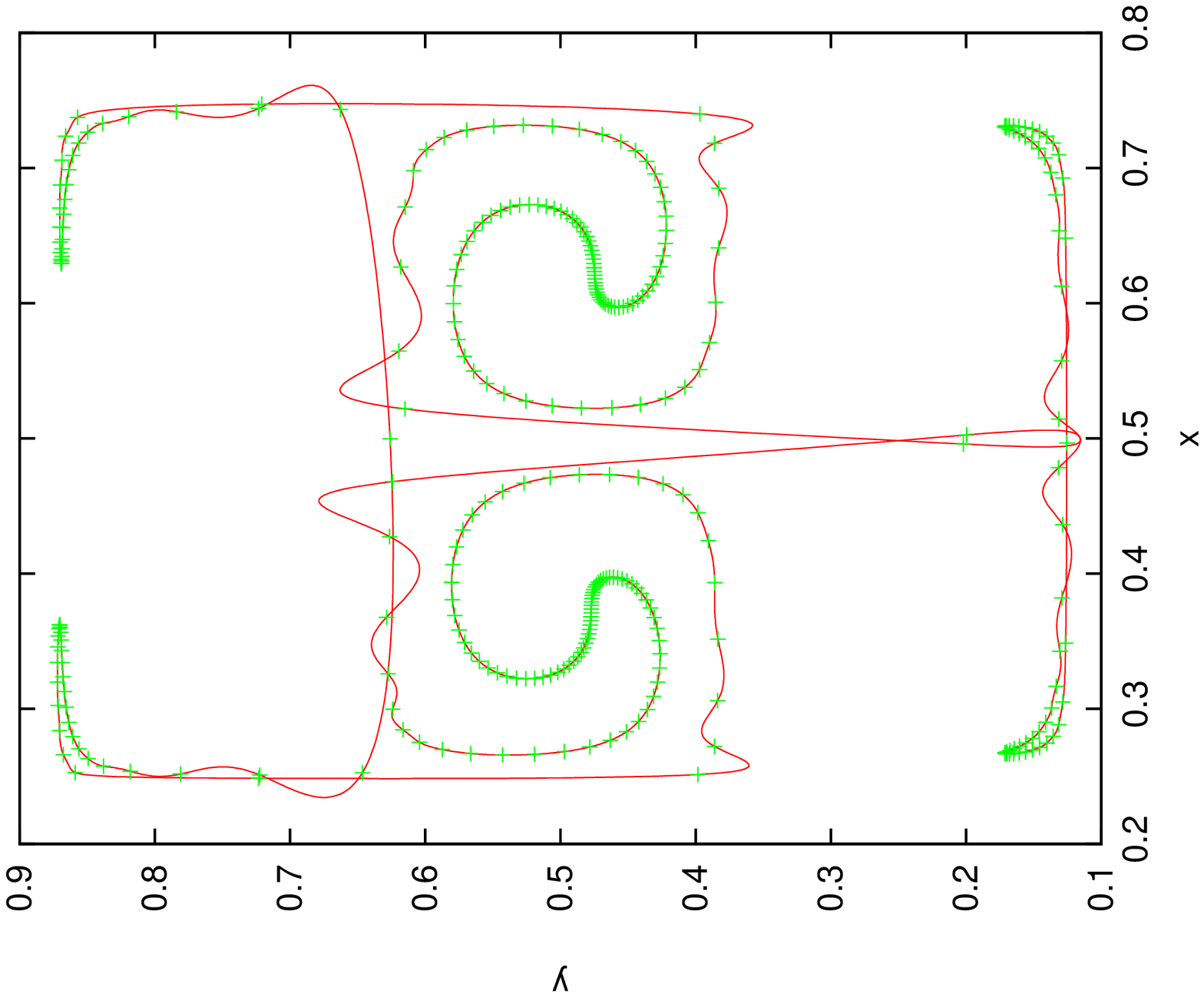}
}\\
$ N =312, t = T/2 $
\end{minipage} \nolinebreak
\begin{minipage}[t][0.23\textheight][t]{0.5\linewidth}
\centering
\rotatebox{270}{
\includegraphics[height=0.2\textheight]{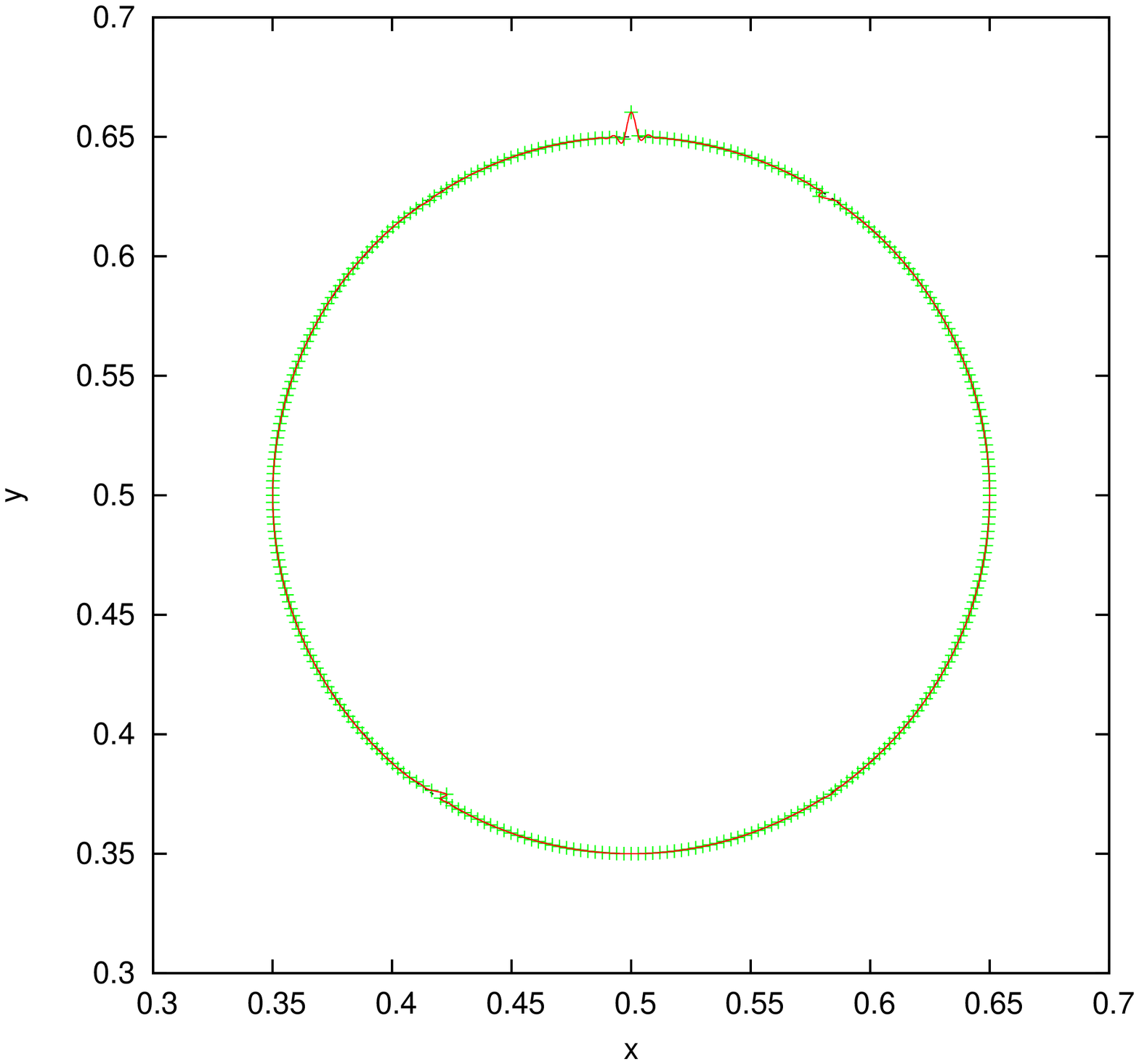}
}\\
$ N = 312, t = T $
\end{minipage}\\
\begin{minipage}[t][0.23\textheight][t]{0.5\linewidth}
\centering
\rotatebox{270}{
\includegraphics[height=0.2\textheight]{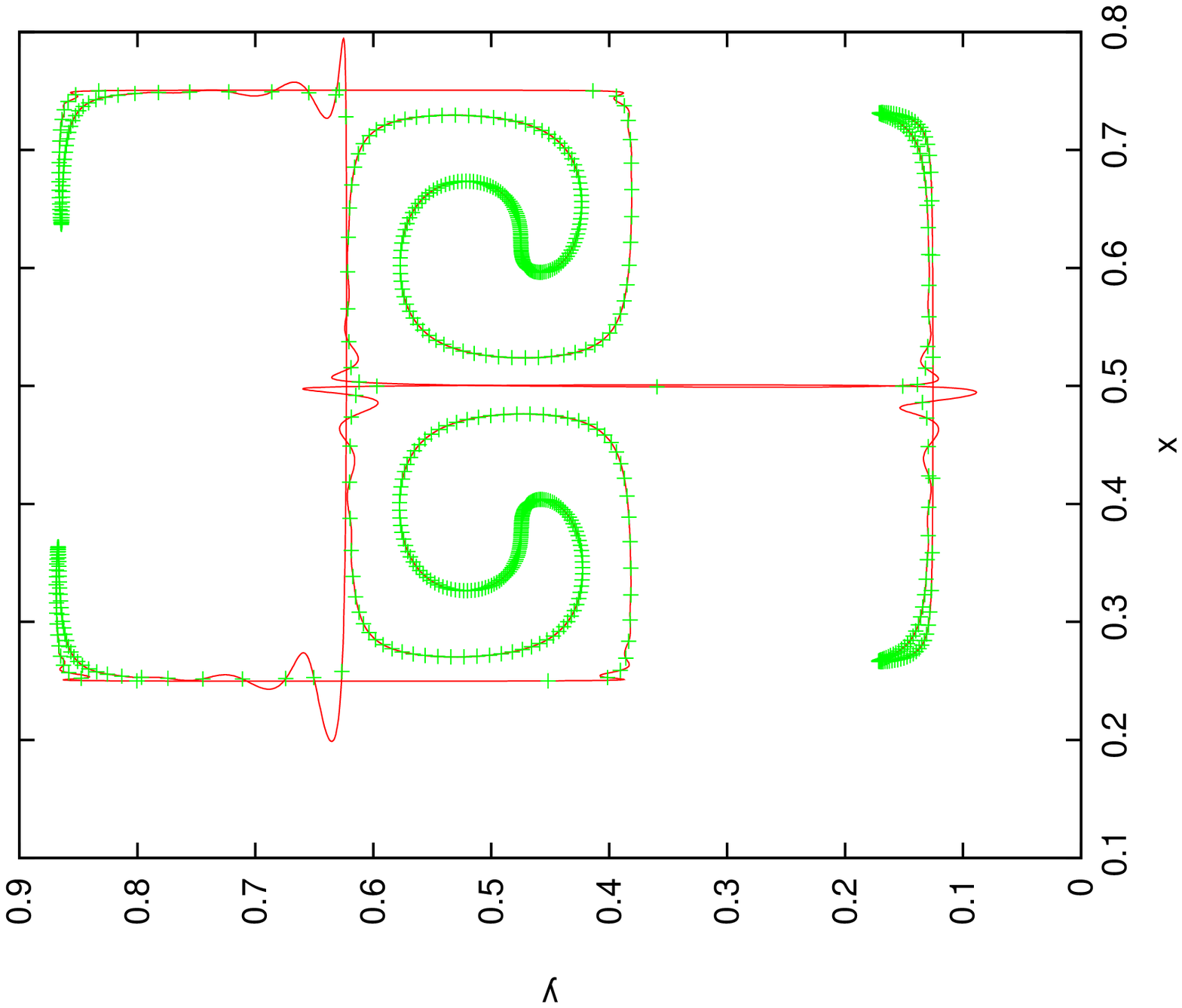}
}\\
{$ N =625, t = T/2$}
\end{minipage} \nolinebreak
\begin{minipage}[t][0.23\textheight][t]{0.5\linewidth}
\centering
\rotatebox{270}{
\includegraphics[height=0.2\textheight]{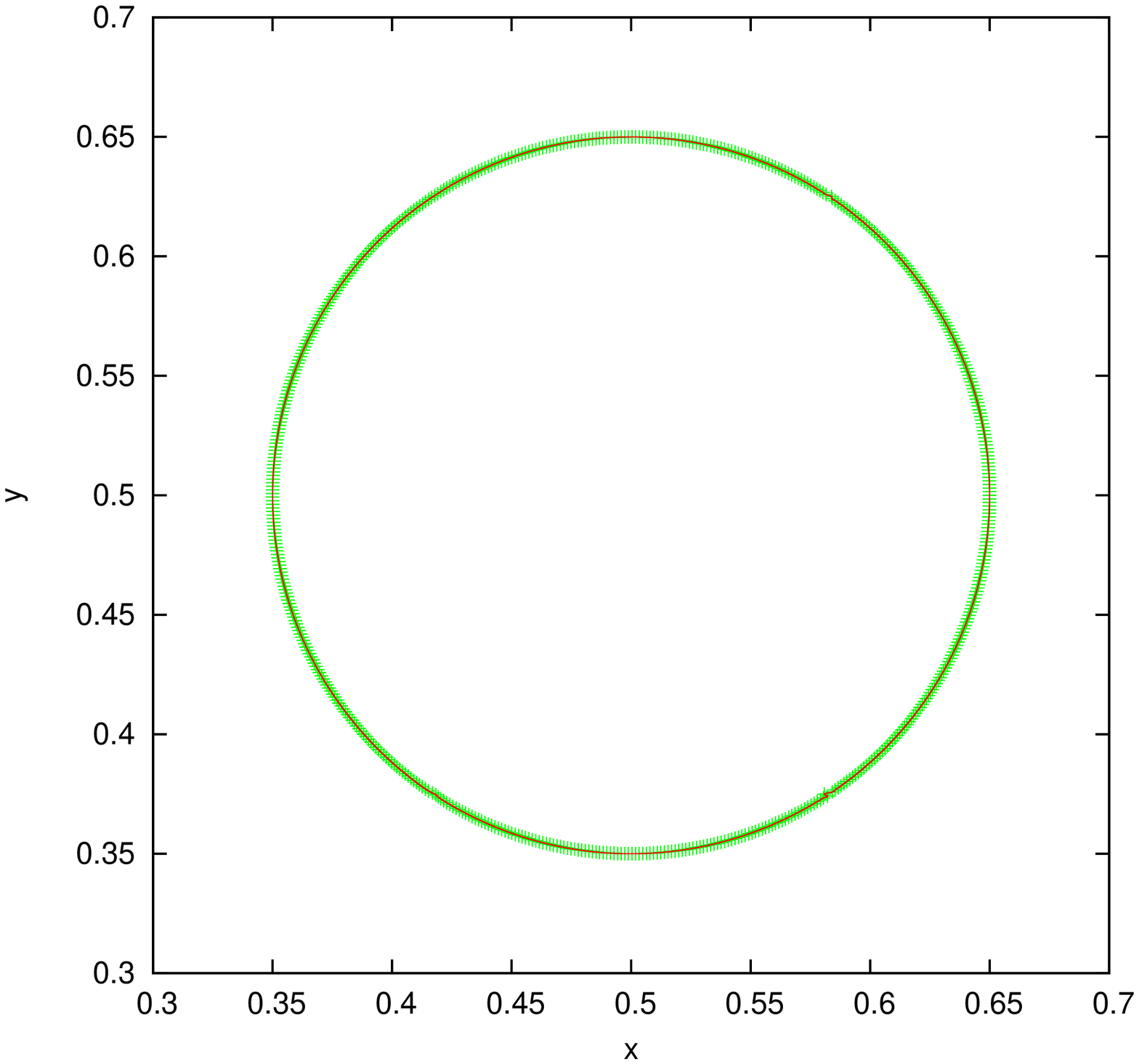}
}\\
$ N = 625, t = T $
\end{minipage}\\
\begin{minipage}[t][0.23\textheight][t]{0.5\linewidth}
\centering
\rotatebox{270}{
\includegraphics[height=0.2\textheight]{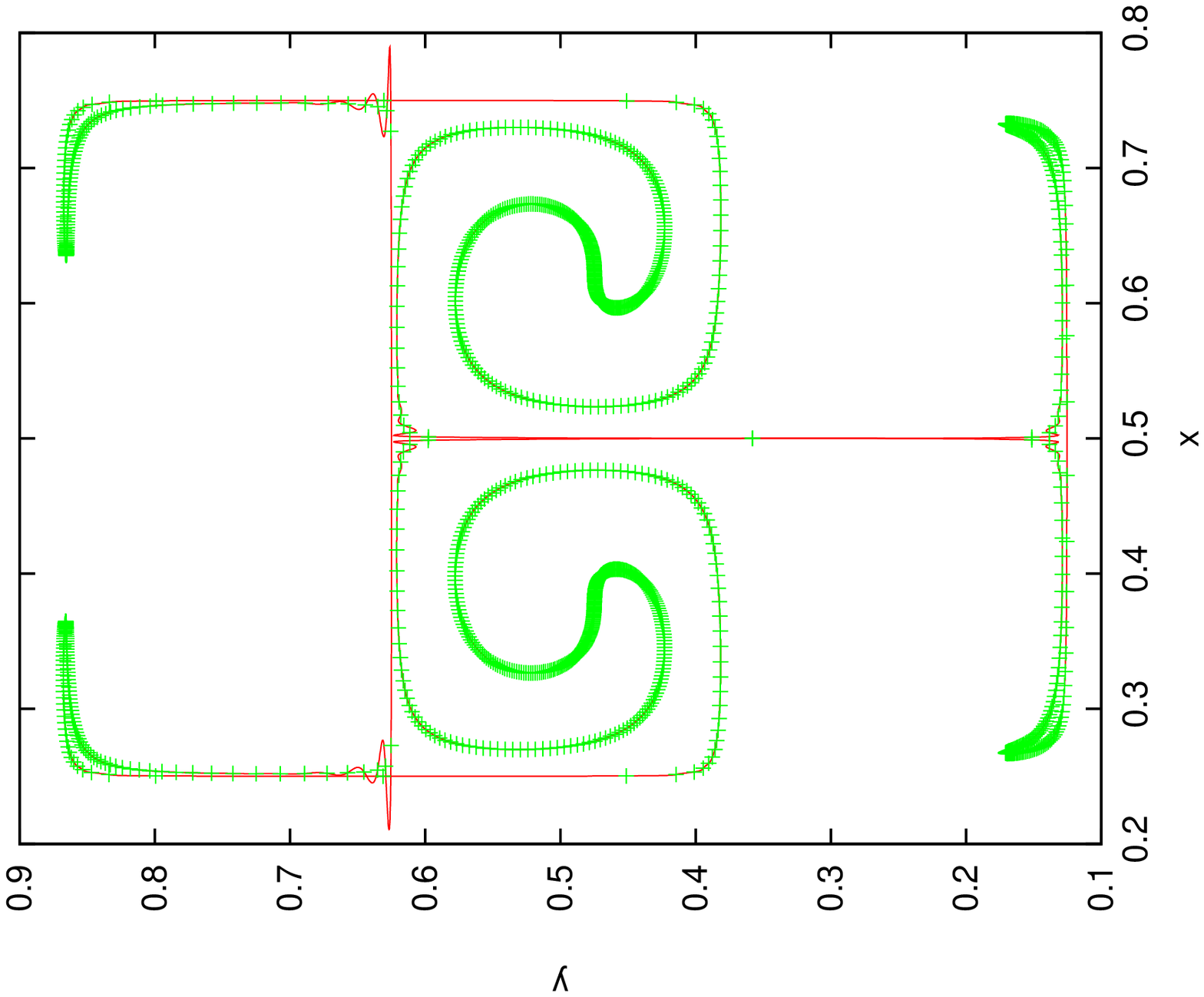}
}\\
$ N = 1250, t = T/2 $
\end{minipage} \nolinebreak
\begin{minipage}[t][0.23\textheight][t]{0.5\linewidth}
\centering
\rotatebox{270}{
\includegraphics[height=0.2\textheight]{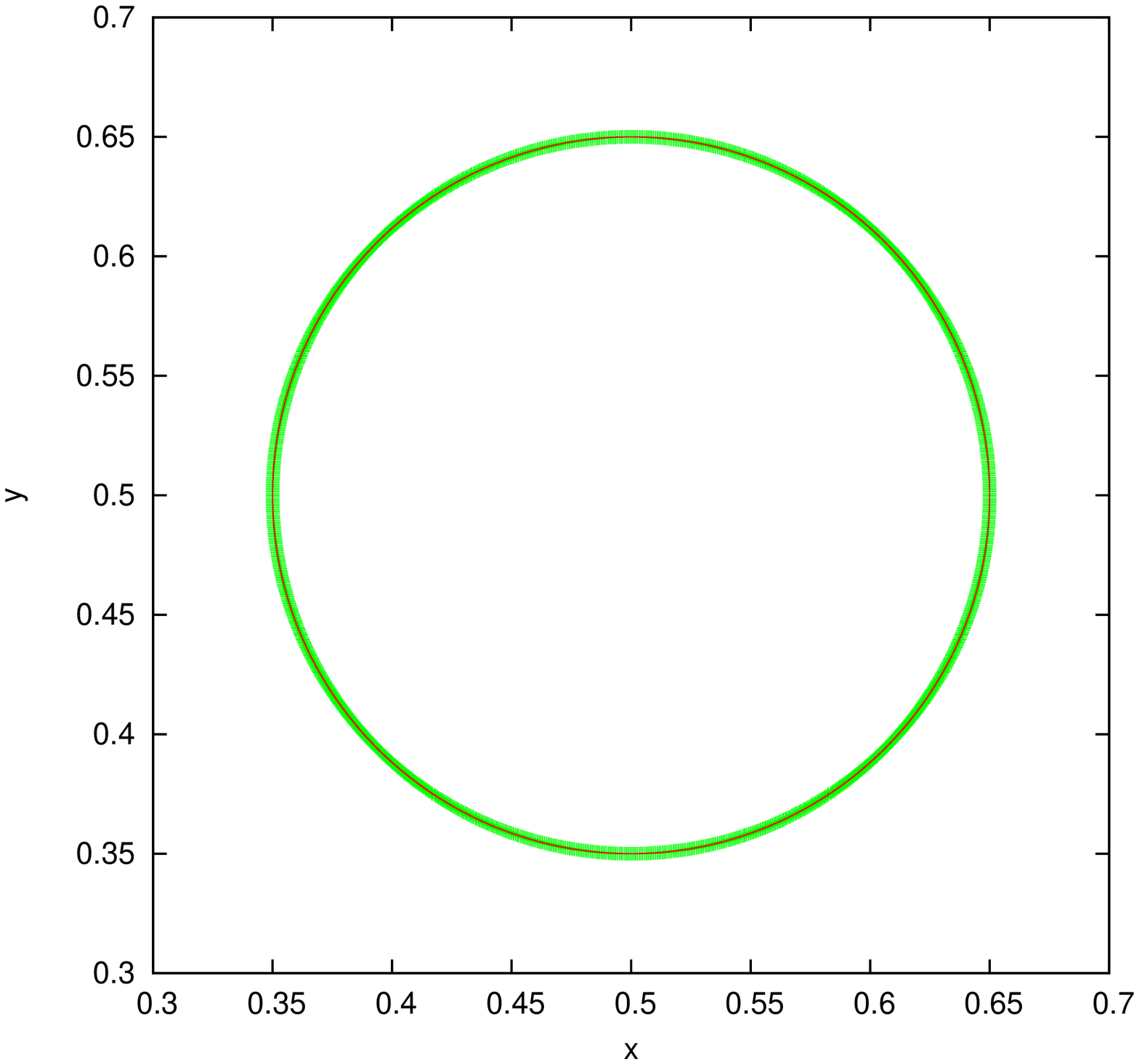}
}\\
$ N = 1250, t = T $
\end{minipage}\\
\caption{Result of the deformation field test for
different resolutions $N $. The order of B-spline interpolation is $ P = 5$. 
The graphs are shown at the maximal deformation, $ t = T/2 $ and
after returning to the initial position $ t = T $. The black dashed line
is the exact solution for $ t = T $. }
\label{fig:deformationP5}
\end{figure}

\begin{figure}
\centering
\begin{minipage}[t][0.23\textheight][t]{0.5\linewidth}
\centering
\rotatebox{270}{
\includegraphics[height=0.2\textheight]{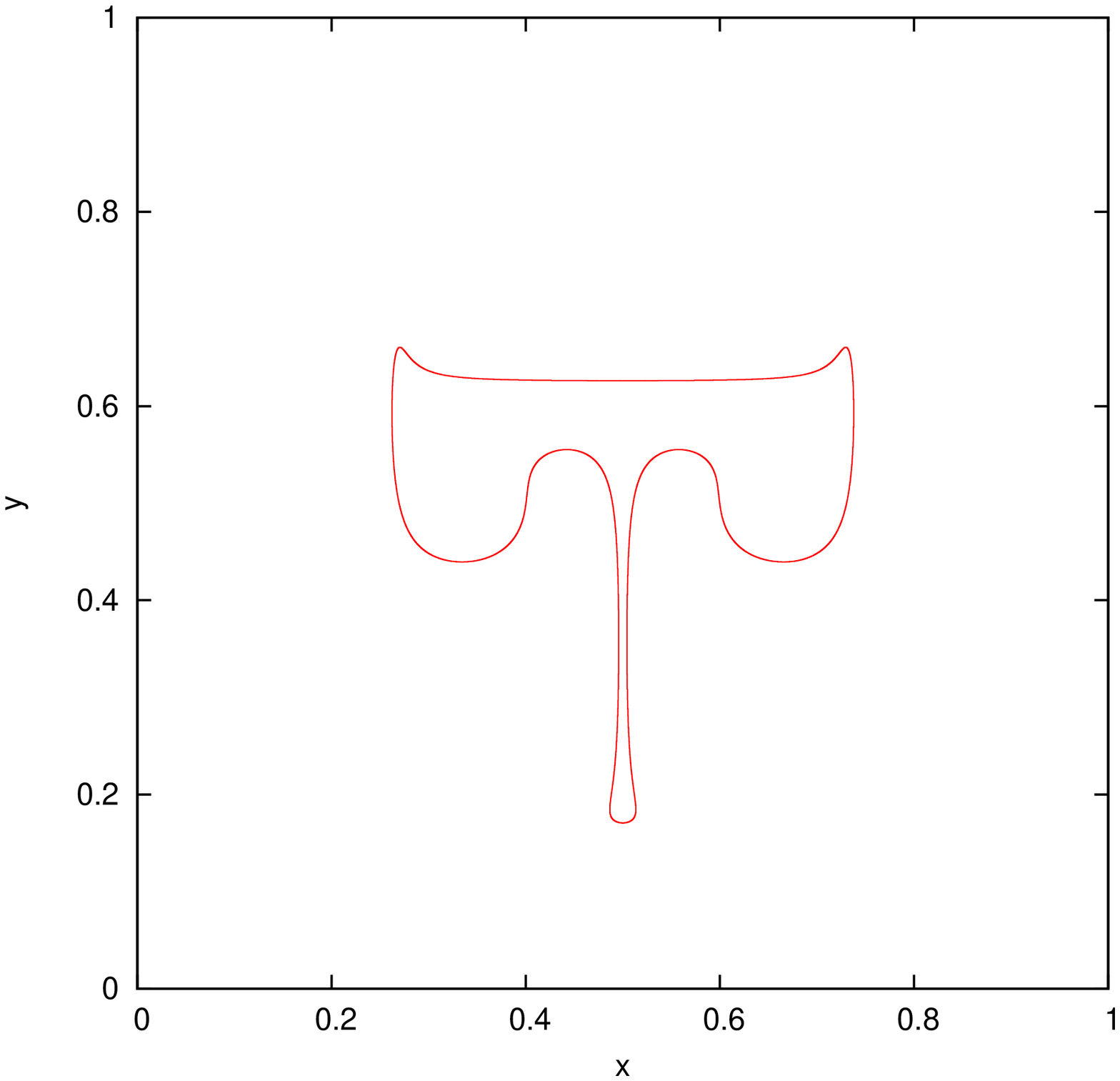}
}\\
{$ t = \frac{1}{8} T $}
\end{minipage} \nolinebreak
\begin{minipage}[t][0.23\textheight][t]{0.5\linewidth}
\centering
\rotatebox{270}{
\includegraphics[height=0.2\textheight]{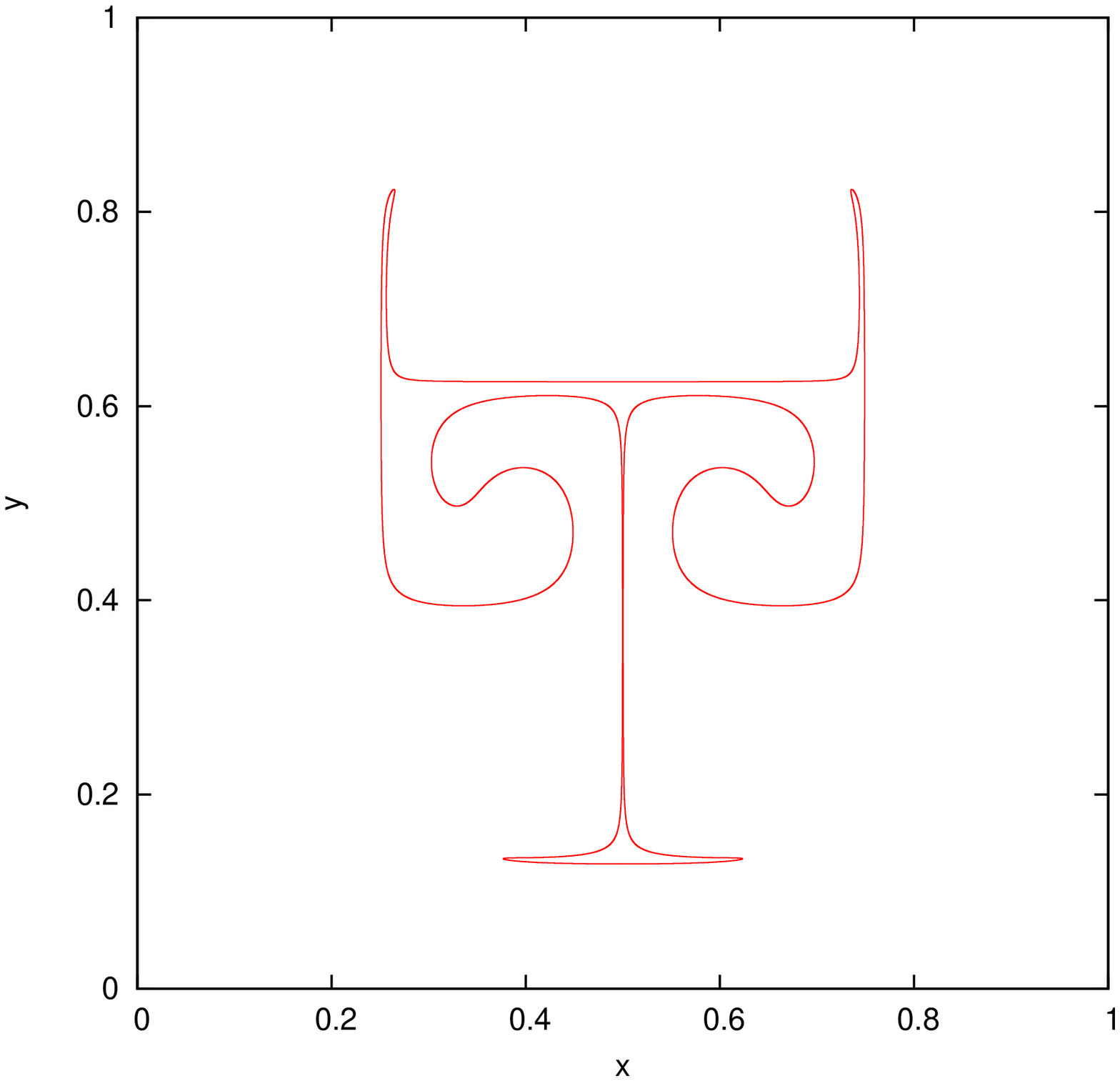}
}\\
$ t = \frac{1}{4} T $
\end{minipage}\\
\begin{minipage}[t][0.23\textheight][t]{0.5\linewidth}
\centering
\rotatebox{270}{
\includegraphics[height=0.2\textheight]{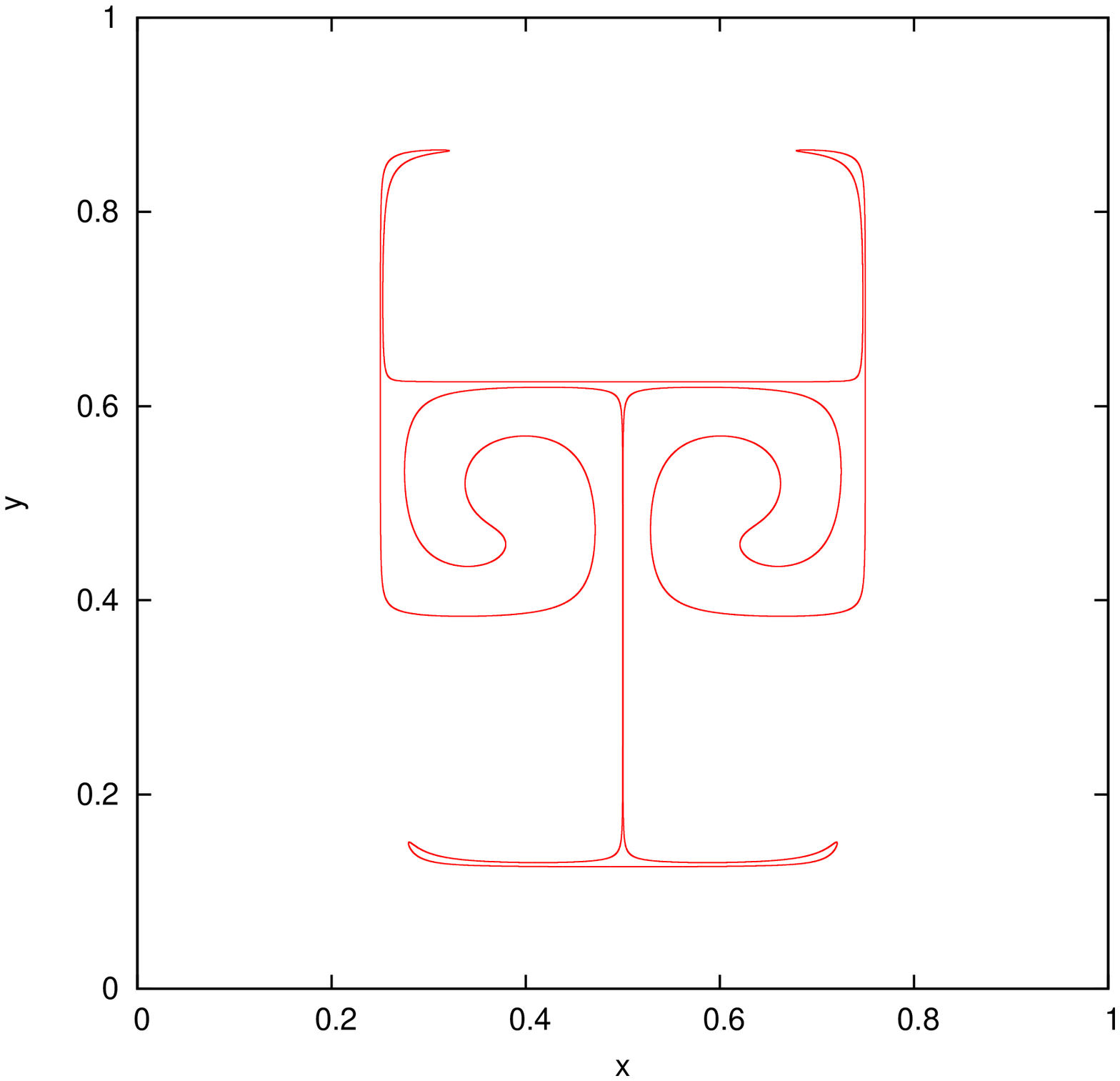}
}\\
$ t = \frac{3}{8} T $
\end{minipage} \nolinebreak
\begin{minipage}[t][0.23\textheight][t]{0.5\linewidth}
\centering
\rotatebox{270}{
\includegraphics[height=0.2\textheight]{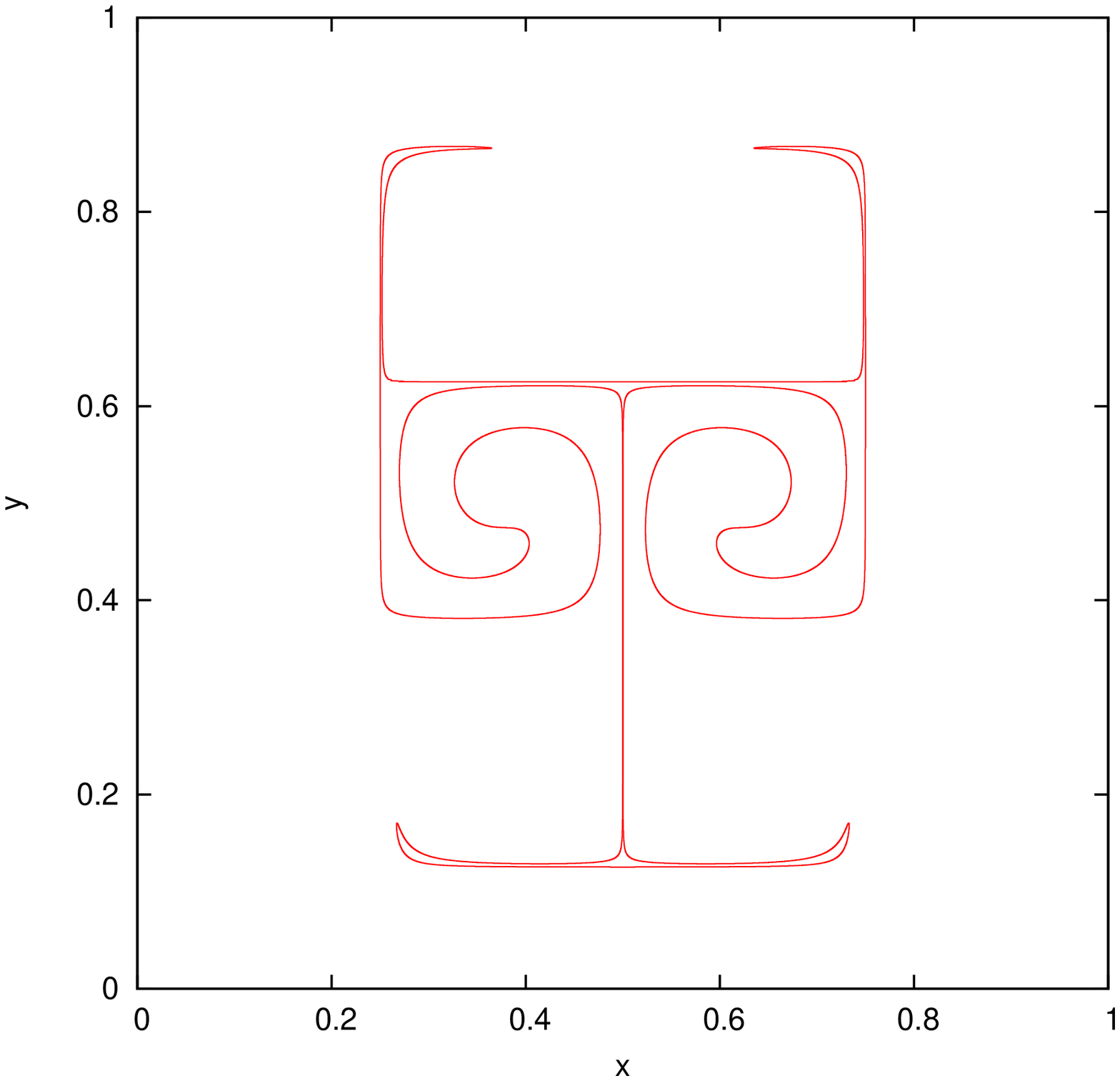}
}\\
$ t = \frac{1}{2} T $
\end{minipage}\\
\begin{minipage}[t][0.23\textheight][t]{0.5\linewidth}
\centering
\rotatebox{270}{
\includegraphics[height=0.2\textheight]{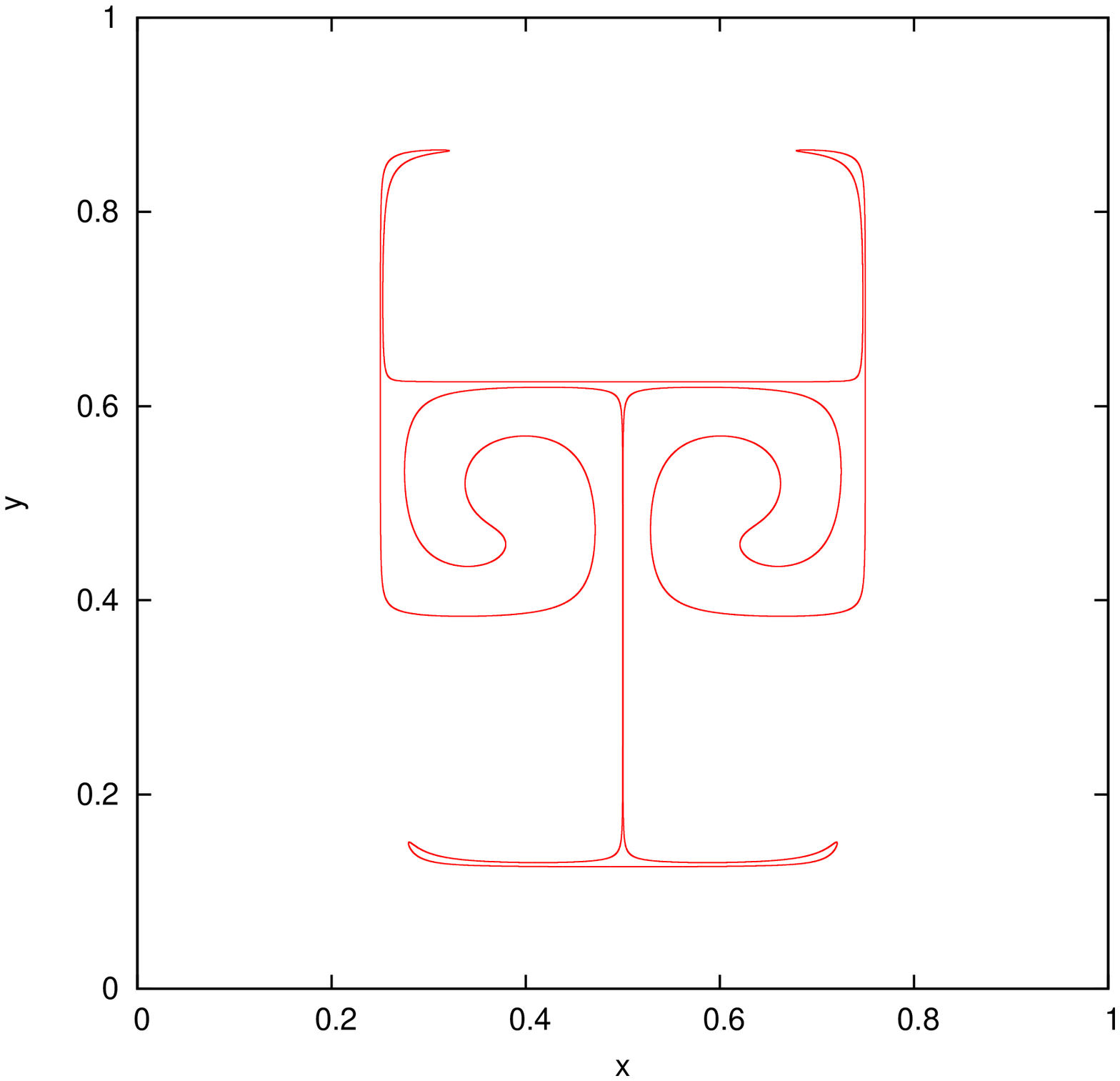}
}\\
{$ t = \frac{5}{8} T $}
\end{minipage} \nolinebreak
\begin{minipage}[t][0.23\textheight][t]{0.5\linewidth}
\centering
\rotatebox{270}{
\includegraphics[height=0.2\textheight]{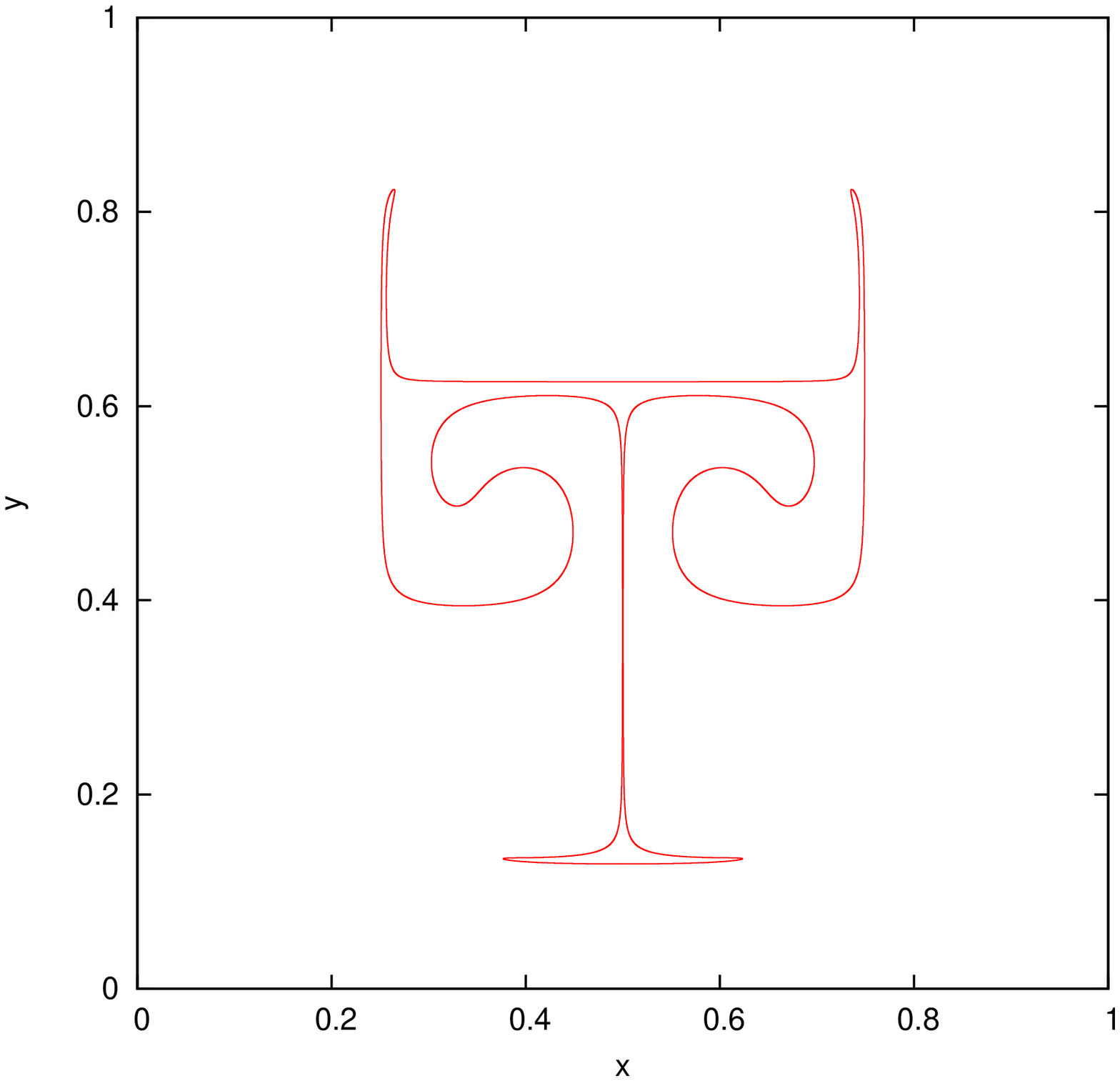}
}\\
$ t = \frac{3}{4} T $
\end{minipage}\\
\begin{minipage}[t][0.23\textheight][t]{0.5\linewidth}
\centering
\rotatebox{270}{
\includegraphics[height=0.2\textheight]{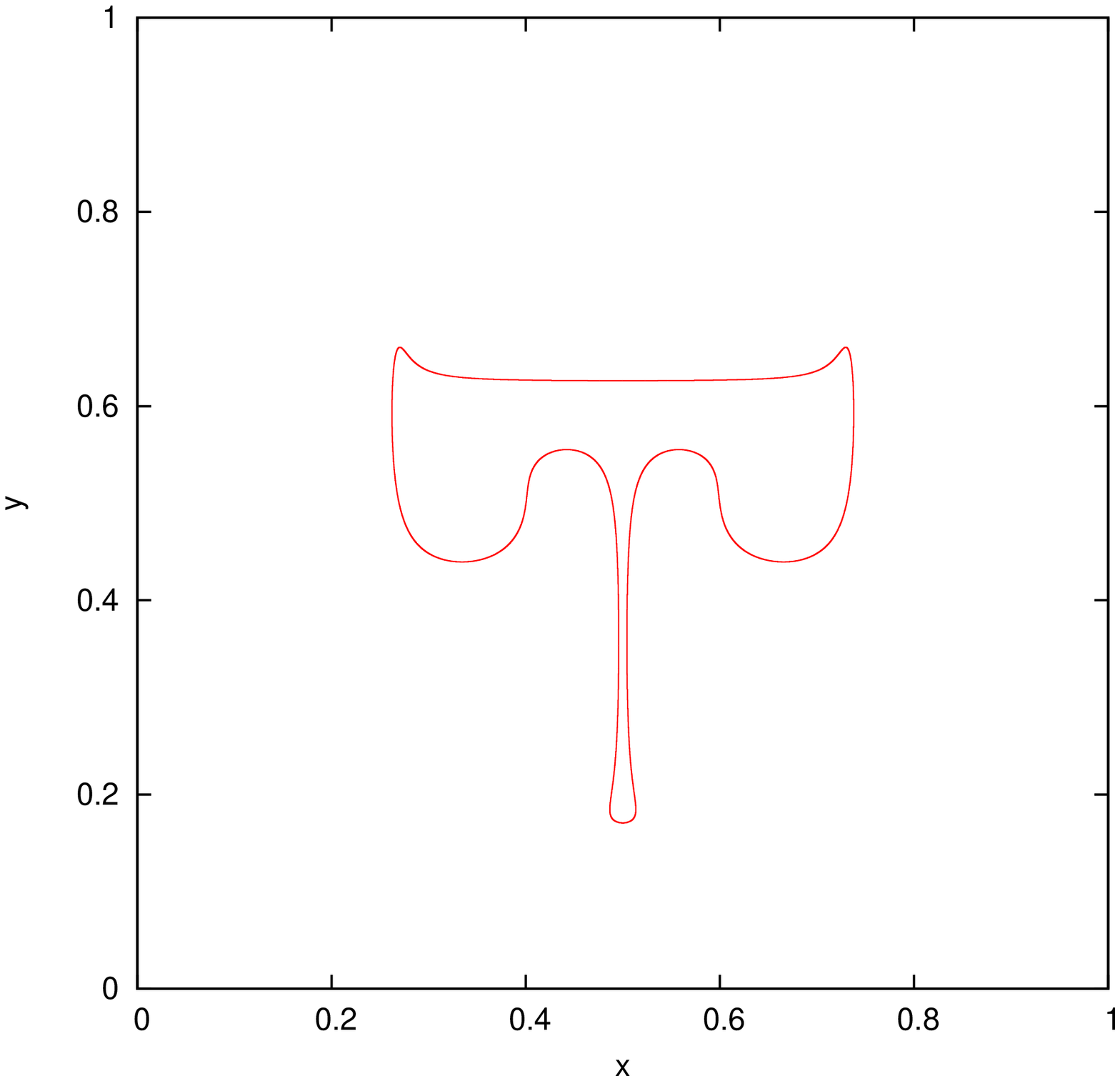}
}\\
$ t = \frac{7}{8} T $
\end{minipage} \nolinebreak
\begin{minipage}[t][0.23\textheight][t]{0.5\linewidth}
\centering
\rotatebox{270}{
\includegraphics[height=0.2\textheight]{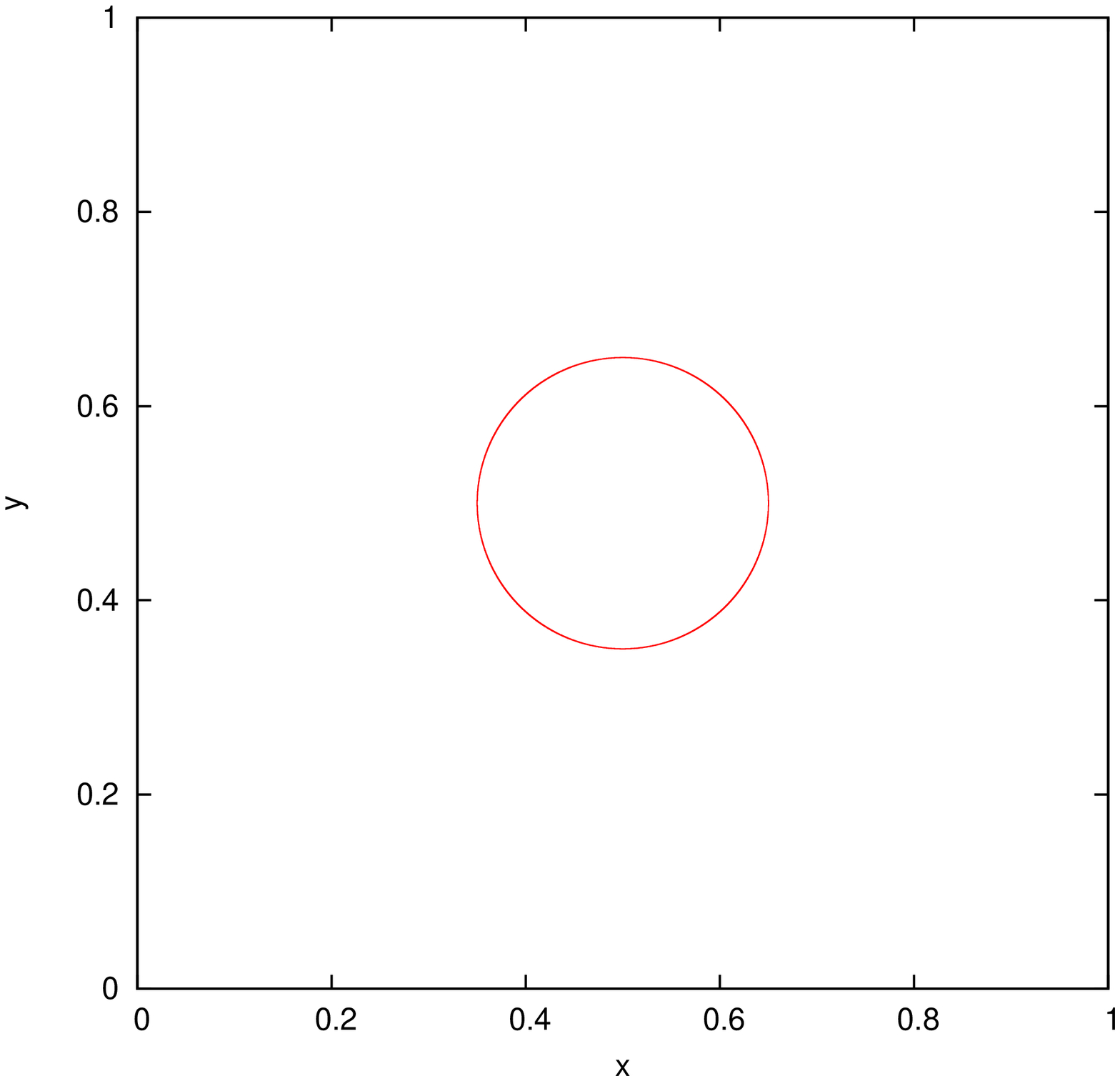}
}\\
$ t =  T $
\end{minipage}\\
\caption{Results of the present method for the deformation field test at 
different points in time, here $ N = 5000 $ and $ P = 5 $. }
\label{fig:deformationLoop}
\end{figure}

\begin{figure}
\centering
\rotatebox{270}{\includegraphics[height=10cm]{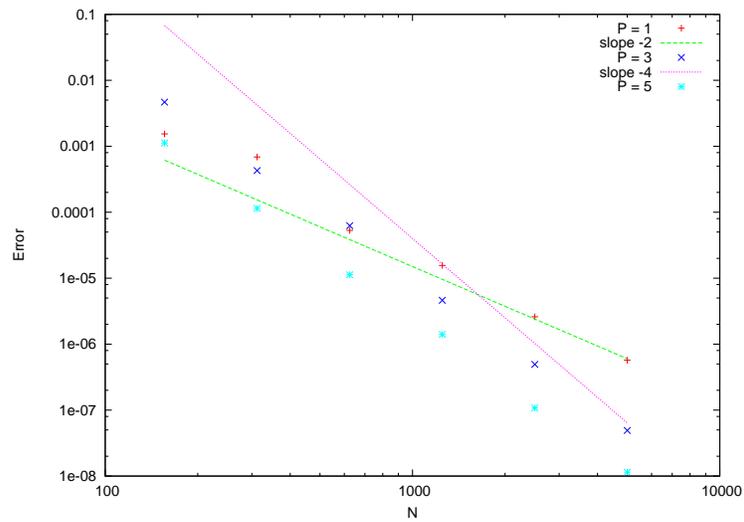}}
\caption{Error decrease for the deformation field test. 
The poorer convergence might be due 
to the appearance of spurious oscillations at regions of low
resolution. }
\label{fig:deformationError}
\end{figure}

The deformation field test \cite{Smolarkiewicz1982} 
uses the following stream function:
\be
\psi(x,y,t) = \frac{1}{n\pi} \cos \left( \frac{ \pi t}{T} \right)
\sin \left( n \pi \left( x + \frac{1}{2} \right) \right) 
\cos \left( n \pi \left( y + \frac{1}{2} \right) \right), 
\ee
where $ n $ is the number of vortexes in the computation domain and
chosen to be $ n = 4 $ to match the geometry used in \cite{ZhangLiu2008,VerschaevePart1},
as was the period with $ T = 2 $. 
The computational domain is a square box of side length one. A
drop of radius $ r_0 = 0.15 $ has its center at 
$ ( x_0 , y_0 ) = (0.5 , 0.5 ) $ at 
time $ t = 0 $. Since the flow is reversed after $ T/2 $ the drop 
returns to its initial position at $ t = T $ assuming its initial shape. 
As initial condition we used the same functions $ \alpha $, resp. $ \beta $, 
as for the Rider-Kothe single vortex test, equations
(\ref{eq:initialConditionA}-\ref{eq:initialConditionB}).
We performed a series of tests with $ P = 1$, $ P = 3 $ and $ P = 5 $. The 
time steps were chosen $ \Delta t = 3.906 \times 10^{-3} $ for $ P = 1 $ and 
$ P = 3 $ and
$ \Delta t = 9.76 \times 10^{-4 } $ for $ P = 4 $. This was, as before,
done in order to make the error contribution by the advection step subdominant. 
This benchmark test is relatively difficult since the interface develops 
regions with very small local radii of curvature. In addition, at some parts
the drop becomes very thin. The results of the present method applied
on this benchmark test can be seen in figures \ref{fig:deformationP1} for
$ P = 1 $, \ref{fig:deformationP3} for
$ P = 3 $, and \ref{fig:deformationP5} for $ P= 5 $. In these
figures we depicted the graph of the interface at maximum deformation $ t = T/2 $ and after the drop has returned to its initial position at $ t = T $ for 
different resolutions. For low resolutions the graph at maximum deformation 
($ t = T/2 $) does only capture the coarse features of the 
solution for all three orders $ P = 1 $, 
$ P = 3 $ and $ P = 5 $. In addition, at regions were the
resolution is low but the curvature high, the numerical solution seems
to develop a kind of Gibbs phenomenon for
$ P = 3 $ and $ P = 5$, as for Zalesak's slotted disk rotation
test. These oscillations become smaller as the resolution increases. 
When the drop has returned to its initial position we observe that the 
interface develops spikes at those regions where the resolution was low at 
maximum deformation. This is due to the fact that as the error is larger in 
these regions a point might fall into the wrong vortex and be traced to a
different location. These spikes become smaller with increasing resolution
$ N $ and increasing order of the interpolating B-spline $ P $. For finer
resolutions, $ N = 1250 $, 
the final position is extremely close to the exact solution. However,
at maximum deformation some wiggles can be observed for both $ P = 3 $ and $ P = 5 $, indicating that measuring the error at maximum deformation might give
a better estimate of the accuracy of the present method
than measuring it at the final position. For $ N = 5000 $ and
$ P = 5 $ the wiggles have disappeared, as can be observed from figure 
\ref{fig:deformationLoop}. Concerning the order of convergence, cf. 
figure \ref{fig:deformationError} and table \ref{tab:deformationError},
the present method seems to converge at a lower speed both for $ P = 3 $ 
and for $ P = 5 $. This
might have its origin in the spurious oscillations which might prevent
the method from converging at the right rate. 
Grid adaptation or remeshing redistributing the points to regions were
needed, as mentioned in section \ref{sec:interpolation}, might
be advantageous for this benchmark test. Nevertheless, an 
order of convergence between three and four 
with respect to the spatial resolution 
for this benchmark test is quite acceptable. 

\begin{table}
\centering
\[
\begin{array}{l|l|l|l}
P & N  & \mbox{Error} & O \\
\hline
1 & 156&1.53946\times 10^{-3} & -\\
  & 312&6.86122\times 10^{-4} & 1.17 \\
  & 625&5.30438\times 10^{-5} & 3.69 \\
  &1250&1.56053\times 10^{-5} & 1.77 \\
  &2500&2.59629\times 10^{-6} & 2.59 \\
  &5000&5.70028\times 10^{-7} & 2.19 \\
\hline
3 & 156&4.67185\times 10^{-3} & - \\
  & 312&4.27590\times 10^{-4} &3.45\\
  & 625&6.24600\times 10^{-5} &2.76\\
  &1250&4.61800\times 10^{-6} &3.75\\
  &2500&4.93238\times 10^{-7} &3.23\\
  &5000&4.91313\times 10^{-8} &3.33\\
\hline
5 & 156&1.11787\times 10^{-3}  & - \\
  & 312&1.14428\times 10^{-4} &3.29\\
  & 625&1.12848\times 10^{-5} &3.34\\
  &1250&1.40122\times 10^{-6} &3.01\\
  &2500&1.07634\times 10^{-7} &3.70\\
  &5000&1.14465\times 10^{-8} &3.23
\end{array}
\]
\caption{Results for deformation field test for $P = 1$,$P = 3 $ and
$ P = 5 $.}
\label{tab:deformationError}
\end{table}

\clearpage

\section{Conclusions \label{sec:conclusions} }

In the present discussion we derived an alternative formulation for
the interface representation for the volume of fluid method. The
interface is represented in a periodic fashion by two functions 
$ \alpha $ and $ \beta $, from which the position of the 
interface can be calculated. These two functions are approximated
by periodic B-spline interpolation which allows a systematic extension 
to higher order accuracy with respect to the grid spacing. 
The advection scheme has been simplified and extended to higher order
accuracy with respect to the time step. Numerical verification 
indicates that the present scheme has indeed the 
order of convergence predicted by the theory.
This allows for very accurate simulations with a limited number of 
knots. However, if the sampling
rate is too small, the present scheme can break down, providing 
a numerical solution far away from the exact one. In addition, for 
discontinuities in the first derivatives at a point on the 
interface, or at regions of the interface
with poor resolution and high curvature, the present method can display 
a kind of Gibbs phenomenon. Taking a lower order B-spline 
interpolation $ P = 1 $ can in this case provide more appealing results. 
A remeshing or adaptive grid approach could increase the efficiency of 
the algorithm. In addition, such an approach might 
furnish a way to simulate topological changes such as coalescence or 
drop break up. The extension to three dimensions is also left for future
research. 

\section{Acknowledgment}

Thanks goes to Claudio Walker for interesting discussions. 
The author is grateful to Bernhard M\"uller for guidance and supervision. 

%\bibliographystyle{abbrv}
%\bibliography{LBM}

\end{document}